\documentclass[prb,preprint,aps,showpacs,superscriptaddress,amsmath,amssymb]{revtex4-1}

\usepackage{amsthm}

\usepackage[pdftex]{graphicx}
\usepackage{color}
\definecolor{dred}{rgb}{0.7,0.0,0.0}
\usepackage{dcolumn}
\usepackage{bm}
\allowdisplaybreaks 
\usepackage{braket}
\usepackage{ulem} 
\usepackage{hyperref}
\hypersetup{colorlinks=true, linkcolor=blue, citecolor=blue, filecolor=blue, urlcolor=blue, pdftitle=, pdfauthor=, pdfsubject=, pdfkeywords=}
\usepackage{amsmath,amssymb,amstext}

\newcommand{\bs}[1]{\ensuremath{\boldsymbol{#1}}}

\definecolor{orange}{rgb}{1,0.5,0}
\definecolor{black}{rgb}{0,0,0}

\newcommand{\pdag}{\phantom{\dagger}}

\newcommand{\be}{\begin{equation}}
\newcommand{\ee}{\end{equation}}
\newcommand{\bea}{\begin{eqnarray}}
\newcommand{\eea}{\end{eqnarray}}

\newcommand{\beq}{\begin{equation}}
\newcommand{\eneq}{\end{equation}}

\begin{document}
%
%
%

\title{Topological Superconductors and Category Theory}

\vskip 10pt
\author{
Andrei Bernevig}
\address{
Department of Physics, Princeton University, Princeton, New Jersey 08544, USA
}
\author{
Titus Neupert}
\address{
 Princeton Center for Theoretical Science, Princeton University, Princeton, New Jersey 08544, USA
}

\date{\today}

\begin{abstract}
We give a pedagogical introduction to topologically ordered states of matter, with the aim of familiarizing the reader with their axiomatic topological quantum field theory description. We introduce basic noninteracting topological phases of matter protected by symmetries, including the Su-Schrieffer-Heeger model and the one-dimensional $p$-wave superconductor. The defining properties of topologically ordered states are illustrated explicitly using the toric code and -- on a more abstract level -- Kitaev's 16-fold classification of two-dimensional topological superconductors. Subsequently, we present a short review of category theory as an axiomatic description of topological order in two-dimensions. Equipped with this structure, we revisit Kitaev's 16-fold way.
\vspace{1.5cm}
\\
These lectures were in parts held at:  
\begin{itemize}
\item
Les Houches Summer School ``Topological Aspects of Condensed Matter Physics", 
4--29~August 2014,
\'{E}cole de Physique des Houches, Les Houches, France
\item
XVIII Training Course in the Physics of Strongly Correlated Systems, 
6--17 October 2014,
International Institute for Advanced Scientific Studies, Vietri sul Mare, Italy
\item
7th School on Mathematical Physics ``Topological Quantum Matter: From Theory to Applications",
25--29 May 2015, 
Universidad de los Andes, Bogot\'{a}, Colombia
\end{itemize}
\end{abstract}

\maketitle

\tableofcontents

\newpage

\section{Introduction to topological phases in condensed matter}

\subsection{The notion of topology}

In these lectures we will learn how to categorize and characterize some phases of matter that have topological attributes. A topological property of a phase, such as boundary modes (in an open geometry), topological response functions, or the character of its excitations, is described by a set of quantized numbers, related to so-called topological invariants of the phase. The quantization immediately implies that topological properties are \emph{universal} (they can be used to label the topological phase) and in some sense \emph{protected}, because they cannot change smoothly when infinitesimal perturbations are added. 
Topological properties, in the sense that we want to discuss them here, can only be defined for 
\begin{itemize}
\item spectrally \emph{gapped} ground states on a manifold without boundary of
\item \emph{local} Hamiltonians at
\item \emph{zero temperature}.
\end{itemize}
The spectral gap allows to define an equivalence class of states, i.e., a phase, with the help of the adiabatic theorem. Two gapped ground states are in the same phase if there exists an adiabatic interpolation between their respective Hamiltonians, such that the spectral gap above the ground state as well as the locality is preserved for all Hamiltonians along the interpolation.

Often it is useful to further modify these rules to define topological phases that are subject to symmetry constraints. We refer to topological states as being protected/enriched by a symmetry group $G$, if the Hamiltonian has a symmetry $G$ and only $G$-preserving interpolations are allowed. Since the $G$-preserving interpolations are a subset of all local interpolations, it is clear that symmetries make a topological classification of Hamiltonians more refined.  

The locality of a Hamiltonian is required to guarantee the quantization of topological response functions and to distinguish topological characterizations depending on the dimensionality of space. If we were not to impose locality, any system could in essence be zero-dimensional and there would be no notion of boundary states (which are localized over short distances) or point-like and line-like excitations etc.

Equipped with this definition of a topological phase, the exploration of topological states of matter above all poses a classification problem. We would like to know how many phases of quantum systems exists, that can be distinguished by their topological properties. We would like to obtain such a classification while imposing any symmetry $G$ that is physically relevant, such as time-reversal symmetry, space-group or point-group symmetries of a crystal, particle-number conservation etc. To identify the right mathematical tools that allow for such a classification and to guarantee its completeness is a subject of ongoing research. Here, we shall focus on aspects of this classification problem, which are well established and understood.

Most fundamental is a distinction between two types of topological states of matter: Those with \emph{intrinsic (long-range entangled) topological order}~\cite{Wen9095} and those without. This notion is also core to the structure of these lecture notes. In this Section, we only discuss phases without intrinsic topological order, while the ensuing two Sections are devoted to states with intrinsic topological order. A definition of intrinsic topological order can be based on several equivalent characterizations of such a phase, of which we give three:
\begin{itemize}
\item Topological ground state degeneracy: On a manifold without boundary, the degeneracy of gapped topologically degenerate ground states depends on the topological properties of the manifold. There are no topologically degenerate ground states if the system is defined on a sphere. The matrix elements of any local operator taken between two distinct topologically degenerate ground states vanishes. 
\item Fractionalized excitations: 
There exist low-energy excitations which are point-like [in two dimensions (2D) or above] or line-like [in three dimensions (3D) or above]. These excitations carry a fractional quantum numbers as compared to the microscopic degrees of freedom that enter the Hamiltonian (for example, a fractional charge), are deconfined and dynamical (i.e, free to move in the low-energy excited states). 
\item Topological entanglement entropy: 
The entanglement entropy between two parts of a system that is in a gapped zero-temperature ground state typically scales with the size of the line/surface that separates the two regions (``area-law entanglement"). Topologically ordered, long-range entangled states have a universal subleading correction to this scaling that is characteristic for the type of topological order. 
\end{itemize}
(Note that these statements, as many universal properties we discuss, are only strictly true in the thermodynamic limit of infinite system size. For example, in a finite system, the ground state degeneracy is lifted by an amount that scales exponential in the system size.) As fractionalized excitations in the above sense may only exist in two or higher dimensions, intrinsic topological order cannot be found in one-dimensional (1D) phases of matter. Further, for intrinsic topological order to occur, interactions are needed in the system.  

Examples of topologically nontrivial phases (both with and without intrinsic topological order) exist in absence of any symmetry. However, most of the phases without intrinsic topological order belong to the so-called symmetry protected topological (SPT) phases. In these cases, the topology is protected by a symmetry. These phases almost always possess topologically protected boundary modes when defined on a manifold with boundary, except if the boundary itself breaks the protecting symmetry (as could be the case with inversion symmetry, for example). 

In contrast, phases with intrinsic topological order are not necessarily equipped with boundary modes, even if the boundary of the manifold preserves the defining symmetries of the phase. If the definition of a phase with intrinsic topological order relies on symmetries, it is named symmetry enriched topological phase (SET).

An alternative characterization of topological properties of a phase uses the entanglement between different subsystems. While we opt not to touch upon this concept here, we want to make contact to the ensuing terminology: All phases with intrinsic topological order are called long-range entangled (LRE). The term short-range entangled (SRE) phase is often used synonymously with ``no intrinsic topological order". (Some authors count 2D phases with nonvanishing thermal Hall conductivity, such as the $p+\mathrm{i}p$ superconductors, but no intrinsic topological order unless gauged, as LRE.) 

In these lecture notes, we will encounter two classifications of a subset of topological phases. The following Subsection introduces the complete classification of non-interacting fermionic Hamiltonians with certain symmetries (which have no intrinsic topological order). Section~\ref{sec: Category theory} is concerned with the unified description of 2D phases with intrinsic topological order in absence of any symmetries.

\subsection{Classification of noninteracting fermion Hamiltonians: The 10-fold way}

We have stated that SPT order in SRE states
manifests itself via the presence of gapless boundary states in an open geometry. In fact, there exists a intimate connection between the topological character of the gapped bulk state and its boundary modes. The latter are protected against local perturbations on the boundary that (i) preserve the bulk symmetry and (ii) induce no intrinsic topological order or spontaneous symmetry breaking in the boundary modes. 
This bulk-boundary correspondence can be used to classify SPT phases. 
Two short-range entangled phases with the same symmetries belong to a different topological class, if the interface between the two phases hosts a state in the bulk gap and this state cannot be moved into the continuum of excited states by any local perturbation that obey (i) and (ii). Equivalently, to change the topological attribute of a gapped bulk state via any smooth changes in the Hamiltonian, the bulk energy gap has to close and reopen. 

In Ref.~\onlinecite{Schnyder08} Schnyder et al. use this bulk-boundary correspondence to classify all noninteracting fermionic Hamiltonians. 
For the topological phases that they discuss, two fundamental symmetries, particle-hole symmetry (PHS) and time-reversal symmetry (TRS), are considered. In the following, we will review the essential results of this classification.~\cite{Schnyder08, Ryu10, Kitaev09}

\subsubsection{Classification with respect to time-reversal and particle-hole symmetry}

Symmetries in quantum mechanics are operators that have to preserve the absolute value of the scalar product of any two vectors in the Hilbert space. They can thus be either unitary operators, preserving the scalar product, or antiunitary operators, turning the scalar product into its complex conjugate (up to a phase).
For a unitary operator to be a symmetry of a given Hamiltonian $H$, the operator has to commute with $H$. Consequently, the Hamiltonian can be block diagonalized, where each block acts on one eigenspace of the unitary symmetry. If $H$ has a unitary symmetry, we block-diagonalize it and then consider the topological properties of each block individually. This way, we do not have to include unitary symmetries (except for the product of TRS and PHS and the omnipresent particle number conservation) in the further considerations, as we will not focus on the burgeoning field of crystalline topological insulators.

\begin{subequations}
A fundamental antiunitary operator in quantum mechanics is the reversal of time $\mathcal{T}$.
Let us begin by recalling its elementary properties.
If a given Hamiltonian $H$ is TRS, that is, 
\begin{equation}
\mathcal{T} H\mathcal{T}^{-1}=+H,
\label{eq: TRS}
\end{equation}
the time-evolution operator at time $t$ should be mapped to the time-evolution operator at $-t$ by the operator $\mathcal{T}$
\begin{equation}
\begin{split}
\mathcal{T}e^{-\mathrm{i}tH}\mathcal{T}^{-1}
=&\,
e^{-\mathcal{T}\mathrm{i}\mathcal{T}^{-1}tH}\\
=&\,
e^{-\mathrm{i}(-t)H}.
\end{split}
\end{equation}
We conclude that the reversal of time is indeed an antiunitary operator $\mathcal{T}\mathrm{i}\mathcal{T}^{-1}=-\mathrm{i}$. It can be represented as $\mathcal{T}=T\mathcal{K}$, where $\mathcal{K}$ denotes complex conjugation and $T$ is a unitary operator. Applying the reversal of time twice on any state must return the same state up to an overall phase factor $e^{\mathrm{i}\phi}$ 
\begin{equation}
e^{\mathrm{i}\phi}\stackrel{!}{=}\mathcal{T}^2=T(T^{\mathsf{T}})^{-1}
\quad\Rightarrow\quad
T=e^{\mathrm{i}\phi}T^{\mathsf{T}}, \quad
T^{\mathsf{T}}=e^{\mathrm{i}\phi}T
.
\label{eq: TRS antiunitary}
\end{equation}
Inserting the two last equations into one another, one obtains $T=e^{2\mathrm{i}\phi} T$, i.e., $e^{2\mathrm{i}\phi}$ has to equal $+1$. We conclude that the time-reversal operator either squares to $+1$ or to $-1$
\begin{equation}
\mathcal{T}^2=+1,\qquad \qquad \mathcal{T}^2=-1.
\end{equation}

The second fundamental antiunitary symmetry considered here is charge conjugation $\mathcal{P}$. 
Its most important incarnation in solid state physics is found in the theory of superconductivity. In an Andreev reflection process, an electron-like quasi particle that enters a superconductor is reflected as a hole-like quasi particle. The charge difference between incident and reflected state is accounted for by adding one Cooper pair to the superconducting condensate. In the mean-field theory of superconductivity, the energies of the electron-like state and the hole-like state are equal in magnitude and have opposite sign, giving rise to the PHS. In this case, rather than being a fundamental physical symmetry of the system like TRS is, PHS emerges due to a redundancy in the mean-field description. 
We define a (single-particle) Hamiltonian $H$ to be PHS if 
\begin{equation}
\mathcal{P} H\mathcal{P}^{-1}=+H.
\label{eq: PHS}
\end{equation}
In order to also reverse the sign of charge, $\mathcal{P}$ has to turn the minimal coupling $\bs{p}-\mathrm{i}e\bs{A}$ into $\bs{p}+\mathrm{i}e\bs{A}$, where $\bs{p}$ is the momentum operator and $\bs{A}$ is the electromagnetic gauge potential.
This is achieved by demanding $\mathcal{P}\mathrm{i}\mathcal{P}^{-1}=-\mathrm{i}$. We conclude that $\mathcal{P}$ is indeed an antiunitary operator that can be decomposed as $\mathcal{P}=P\mathcal{K}$, where $P$ is a unitary operator. As a consequence, the reasoning of Eq.~\eqref{eq: TRS antiunitary} also applies to $\mathcal{P}$
and we conclude that the charge conjugation operator either squares to $+1$ or to $-1$
\begin{equation}
\mathcal{P}^2=+1,\qquad \qquad \mathcal{P}^2=-1.
\end{equation}

In the case where the operators $\mathcal{T}$ and $\mathcal{P}$ are both symmetries of $H$, their product is also a symmetry of $H$. 
We call this product chiral transformation $C:=\mathcal{T}\mathcal{P}$. It is a unitary operator. 
The Hamiltonian $H$ transforms under the chiral symmetry as 
\begin{equation}
C HC^{-1}=+H.
\end{equation}
\end{subequations}
(It is important to note that both $\mathcal{P}$ and $C$ anticommute rather than commute with the single-particle first-quantized Hamiltonian $\mathcal{H}_{\alpha,\alpha'}$ that we will introduce below.)
Observe that a Hamiltonian can have a chiral symmetry, even if it possesses neither of PHS and TRS.
We can now enumerate all combinations of the symmetries $\mathcal{P}$, $\mathcal{T}$, and $C$ that a Hamiltonian can obey, accounting for the different signs of $\mathcal{T}^2$ and $\mathcal{P}^2$. There are in total ten such symmetry classes, listed in Tab.~\ref{table: symm classes}.
The main result of Schnyder et al. in Ref.~\onlinecite{Schnyder08} is to establish how many distinct phases with protected edge modes exist on the $(d-1)$-dimensional boundary of a phase in $d$ dimensions. We find three possible cases: If there is only one (topologically trivial) phase, the entry $\varnothing$ is found in Tab.~\ref{table: symm classes}. If there are exactly two distinct phases (one trivial and one topological phase), $\mathbb{Z}_2$ is listed. Finally, if there exists a distinct topological phase for every integer, $\mathbb{Z}$ is listed.

\begin{table}[t]
\caption{
Symmetry classes of noninteracting fermionic Hamiltonians from Refs.~\onlinecite{Ryu10} and~\onlinecite{Kitaev09}.
The columns contain from left to right: Cartan's name for the symmetry class;
the square of the time reversal operator, the particle-hole operator, and the chiral operator ($\varnothing$ means the symmetry is not present); the group of topological phases that a Hamiltonian with the respective symmetry can belong to for the dimensions $d=1,\cdots,8$ of space.
The first two rows are called ``complex classes'', while the lower eight rows are the ``real classes''. The homotopy groups of the former show a periodicity with period 2 in $d$, while those of the latter have a period 8 in $d$ (Bott periodicity). 
}

\begin{center} 
\begin{tabular}{| l | c c c  | c c c c c c c c c |}
\hline

&$\mathcal{T}^2$ &$\mathcal{P}^2$&$C^2$&d&
$1$&$2$&$3$&$4$&$5$&$6$&$7$&$8$\\
\hline
\hline

A
&$\varnothing$ &$\varnothing$&$\varnothing$&
&
$\varnothing$&$\mathbb{Z}^{\ }$&$\varnothing$&$\mathbb{Z}^{\ }$&$\varnothing$&$\mathbb{Z}^{\ }$&$\varnothing$&$\mathbb{Z}^{\ }$\\
\hline
AIII
&$\varnothing$ &$\varnothing$&$+$&
&
$\mathbb{Z}^{\ }$&$\varnothing$&$\mathbb{Z}^{\ }$&$\varnothing$&$\mathbb{Z}^{\ }$&$\varnothing$&$\mathbb{Z}^{\ }$&$\varnothing$\\
\hline
\hline

AII
&$-$ &$\varnothing$&$\varnothing$&
&
$\varnothing$&$\mathbb{Z}^{\ }_2$&$\mathbb{Z}^{\ }_2$&$\mathbb{Z}^{\ }$&$\varnothing$&$\varnothing$&$\varnothing$&$\mathbb{Z}^{\ }$\\
\hline

DIII
&$-$ &$+$&$+$&
&
$\mathbb{Z}^{\ }_2$&$\mathbb{Z}^{\ }_2$&$\mathbb{Z}^{\ }$&$\varnothing$&$\varnothing$&$\varnothing$&$\mathbb{Z}^{\ }$&$\varnothing$\\

D
&$\varnothing$ &$+$&$\varnothing$&
&
$\mathbb{Z}^{\ }_2$&$\mathbb{Z}^{\ }$&$\varnothing$&$\varnothing$&$\varnothing$&$\mathbb{Z}^{\ }$&$\varnothing$&$\mathbb{Z}^{\ }_2$\\

BDI
&$+$ &$+$&$+$&
&
$\mathbb{Z}^{\ }$&$\varnothing$&$\varnothing$&$\varnothing$&$\mathbb{Z}^{\ }$&$\varnothing$&$\mathbb{Z}^{\ }_2$&$\mathbb{Z}^{\ }_2$\\
\hline

AI
&$+$ &$\varnothing$&$\varnothing$&
&
$\varnothing$&$\varnothing$&$\varnothing$&$\mathbb{Z}^{\ }$&$\varnothing$&$\mathbb{Z}^{\ }_2$&$\mathbb{Z}^{\ }_2$&$\mathbb{Z}^{\ }$\\
\hline

CI
&$+$ &$-$&$+$&
&
$\varnothing$&$\varnothing$&$\mathbb{Z}^{\ }$&$\varnothing$&$\mathbb{Z}^{\ }_2$&$\mathbb{Z}^{\ }_2$&$\mathbb{Z}^{\ }$&$\varnothing$\\

C
&$\varnothing$ &$-$&$\varnothing$&
&
$\varnothing$&$\mathbb{Z}^{\ }$&$\varnothing$&$\mathbb{Z}^{\ }_2$&$\mathbb{Z}^{\ }_2$&$\mathbb{Z}^{\ }$&$\varnothing$&$\varnothing$\\

CII
&$-$ &$-$&$+$&
&
$\mathbb{Z}^{\ }$&$\varnothing$&$\mathbb{Z}^{\ }_2$&$\mathbb{Z}^{\ }_2$&$\mathbb{Z}^{\ }$&$\varnothing$&$\varnothing$&$\varnothing$\\

\hline
\end{tabular}
\end{center}
\label{table: symm classes} 
\end{table}

\subsubsection{Flatband Hamiltonians and homotopy groups }
\label{sec: homotopy groups}

There are several approaches to obtain the entries $\mathbb{Z}_2$ and $\mathbb{Z}$ in Tab.~\ref{table: symm classes}.
For one, the theory of Anderson localization can be employed to determine in which spatial dimensions boundaries can host localization-protected 
states (the topological surface states) under a given symmetry. This was done by Schnyder et al.\ in Ref.~\onlinecite{Schnyder08}.
Kitaev, on the other hand, derived the table using the algebraic structure of Clifford algebras in the various dimensions and symmetry classes.~\cite{Kitaev09} In mathematics, this goes under the name K-theory. 

Here, we want to give a flavor of the mathematical structure behind the table by considering two examples. 
To keep matters simple, we shall restrict ourselves to the situation where the system is translationally invariant and periodic boundary conditions are imposed. 
\begin{subequations}
In second quantization, the Hamiltonian $H$ has the Bloch representation
\begin{equation}
H=\int \mathrm{d}^d\bs{k} \psi^\dagger_{\alpha}(\bs{k})\mathcal{H}_{\alpha,\alpha'}(\bs{k}) \psi^{\ }_{\alpha'}(\bs{k}),
\end{equation} 
where 
$\psi^\dagger_{\alpha}(\bs{k})$
creates a fermion of flavor $\alpha=1,\cdots,N$ at momentum $\bs{k}$ in the Brillouin zone (BZ) and the summation over $\alpha$ and $\alpha'$ is implicit.
The flavor index may represent orbital, spin, or sublattice degrees of freedom. 
Energy bands are obtained by diagonalizing the $N\times N$ matrix $\mathcal{H}(\bs{k})$ at every momentum $\bs{k}\in \mathrm{BZ}$ with the aid of a unitary transformation $U(\bs{k})$
\begin{equation}
U^\dagger(\bs{k})
\mathcal{H}(\bs{k})
U(\bs{k})
=\mathrm{diag}\left[\varepsilon^{\ }_{m+n}(\bs{k}),\cdots, \varepsilon^{\ }_{n+1}(\bs{k}), 
\varepsilon^{\ }_{n}(\bs{k}),\cdots,\varepsilon^{\ }_{1}(\bs{k})\right],
\label{eq: diagonalize H with unitary}
\end{equation}
\end{subequations}
where the energies are arranged in descending order on the righthand side and $n,m\in \mathbb{Z}$ such that $n+m=N$. 
So as to start from an insulating ground state, we assume that there exists an energy gap between the bands $n$ and $n+1$
and that the chemical potential $\mu$ lies in this gap
\begin{equation}
\varepsilon^{\ }_{n}(\bs{k})<\mu<\varepsilon^{\ }_{n+1}(\bs{k}),\qquad \forall \bs{k}\in\mathrm{BZ}.
\end{equation} 
The presence of the gap allows us to adiabatically deform the Bloch Hamiltonian $\mathcal{H}(\bs{k})$
to the flatband Hamiltonian
\begin{subequations}
\begin{equation}
\mathcal{Q}(\bs{k})
:=
U(\bs{k})
\begin{pmatrix}
\openone^{\ }_{m}&0\\
0&-\openone^{\ }_{n}
\end{pmatrix}
U^\dagger(\bs{k})
\label{eq: flatband projector H}
\end{equation}
\end{subequations}
that assigns the energy $-1$ and $+1$ to all states in the bands below and above the gap, respectively.
This deformation preserves the eigenstates, but removes the nonuniversal information about energy bands from the Hamiltonian.

In other words, the degenerate eigenspaces of the eigenvalues $\pm1$ of $\mathcal{Q}(\bs{k})$ reflect the partitioning of the single-particle Hilbert space introduced by the spectral gap in the spectrum of $\mathcal{H}(\bs{k})$.
The degeneracy of its eigenspaces equips $\mathcal{Q}(\bs{k})$ with an extra $\mathrm{U}(n)\times \mathrm{U}(m)$ gauge symmetry:
While the $(n+m)\times(n+m)$ matrix $U(\bs{k})$ of Bloch eigenvectors that diagonalizes $\mathcal{Q}(\bs{k})$ is an element
of $\mathrm{U}(n+m)$ for every $\bs{k}\in \mathrm{BZ}$, we are free to change the basis for its lower and upper bands by a $\mathrm{U}(n)$ and $\mathrm{U}(m)$ transformation, respectively. Hence $\mathcal{Q}(\bs{k})$ is an element of the space 
$C^{\ }_0:=\mathrm{U}(n+m)/[\mathrm{U}(n)\times\mathrm{U}(m)]$ defining a map
\begin{equation}
\mathcal{Q}:\quad  \mathrm{BZ}\to C^{\ }_{0}.
\end{equation}
The group of topologically distinct maps $\mathcal{Q}$, or, equivalently, the number of topologically distinct Hamiltonians $\mathcal{H}$, 
is given by the homotopy group
\begin{equation}
\pi^{\ }_d\left(C^{\ }_0\right)
\end{equation}
for any dimension $d$ of the BZ.
(The homotopy group is the group of equivalence classes of maps from the $d$-dimensional \emph{sphere} to a target space, in this case $C_0$. Even though the BZ is a $d$-dimensional \emph{torus}, it turns out that this difference between torus and sphere does not affect the classification as discussed here.)

For example, in $d=2$ we have $\pi^{\ }_2\left(C^{\ }_0\right)=\mathbb{Z}$.
A physical example of a family of Hamiltonians that exhausts the topological sectors of this group 
is found in the integer quantum Hall effect. The incompressible ground state with $r\in \mathbb{N}$ filled Landau levels is topologically distinct from the ground state with $\mathbb{N}\ni r'\neq r$ filled Landau levels. 
Two different patches of space with $r$ and $r'$ filled Landau levels have $|r-r'|$ gapless edge modes running at their interface, reflecting the bulk-boundary correspondence of the topological phases. 
In contrast, $\pi^{\ }_3\left(C^{\ }_0\right)=\mathbb{Z}^{\ }_1$ renders all noninteracting fermionic Hamiltonians in 3D space topologically equivalent to the vacuum, if no further symmetries besides the U(1) charge conservation are imposed.

As a second example, let us discuss a Hamiltonian that has only chiral symmetry and hence belongs to the symmetry class AIII.
The chiral symmetry implies a spectral symmetry of $\mathcal{H}(\bs{k})$. 
If gapped, $\mathcal{H}(\bs{k})$ must have an even number of bands $N=2n$, $n\in \mathbb{Z}$.
When represented in the eigenbasis of the chiral symmetry operator $C$, the spectrally flattened Hamiltonian $\mathcal{Q}(\bs{k})$ and the chiral symmetry operator have the representations
\begin{subequations}
\begin{equation}
\mathcal{Q}(\bs{k})=
\begin{pmatrix}
0&q(\bs{k})\\
q^\dagger(\bs{k})&0
\end{pmatrix},
\qquad
C=
\begin{pmatrix}
\openone^{\ }_n&0\\
0&-\openone^{\ }_n
\end{pmatrix},
\label{eq: def of q chiral}
\end{equation}
respectively.
From $\mathcal{Q}(\bs{k})^2=1$, one concludes that $q(\bs{k})$ can be an arbitrary unitary matrix. We are thus led to consider the homotopy group $\pi^{\ }_d(C^{\ }_1)$ of the mapping
\begin{equation}
q:\quad  \mathrm{BZ}\to C^{\ }_{1}=\mathrm{U}(n).
\end{equation}
\end{subequations}
For example, in $d=1$ spatial dimensions $\pi^{\ }_3(C^{\ }_1)=\mathbb{Z}$. A tight-binding model with non-trivial topology that belongs to this symmetry class will be discussed in 
Sec.~\ref{sec: SSH model}.

With these examples, we have discussed the two complex classes A and AIII.
In the real classes, which have at least one antiunitary symmetry, it is harder to obtain the constraints on the spectrally flattened Hamiltonian $\mathcal{Q}(\bs{k})$. The origin for this complication is that the antiunitary operators representing time-reversal and particle-hole symmetry relates $\mathcal{Q}(\bs{k})$ and $\mathcal{Q}(-\bs{k})$ rather than acting locally in momentum space.

\subsubsection{Topological invariants}
\label{sec: topological invariants}

Given a gapped noninteracting fermionic Hamiltonian with certain symmetry properties in $d$-dimensional space, one can use Tab.~\ref{table: symm classes} to conclude whether the system can potentially be in a topological phase. However, to understand in which topological sector the system is, we have to do more work. To obtain this information, one computes topological invariants or topological quantum numbers of the ground state. Such invariants are automatically numbers in the  group of possible topological phases ($\mathbb{Z}$ or $\mathbb{Z}_2$). For many of them, a variety of different-looking but equivalent representations are known. 

To give concrete examples, we shall discuss the invariants for all $\mathbb{Z}$ topological phases found in Tab.~\ref{table: symm classes}. These are called Chern numbers in the symmetry classes without chiral symmetry and winding numbers in the classes with chiral symmetry. 

In physics, topological attributes refer to \emph{global} properties of a physical system that is made out of \emph{local} degrees of freedom and might only have local, i.e., short-ranged, correlations. The distinction between global and local properties parallels the distinction between topology and geometry in mathematics, where the former refers to global structure, while the latter refers to local structure of objects. 
In differential geometry, a bridge between topology and geometry is given by the Gauss-Bonnet theorem. It states that for compact 2D Riemannian manifolds $M$ without boundary, the integral over the Gaussian curvature $F(\bs{x})$ of the manifold is (i) integer and (ii) a topological invariant
\begin{equation}
2(1-g)=\frac{1}{2\pi}\int_M \mathrm{d}^2\bs{x} \, F(\bs{x}).
\label{eq: Gauss-bonnet}
\end{equation}
Here, $g$ is the genus of $M$, e.g., $g=0$ for a 2D sphere and $g=1$ for a 2D torus.
The Gaussian curvature $F(\bs{x})$ can be defined as follows. Attach to every point on $M$ the tangential plane, a 2D vector space. Take some vector from the tangential plane at a given point on $M$ and parallel transport it around an infinitesimal closed loop on $M$. The angle mismatch of the vector before and after the transport is proportional to the Gaussian curvature enclosed in the loop. 

In the physical systems that we want to describe, the manifold $M$ is the BZ and the analogue of 
the tangent plane on $M$ is a space spanned by the Bloch states of the occupied bands at a given momentum $\bs{k}\in\mathrm{BZ}$. 
The Gaussian curvature of differential geometry is now generalized to a curvature form, called Berry curvature $\mathrm{F}$. In our case, it is given by an $n\times n$ matrix of differential forms that is defined via the Berry connection
 $\mathrm{A}$ as 
 \begin{subequations}
\begin{eqnarray}
\mathrm{F}&:=&F^{\ }_{ij}(\bs{k})\, \mathrm{d}k^{\ }_i\wedge \mathrm{d}k^{\ }_j\\
F^{\ }_{ij}(\bs{k})&:=&\partial^{\ }_{i}A^{\ }_j(\bs{k})-\partial^{\ }_{j}A^{\ }_i(\bs{k})+[A^{\ }_i(\bs{k}),A^{\ }_j(\bs{k})],
\qquad i,j=1,\cdots, d,
\\
\mathrm{A}&:=&A^{\ }_i(\bs{k})\, \mathrm{d}k^{\ }_i,
\\
A^{(ab)}_i(\bs{k})&:=&\sum_{\alpha=1}^NU^\dagger_{a\alpha}(\bs{k})\partial^{\ }_{i}U^{\ }_{\alpha b}(\bs{k}),
\qquad a,b=1,\cdots,n,\qquad i=1,\cdots, d.
\end{eqnarray}
\end{subequations}
(Two different conventions for the Berry connection are commonly used: Either it is purely real or purely imaginary. Here we choose the latter option.)
The unitary transformation $U(\bs{k})$ that diagonalizes the Hamiltonian was defined in Eq.~\eqref{eq: diagonalize H with unitary}, both $A^{\ }_i(\bs{k})$ and $F^{\ }_{ij}(\bs{k})$ are $n\times n$ matrices, 
we write $\partial^{\ }_i\equiv \partial/\partial k^{\ }_i$ and the sum over repeated spatial coordinate components $i,j$ is implicit.

\begin{subequations}
Under a local $\mathrm{U}(n)$ gauge transformation in momentum space that acts on the states of the lower bands and is parametrized by the $n\times n$ matrix $G(\bs{k})$
\begin{equation}
U^{\ }_{\alpha a}(\bs{k}) \longrightarrow
U^{\ }_{\alpha b}(\bs{k}) G^{\ }_{b a}(\bs{k}),
\qquad \alpha =1,\cdots, N,\qquad a=1,\cdots, n,
\label{eq: gauge trafo}
\end{equation}
the Berry connection $\mathrm{A}$ changes as
\begin{equation}
\mathrm{A}\longrightarrow G^\dagger \mathrm{A} G + G^\dagger \mathrm{d} G,
\end{equation}
while the Berry curvature $\mathrm{F}$ changes covariantly
\begin{equation}
\mathrm{F}\longrightarrow G^\dagger \mathrm{F} G,
\end{equation}
leaving its trace invariant.
\end{subequations}

\paragraph{Chern numbers}
For the spatial dimension $d=2$, the generalization of the Gauss-Bonnet theorem~\eqref{eq: Gauss-bonnet} in algebraic topology was found by Chern to be
\begin{equation}
\begin{split}
2 \mathsf{C}^{(1)}:=&\,\frac{\mathrm{i}}{2\pi}\int\limits_{\mathrm{BZ}}\,\mathrm{tr}\,\mathrm{F}\\
=&\,2\frac{\mathrm{i}}{2\pi}\int\limits_{\mathrm{BZ}}\,\mathrm{d}^2\bs{k}\,\mathrm{tr}\,F^{\ }_{12}.
\end{split}
\label{eq: Chern number 1}
\end{equation}
This defines a gauge-invariant quantity, the first Chern number $\mathsf{C}^{(1)}$. Remarkably, $\mathsf{C}^{(1)}$ can only take integer values.
In order to obtain a topological invariant for any even dimension $d=2s$ of space, we can use the $s$-th power of the local Berry curvature form $\mathrm{F}$ (using the wedge product) to build a gauge invariant $d$-form that can be integrated over the BZ to obtain scalar. Upon taking the trace, this scalar is invariant under the gauge transformation~\eqref{eq: gauge trafo} and defines the $s$-th Chern number
\begin{equation}
2 \mathsf{C}^{(s)}
:=
\frac{1}{s!}
\left(\frac{\mathrm{i}}{2\pi}\right)^s \int\limits_{\mathrm{BZ}}\mathrm{tr}\left[\mathrm{F}^s\right],
\label{eq: Chern numbers for any d} 
\end{equation} 
where $\mathrm{F}^{s}=\mathrm{F}\wedge\cdots\wedge\mathrm{F}$.
As with the case $s=1$ that we have exemplified above, $\mathsf{C}^{(s)}$ is integer for any $s=1,2,\cdots$.

From inspection of Tab.~\ref{table: symm classes} we see that symmetry classes without chiral symmetry may have integer topological invariants $\mathbb{Z}$ only when the dimension $d$ of space is even. In fact, all the integer invariants of these classes are given by the Chern number $\mathsf{C}^{(d/2)}$ of the respective dimension. 

\paragraph{Winding numbers}
Let us now consider systems with chiral symmetry $C$.
To construct their topological invariants as a natural extension of the above, we consider a different representation of the Chern numbers $\mathsf{C}^{(s)}$.
In terms of the flatland projector Hamiltonian $\mathcal{Q}(\bs{k})$ that was defined in Eq.~\eqref{eq: flatband projector H},
we can write
\begin{equation}
\mathsf{C}^{(s)}
\propto\varepsilon^{\ }_{i_1\cdots i_{d}}\int\limits_{\mathrm{BZ}}
\mathrm{d}^d\bs{k}\,
\mathrm{tr}\left[\mathcal{Q}(\bs{k})\partial^{\ }_{i_1}\mathcal{Q}(\bs{k})\cdots\partial^{\ }_{i_d}\mathcal{Q}(\bs{k})\right],
\qquad d=2s.
\label{eq: Chern in terms of H}
\end{equation}
The form of Eq.~\eqref{eq: Chern in terms of H} allows to interpret $\mathsf{C}^{(s)}$ as the \emph{winding number} of the unitary transformation $\mathcal{Q}(\bs{k})$ over the compact BZ.
One verifies that $\mathsf{C}^{(s)}=0$ for symmetry classes with chiral symmetry by inserting $CC^\dagger$ at some point in the expression and anticommuting $C$ with all $\mathcal{Q}$, using the cyclicity of the trace. After $2s+1$ anticommutations, we are back to the original expression up to an overall minus sign and found $\mathsf{C}^{(s)}=-\mathsf{C}^{(s)}$. Hence, all systems with chiral symmetry have vanishing Chern numbers. 

In odd dimensions of space, we can define an alternative topological invariant for systems with chiral symmetry by modifying Eq.~\eqref{eq: Chern in terms of H} and using the chiral operator $C$
\begin{equation}
\begin{split}
\mathsf{W}^{(s)}
:=&\,
\frac{(-1)^s s!}{2(2s+1)!}\left(\frac{\mathrm{i}}{2\pi}\right)^{s+1}
\varepsilon^{\ }_{i_1\cdots i_{d}}\int\limits_{\mathrm{BZ}}
\mathrm{d}^d\bs{k}\,
\mathrm{tr}\left[C\mathcal{Q}(\bs{k})\partial^{\ }_{i_1}\mathcal{Q}(\bs{k})\cdots\partial^{\ }_{i_d}\mathcal{Q}(\bs{k})\right]\\
=&\,
\frac{(-1)^s s!}{(2s+1)!}\left(\frac{\mathrm{i}}{2\pi}\right)^{s+1}
\varepsilon^{\ }_{i_1\cdots i_{d}}\int\limits_{\mathrm{BZ}}
\mathrm{d}^d\bs{k}\,
\mathrm{tr}\left[q^\dagger(\bs{k})\partial^{\ }_{i_1}q(\bs{k})\partial^{\ }_{i_2}q^\dagger(\bs{k})\cdots\partial^{\ }_{i_d}q(\bs{k})\right]
,
\qquad d=2s+1.
\end{split}
\label{eq: chiral winding number}
\end{equation}
Upon anticommuting the chiral operator $C$ once with all matrices $\mathcal{Q}$ and using the cyclicity of the trace, one finds that the expression for $\mathsf{W}^{(s)}$ vanishes for even dimensions. 
The second line of Eq.~\eqref{eq: chiral winding number} allows to interpret $\mathsf{W}^{(s)}$ as the \emph{winding number} of the unitary off-diagonal part $q(\bs{k})$ of the chiral Hamiltonian that was defined in Eq.~\eqref{eq: def of q chiral}.
With Eq.~\eqref{eq: chiral winding number} we have given topological invariants for all entries $\mathbb{Z}$ in odd dimensions $d$ in Tab.~\ref{table: symm classes}.

In summary, we have now given explicit formulas for the topological invariants for all entries $\mathbb{Z}$ in Tab.~\ref{table: symm classes} for systems with translational invariance.
It is important to remember that the classification of  Tab.~\ref{table: symm classes} is restricted to systems without interactions. If interactions are allowed, that neither spontaneously nor explicitly break the defining symmetry of a symmetry class, one of two things can happen:
i) Two phases which are distinguished by a noninteracting invariant like $\mathsf{W}^{(0)}$ might, sometimes but not always, be connected adiabatically (i.e., without a closing of the spectral gap) by turning on strong interactions.
ii) Interactions can enrich the classification of Tab.~\ref{table: symm classes} by inducing new phases with topological response functions that are distinct from those of the noninteracting phases. 
We will given an example for the scenario i) in Sec~\ref{sec: reduction}.

Besides, interactions can strongly modify the topological boundary modes of the noninteracting systems to the extend that they can be gapped without breaking the protective symmetries, but at the expense of introducing topological order on the boundary. 

\subsection{The Su-Schrieffer-Heeger model}
\label{sec: SSH model}

The first example of a topological band insulator that we consider here is also the simplest: The  
 Su-Schrieffer-Heeger model~\cite{Su79} describes a 1D chain of atoms with one (spinless) electronic orbital each at half filling. 
The model was originally proposed to describe the electronic structure of polyacetylene.
This 1D organic molecule features a Peierls instability by which the hopping integral between consecutive sites is 
alternating between strong and weak. This enlarges the unit cell to contain two sites $A$ and $B$.
The second-quantized mean-field Hamiltonian reads
\begin{equation}
H=t\sum_{i=1}^N\left[
(1-\delta)c^\dagger_{A,i}c_{B,i}
+
(1+\delta)c^\dagger_{B,i}c_{A,i+1}
+
\mathrm{h.c.}
\right].
\label{eq: SSH lattice Hamiltonian}
\end{equation}
Here, $c^\dagger_{A,i}$ and $c^\dagger_{B,i}$ create an electron in the $i$-th unit cell on sublattice 
$A$ and $B$, respectively. 
If we identify $i=N+1\equiv 1$, periodic boundary conditions are implemented. 
The corresponding Bloch Hamiltonian 
\begin{subequations}
\begin{eqnarray}
H&=&
t\sum_{k\in\mathrm{BZ}}\sum_{\alpha=A,B}
c^\dagger_{\alpha, k}
h_{\alpha\beta,k}
c_{\beta,k}\\
h_{k}
&=&
\begin{pmatrix}
0&(1-\delta)+(1+\delta) e^{-\mathrm{i}k}
\\
(1-\delta)+(1+\delta) e^{\mathrm{i}k}&0
\end{pmatrix}
\\
&=&
\sigma_x\left[(1-\delta)+(1+\delta)\, \cos\, k\right]
+
\sigma_y(1+\delta)\, \sin \,k,
\end{eqnarray}
where $\sigma_x$ and $\sigma_y$ are the first two Pauli matrices acting on the sublattice index, $t$ is the nearest-neighbor hopping integral, and $\delta$ is a dimensionless parametrization of the strong-weak dimerization of bonds. 
\label{eq: SSH Hamiltonian}
\end{subequations}

We observe that Hamiltonian~\eqref{eq: SSH Hamiltonian} has time-reversal symmetry $\mathcal{T}=\mathcal{K}$, chiral symmetry $C=\sigma_z$ and thus also particle-hole symmetry $\mathcal{P}=\sigma_z\mathcal{K}$. This places it in class BDI of Tab.~\ref{table: symm classes} with a $\mathbb{Z}$ topological characterization. Observe that breaking the time-reversal symmetry would not alter the topological properties, as long as the chiral symmetry is intact. The model would then belong to class AIII, which also features a $\mathbb{Z}$ classification. Hence, it is the chiral symmetry that is crucial to protecting the topological properties of Hamiltonian~\eqref{eq: SSH Hamiltonian}. Notice that generic longer-range hopping (between sites of the same sublattice) breaks the chiral symmetry.

 \begin{figure}\begin{center}
\includegraphics[page=1,width=0.60\textwidth]{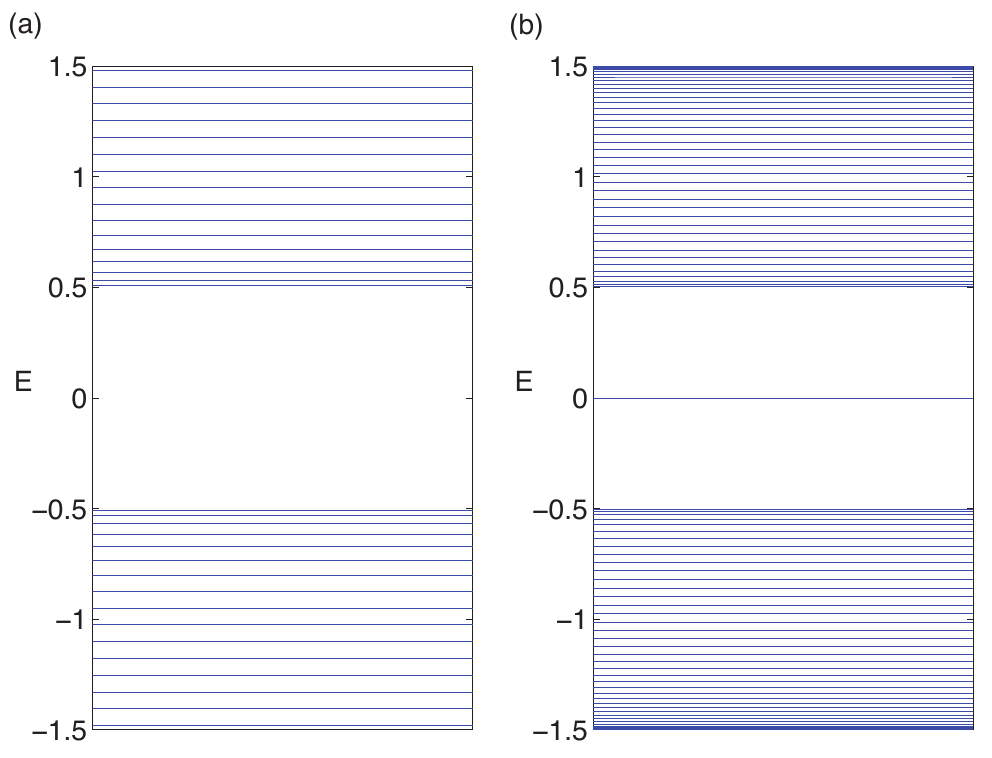}\\
\caption{Energy spectra for the Su-Schrieffer-Heeger model with open boundary conditions (a) in the trivial phase and (b) in the nontrivial topological phase with a zero energy mode on each boundary point.}
\label{fig:SSHopenspectrum}
\end{center}
\end{figure}

What are the  different topological sectors that can be accessed by tuning the parameter $\delta$ in the Su-Schrieffer-Heeger model? We observe that the dispersion
\begin{equation}
\varepsilon^2_{k}
=2\left[(1+\delta^2)+(1-\delta^2)\,\cos\,k\right]
\end{equation}
is gapless for $\delta=0$, hinting that this is the boundary between two distinct phases $\delta>0$ and $\delta<0$.
As we are interested in understanding the topological properties of these phases, we can analyze them for any convenient value of the parameter $\delta$ and then conclude that they are the same in the entire phase by adiabaticity. We consider the Hamiltonian~\eqref{eq: SSH lattice Hamiltonian} with \emph{open} boundary conditions and choose the representative parameters
\begin{itemize}
\item $\delta=+1$ : The operators $c^\dagger_{1,A}$ and $c^\dagger_{N,B}$ do not appear in the Hamiltonian for the open chain. Hence, there exists a state at either end of the open chain that can be occupied or unoccupied at no cost of energy. Thus, either end of the chain supports a localized topological end state (see Fig.~\ref{fig:SSHopenspectrum}). Away from $\delta=+1$, as long as $\delta>0$, the end states start to overlap and split apart in energy by an amount that is exponentially small in the length $N$ of the chain. 
We can back up this observation by evaluating the topological invariant~\eqref{eq: chiral winding number} for this phase. The off-diagonal projector is $q_k=e^{-\mathrm{i}k}$ and its winding number evaluates to
\begin{equation}
\mathsf{W}^{(0)}=\frac{\mathrm{i}}{2\pi}
\int\mathrm{d}k\,e^{\mathrm{i}k}(-\mathrm{i})e^{-\mathrm{i}k}=1.
\end{equation}
\item $\delta=-1$ : In this case strong bonds form between the two sites in every unit cell and no topological end states appear. 
Correspondingly, as the off-diagonal projector $q_k=1$ is independent of $k$, we conclude that the winding number vanishes $\mathsf{W}^{(0)}=0$.
\end{itemize}

One can visualize the winding number of a two-band Hamiltonian that has the form $h_k=\bs{d}_k\cdot\bs{\sigma}$ in the following way. If the Hamiltonian has chiral symmetry, we can choose this symmetry to be represented by $C=\sigma_z$ without loss of generality. Then $\bs{d}_k$ has to lie in the $x$-$y$-plane for every $k$ and may not be zero if the phase is gapped. The winding number $\mathsf{W}^{(0)}$ measures how often $\bs{d}_k$ winds around the origin in the $x$-$y$-plane as $k$ changes from $0$ to $2\pi$.

Besides the topological end states, the Su-Schrieffer-Heeger model also features topological domain wall states between a region with $\delta>0$ and $\delta<0$. Such topological midgap modes have to appear pairwise in any periodic geometry. As the system is considered at half filling, each of these modes binds half an electron charge. This is an example of charge fractionalization at topological defects. It is important to remember that these defects are not dynamical, but are rigidly fixed external perturbations. Therefore, this form of fractionalization is not related to intrinsic topological order.

\subsection{The one-dimensional $p$-wave superconductor}
\label{sec: majorana wire}

In the Su-Schrieffer-Heeger model, particle-hole symmetry (and with it the chiral symmetry) is in some sense fine-tuned, as it is lost if generic longer-range hoppings are considered. In superconductors, particle-hole symmetry arises more naturally as a symmetry that is inherent in the redundant description of mean-field Bogoliubov-deGennes Hamiltonians. 

Here, we want to consider the simplest model for a topological superconductor that has been studied by Kitaev in Ref.~\onlinecite{Kitaev01}. The setup is again a 1D chain with one orbital for spinless fermion on each site. Superconductivity is encoded in pairing terms $c^\dagger_ic^\dagger_{i+1}$ that do not conserve particle number. The Hamiltonian is given by
\begin{equation}
H=\sum_{i=1}^N\left[
-t\left(c^\dagger_i c_{i+1}+c^\dagger_{i+1} c_{i}\right)
-\mu c^\dagger_ic_i
+\Delta c^\dagger_{i+1} c^\dagger_{i}
+\Delta^* c_{i} c_{i+1}
\right].
\label{eq: Kitaev Chain Ham in cs}
\end{equation}
Here, $\mu$ is the chemical potential and $\Delta$ is the superconducting order parameter, which we will decompose into its amplitude $|\Delta|$ and complex phase $\theta$, i.e., $\Delta=|\Delta|e^{\mathrm{i}\theta}$.

The fermionic operators $c^\dagger_i$ obey the algebra
\begin{equation}
\{c^\dagger_i,c_j\}=\delta_{i,j},
\end{equation}
with all other anticommutators vanishing. We can chose to trade the operators $c^\dagger_i$ and $c_i$ on every site $i$ for two other operators $a_i$ and $b_i$ that are defined by
\begin{equation}
a_i=e^{-\mathrm{i}\theta/2}c_i+e^{\mathrm{i}\theta/2}c^\dagger_i,
\qquad
b_i=\frac{1}{\mathrm{i}}\left(e^{-\mathrm{i}\theta/2}c_i-e^{\mathrm{i}\theta/2}c^\dagger_i\right).
\end{equation} 
These so-called Majorana operators obey the algebra
\begin{equation}
\{a_i,a_j\}=\{b_i,b_j\}=2\delta_{ij},\qquad \{a_i,b_j\}=0\qquad \forall i,j.
\label{eq: Majorana ac realtions}
\end{equation}
In particular, they square to 1
\begin{equation}
a_i^2=b_i^2=1,
\end{equation}
and are self-conjugate
\begin{equation}
a_i^\dagger=a_i,\qquad
b_i^\dagger=b_i.
\label{eq: self-conjugacy Majorana operators}
\end{equation}
In fact, we can always break up a complex fermion operator on a lattice site into its real and imaginary Majorana components though it may not always be a useful representation. As an aside, note that the Majorana anti-commutation relation in Eq.~\eqref{eq: Majorana ac realtions} is the same as that of the generators of a Clifford algebra where the generators all square to $+1.$ Thus, mathematically one can think of the operators $a_i$ (or $b_i$) as matrices forming by themselves the representation of Clifford algebra generators.

 \begin{figure}\begin{center}
\includegraphics[page=1,width=0.60\textwidth]{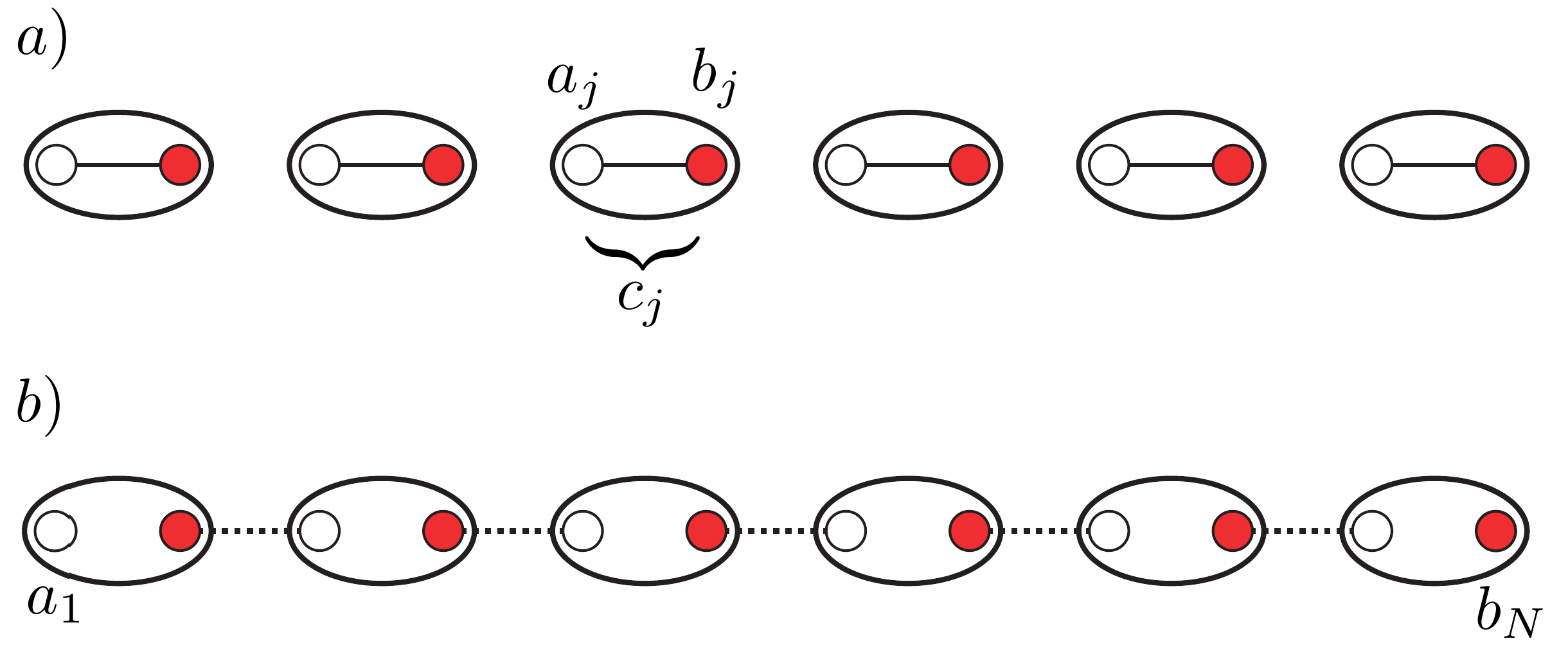}\\
\caption{ Schematic illustration of the lattice p-wave superconductor Hamiltonian in the (a) trivial limit (b) non-trivial limit. The white (empty) and red (filled) circles represent the Majorana fermions making up each physical site (oval). The fermion operator on each physical site ($c_j$) is split up into two Majorana operators ($a_{j}$ and $b_{j}$). In the non-trivial phase the unpaired Majorana fermion states at the end of the chain are labelled with $a_1$ and $b_N$. These are the states which are continuously connected to the zero-modes in the non-trivial topological superconductor phase.}\label{fig:majoranaChain}\end{center}
\end{figure}

When rewritten in the Majorana operators, Hamiltonian~\eqref{eq: Kitaev Chain Ham in cs} takes (up to a constant) the form
\begin{equation}
H=\frac{\mathrm{i}}{2}
\sum_{i=1}^N\left[
-\mu a_i\, b_i+
(t+|\Delta|)b_i\,a_{i+1}
+(-t+|\Delta|)a_i\,b_{i+1}
\right].
\label{eq: Kitaev Chain Ham in as and bs}
\end{equation}
After imposing periodic boundary conditions, it is again convenient to study the system in momentum space. When defining the Fourier transform of the Majorana operators
$a_i=\sum_i e^{\mathrm{i}k i} a_k$
we note that the the self-conjugate property~\eqref{eq: self-conjugacy Majorana operators} that is local in position space translates into $a^\dagger_k=a_{-k}$ in momentum space (and likewise for the $b_k$). The momentum space representation of the Hamiltonian is
\begin{subequations}
\begin{eqnarray}
H&=&
\sum_{k\in\mathrm{BZ}}\sum_{\alpha=A,B}
\left(a_k\ \ b_k\right)
h_{k}
\begin{pmatrix}
a_{-k}\\
b_{-k}
\end{pmatrix}
\\
h_{k}
&=&
\begin{pmatrix}
0&-\frac{\mathrm{i}\mu}{2}+\mathrm{i}t\,\cos\,k+|\Delta|\,\sin\,k
\\
\frac{\mathrm{i}\mu}{2}-\mathrm{i}t\,\cos\,k+|\Delta|\,\sin\,k&0
\end{pmatrix}
\\
&=&
\sigma_x|\Delta|\,\sin\,k
+
\sigma_y\left(\frac{\mu}{2}-t \,\cos\,k\right),
\end{eqnarray}
\label{eq: 1D p-wave Hamiltonian}
\end{subequations}
While this Bloch Hamiltonian is formally very similar to that of the Su-Schrieffer-Heeger model~\eqref{eq: SSH Hamiltonian}, we have to keep in mind that it acts on entirely different single-particle degrees of freedom, namely in the space of Majorana operators instead of complex fermionic operators. 
As with the case of the Su-Schrieffer-Heeger model, the Hamiltonian~\eqref{eq: 1D p-wave Hamiltonian} has a time-reversal symmetry $\mathcal{T}=\sigma_z\mathcal{K}$ and a particle-hole symmetry $P=\mathcal{K}$ which combine to the chiral symmetry $C=\sigma_z$. Hence, it belongs to symmetry class BDI as well. For the topological properties that we explore below, only the particle-hole symmetry is crucial. If time-reversal symmetry is broken, the model changes to symmetry class D, which still supports a $\mathbb{Z}_2$ topological grading.

To determine its topological phases, we notice that Hamiltonian~\eqref{eq: 1D p-wave Hamiltonian} is gapped except for $|t|=|\mu/2|$. We specialize again on convenient parameter values on either side of this potential topological phase transition
\begin{itemize}
\item $\mu=0$, $|\Delta|=t$ : The Bloch matrix $h_k$ takes exactly the same form as that of the Su-Schrieffer-Heeger model~\eqref{eq: SSH Hamiltonian} for the parameter choice $\delta=+1$. We conclude that the Hamiltonian~\eqref{eq: 1D p-wave Hamiltonian} is in a topological phase. 
The Hamiltonian reduces to 
\begin{equation}
H=\mathrm{i}t\sum_{j}b_{j}a_{j+1}.
\end{equation}\noindent 
A pictorial representation of this Hamiltonian is shown in Fig.~\ref{fig:majoranaChain}~b). With open boundary conditions it is clear that the Majorana operators $a_{1}$ and $b_{N}$ are not coupled to the rest of the chain and are `unpaired'. In this limit the existence of two Majorana zero modes localized on the ends of the chain is manifest. 
\item $\Delta=t=0$, $\mu<0$ : This is the topologically trivial phase. The Hamiltonian is independent of $k$ and we conclude that the winding number vanishes $\mathsf{W}^{(0)}=0$.
In this case the Hamiltonian reduces to 
\begin{eqnarray}
H&=&-\mu\frac{\mathrm{i}}{2}\sum_{j}a_{j}b_{j}.
\end{eqnarray}\noindent 
In its ground state the Majorana operators on each physical site are coupled but the Majorana operators between each physical site are decoupled.
In terms of the physical complex fermions, it is the ground state with either all sites occupied or all sites empty.
 A representation of this Hamiltonian is shown in Fig.~\ref{fig:majoranaChain}~a). The Hamiltonian in the physical-site basis is in the atomic limit, which is another way to see that the ground state is trivial. If the chain has open boundary conditions there will be no low-energy states on the end of the chain if the boundaries are cut between \emph{physical} sites. That is, we are not allowed to pick boundary conditions where a physical complex fermionic site is cut in half. 
\end{itemize}
These two limits give the simplest representations of the trivial and non-trivial phases. By tuning away from these limits the Hamiltonian will have some mixture of couplings between Majorana operators on the same physical site, and operators between physical sites. However, since the two Majorana modes are localized at different ends of a gapped chain, the coupling between them will be exponentially small in the length of the wire and they will remain at zero energy. In fact, in the non-trivial phase the zero modes  will not be destroyed until the bulk gap closes at a critical point.

It is important to note that these zero modes count to a different many-body ground state degeneracy than the end modes of the Su-Schrieffer-Heeger model.
 The difference is rooted in the fact that one cannot build a fermionic Fock space out of an odd number of Majorana modes, because they are linear combinations of particles and holes. Rather, we can define a \emph{single} fermionic operator out of \emph{both} Majorana end modes $a_1$ and $b_{N}$ as $c^\dagger:=a_1+\mathrm{i}b_{N}$. The Hilbert space we can build out of $a_1$ and $b_N$ is hence inherently nonlocal.
This \emph{nonlocal} state can be either occupied or empty giving rise to a two-fold degenerate ground state of the chain with two open ends. (In contrast, the topological Su-Schrieffer-Heeger chain has a four-fold degenerate ground state with two open ends, because it has one fermionic mode on each end.) 
The Majorana chain thus displays a different form of fractionalization than the Su-Schrieffer-Heeger chain. For the latter, we observed that the topological end modes carry fractional charge. In the Majorana chain, the end modes are a fractionalization of a fermionic mode into a superposition of particle and hole (and have no well defined charge anymore), but the states $|0\rangle$ (with $c|0\rangle=0$) and $c^\dagger|0\rangle$ do have distinct fermion parity. 
The nonlocal fermionic mode formed by two Majorana end modes is envisioned to work as a qubit (a quantum-mechanical two-level system) that stores quantum information (its state) in a way that is protected against local noise and decoherence. 

\subsection{Reduction of the 10-fold way classification by interactions: $\mathbb{Z}\to\mathbb{Z}_8$ in class BDI}
\label{sec: reduction} 

When time-reversal symmetry $\mathcal{T}=\mathcal{K}$ is present, the model considered in Sec.~\ref{sec: majorana wire} belongs to class BDI of the classification of noninteracting fermionic Hamiltonians in Tab.~\ref{table: symm classes} with a $\mathbb{Z}$ topological characterization. We want to explore how interactions alter this classification, following a calculation by Fidkowski and Kitaev from Ref.~\onlinecite{Fidkowski10}.
To this end, we consider a collection of $n$ identical 1D topological Majorana chains in class BDI and only consider their Majorana end modes on one end, which we denote by $\gamma_1,\cdots, \gamma_n$. 
We will take the point of view that if we can gap the edge, we can continue the bulk to a trivial state (insulator). This is not entirely a correct point of view in general (see 2D topologically ordered states such as the toric code discussed in the next Section), but works for our purposes. 
Given some integer $n$, we ask whether we can couple the Majorana modes locally on one end such that no gapless degrees of freedom are left on that end and the ground state with open boundary conditions becomes singly degenerate. To remain in class BDI, we only allow couplings that respect time-reversal symmetry. 
Let us first derive the action of $\mathcal{T}$ on the Majorana modes.  The complex fermion operators are left invariant under time-reversal $\mathcal{T}c\mathcal{T}^{-1}=c$. Hence,
\begin{equation}
\mathcal{T}(a+\mathrm{i}b)\mathcal{T}^{-1}=\mathcal{T}a\mathcal{T}^{-1}-\mathrm{i} \mathcal{T}b\mathcal{T}^{-1}
\stackrel{!}{=}a+\mathrm{i}b
\quad\Rightarrow\quad
\mathcal{T}a\mathcal{T}^{-1}=a,\qquad
\mathcal{T}b\mathcal{T}^{-1}=-b.
\end{equation}
Thus, when acting on the modes localized on the left end of the wire (which transform like the $a$'s), time-reversal symmetry leaves the Majorana operators invariant.

\begin{figure}\begin{center}
\includegraphics[width=0.7\textwidth]{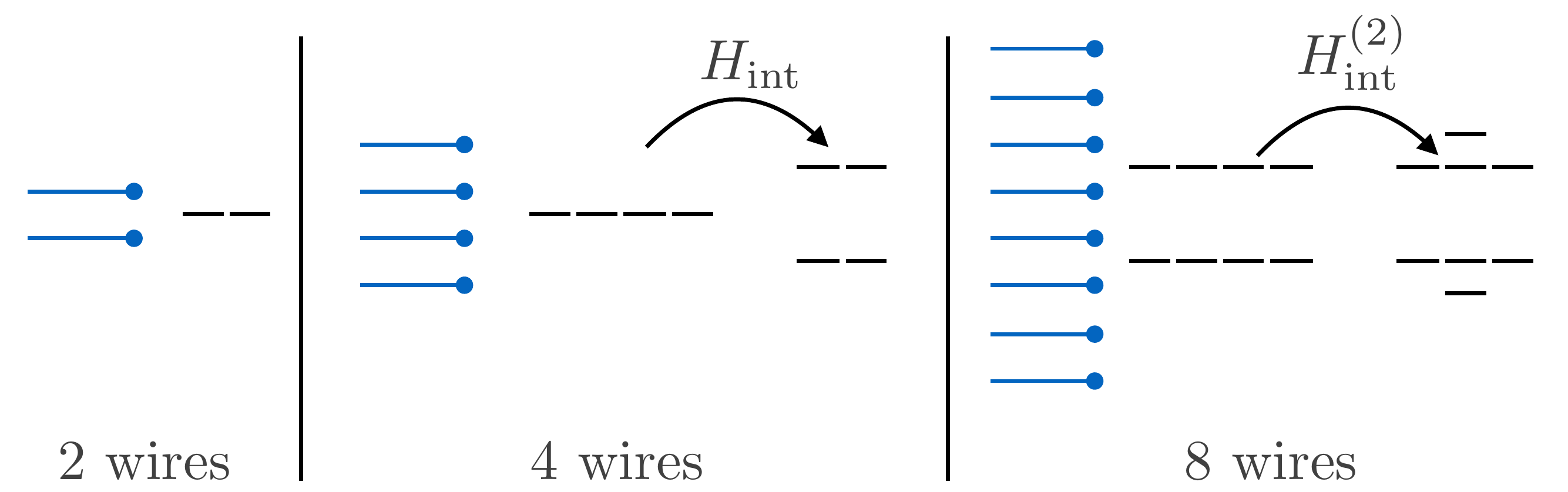}\\
\caption{Schematic illustration of the many body energy levels for 2, 4, and 8 wires with Majorana end states as well as the (partial) lifting of their degeneracy by the Hamiltonians in Eqs.~\eqref{eq: int Hamiltonian 4 wires} and~\eqref{eq: Hamiltonian 8 wires}.}
\label{fig:8 wires spectrum}
\end{center}
\end{figure}

The most naive coupling term that would gap out two Majoranas is $\mathrm{i} \gamma_1 \gamma_2$. This is because two Majoranas can form a local Hilbert space (unlike just one Majorana), and this local Hilbert space can be split unless some other symmetry prevents it from being split. However, time-reversal symmetry forbids these hybridization terms, for it sends $\mathrm{i} \gamma_1 \gamma_2  \rightarrow - \mathrm{i} \gamma_1 \gamma_2$. In spinful systems, another symmetry which can do this is $M \mathcal{T}$, where $M$ is a mirror operator (which in spinful systems squares to $-1$ $M^2=-1$) and $\mathcal{T}$ is the usual time-reversal operator $\mathcal{T}^2=-1$, such that $(M\mathcal{T})^2= M^2 \mathcal{T}^2=1$ and hence $M\mathcal{T}$ acts like spinless time reversal.~\cite{Fang} Realizing that such a term is not allowed is the end of the story for noninteracting systems, yielding the classification $\mathbb{Z}$. Lets find out what interactions do to this system. The steps that we will now outline are summarized in Fig.~\ref{fig:8 wires spectrum}.

We saw that two Majorana end states cannot be gapped: the only possible interacting or noninteracting Hamiltonian is $\mathrm{i}a_1a_2$. Three Majoranas clearly cannot be gapped either, as it is an odd number. Let us thus add two more Majorana end states into the mix. Any one-body term still is disallowed but the term
\beq
H_{\mathrm{int}} = a_1 a_2 a_3 a_4
\eneq 
can be present.
We can now form two complex fermions, $c_1 = (a_1+ \mathrm{i} a_2)/\sqrt{2}, c_2 = (a_3+ \mathrm{i} a_4)/\sqrt{2}$. In terms of these two fermions, the Hamiltonian reads
\beq
H_{\mathrm{int}} = -\left(n_1- \frac{1}{2}\right) \left(n_2 - \frac{1}{2}\right) ,
\label{eq: int Hamiltonian 4 wires}
\eneq 
where $n_1= c_1^\dagger c_1$ and $n_2=c_2^\dagger c_2$ are the occupation numbers. The Hamiltonian is diagonal in the eigenbasis $|n_1n_2\rangle$ of the occupation number operators, and the states $\ket{11}, \ket{00}$ are degenerate at energy $-1/4$, while the states $\ket{01},\ket{10}$ are degenerate at energy $+1/4$. The original noninteracting system of four Majorana fermions had a degeneracy of $2^2=4$. The interaction, however, has lifted this degeneracy, but not all the way to a single nondegenerate ground state. Irrespective of the sign of the interaction, it leaves the states doubly degenerate on one edge, and hence cannot be adiabatically continued to the trivial state of single degeneracy. However, if we add four more Majoranas wires so that we have $n=8$ Majoranas, we can build an interaction which creates a singly degenerate ground state. We can understand this as follows: Add two interactions
\begin{subequations}
\beq
H_{\mathrm{int}}^{(1)} = -\alpha( a_1 a_2 a_3 a_4 + a_5 a_6 a_7 a_8)
\eneq  These create two doublets, one in $c_1, c_2$ defined above, and one in $c_3= (a_5+ \mathrm{i} a_6)/\sqrt{2}, c_4 = (a_7+ \mathrm{i}a_8)/\sqrt{2}$. We couple these doublets via the interaction
\beq
H_{\mathrm{int}}^{(2)} = \sum_{i=x,y,z} \beta (c_1^\dagger \ \ c_2^\dagger) \sigma_i  \left(\begin{array}{cc}
  c_1     \\
  c_2       
\end{array}
\right) (c_3^\dagger\ \  c_4^\dagger) \sigma_i  \left(\begin{array}{cc}
  c_3     \\
  c_4       
\end{array}
\right).
 \eneq 
 \label{eq: Hamiltonian 8 wires}
 \end{subequations}
Representing each of the doublets as a spin-1/2 $\bs{S}$, this interaction is nothing but an $\bs{S}\cdot \bs{S}$ term. If we take $0<\beta \ll \alpha$, then we can approximate the interaction $\beta$ by its action within the two ground state doublets. As such, this interaction creates a singlet and a triplet (in that doublet) and for the right sign of $\beta$, we can put the singlet below the triplet, thereby creating a unique ground state 
\begin{equation}
\frac{1}{\sqrt{2}}\left(|0110\rangle-|1001\rangle\right),
\end{equation}
in terms of the occupation number states $|n_1n_2n_3n_4\rangle$.
This unique ground state can be adiabatically continued to the atomic limit. In this way the noninteracting $\mathbb{Z}$ classification of class BDI breaks down to $\mathbb{Z}_8$ if interactions are allowed.

\clearpage

\section{Examples of topological order}

So far, we have been concerned with symmetry protected topological states and considered examples that were motivated by the 
topological classification of free fermion Hamiltonians. The topological properties of these systems are manifest by the presence of protected boundary modes.  

In this Section, we want to familiarize ourselves with the concept of intrinsic topological order by ways of two examples. We will study the connections between different characterizations of topological order, such as fractionalized excitations in the bulk and the topological ground state degeneracy. Our examples will be in 2D space, as topologically ordered states do not exist in 1D and are best understood in 2D. Our first example, the toric code, has Abelian anyon excitations, while the second example, the chiral $p$-wave superconductor, features non-Abelian anyons. 

\subsection{The toric code}

The first example of a topologically ordered state is an exactly soluble model with vanishing correlation length. The significance of having zero correlation length is the following. The correlation functions of local operators decay exponentially in gapped quantum ground states in 1D and 2D with a characteristic length scale given by the correlation length $\xi$.~\cite{Hastings06} In contrast, topological properties are encoded in quantized expectation values of nonlocal operators (for example the Hall conductivity) or the degeneracy of energy levels (such as the end states of the Su-Schrieffer-Heeger model). In finite systems, such quantizations and degeneracies are generically only exact up to corrections that are of order $e^{-L/\xi}$, where $L$ is the linear system size. Models with zero correlation length are free from such exponential finite-size corrections and thus expose the topological features already for the smallest possible system sizes. The down-side is that their Hamiltonians are rather contrived.

\begin{figure}[t]
\begin{center}
\includegraphics[page=6, width=0.4\linewidth]{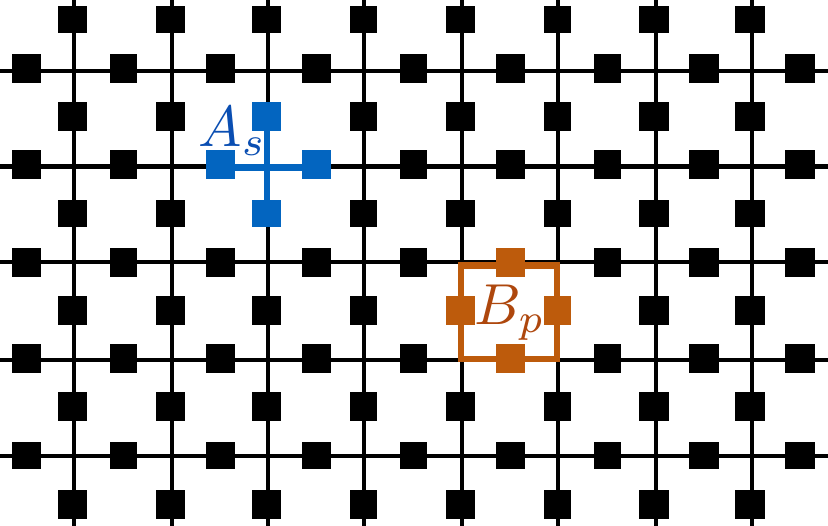}
\caption{
The toric code model is defined on a square lattice with spin-1/2 degrees of freedom on every bond (black squares). The operator $A_s$ acts with $\sigma_x$ on all four spins one the bonds that are connected to a lattice site (a star $s$).
The operator $B_p$ acts with $\sigma_z$ on all four spins around a plaquette $p$.
}
\label{fig: toric code lattice}
\end{center}
\end{figure}

We define the toric code model~\cite{Kitaev97} on a square lattice with a spin-1/2 degree of freedom on every \emph{bond} $j$ (see Fig.~\ref{fig: toric code lattice}).
The four spins that sit on the bonds emanating from a given site of the lattice are referred to as a star $s$.
The four spins that sit on the bonds surrounding a square of the lattice are called  a plaquette $p$.
We define two sets of operators 
\begin{equation}
A_s:=\prod_{j\in s}\sigma_j^x,
\qquad
B_p:=\prod_{j\in p}\sigma_j^z,
\end{equation}
that act on the spins of a given star $s$ and plaquette $p$, respectively. Here, $\sigma^{x,z}_j$ are the respective Pauli matrices acting on the spin on bond $j$. 

These operators have two crucial properties which are often used to construct exactly soluble models for topological states of matter
\begin{enumerate}
\item All of the $A_s$ and $B_p$ commute with each other. This is trivial for all cases except for the commutator of $A_s$ with $B_p$ if $s$ and $p$ have spins in common. However, any star shares with any plaquette an even number of spins (edges), so that commuting $A_s$ with $B_p$ involves commuting an even number of  $\sigma^z$ with $\sigma^x$, each of which comes with a minus sign.
\item The operators
\begin{equation}
\frac{1-B_p}{2},\qquad\frac{1-A_s}{2}
\end{equation}
are projectors. The former projects out plaquette states with an even number of spins polarized in the positive $z$-direction. The latter projects out stars with an even number of spins in the positive $x$-direction.
\end{enumerate}

\subsubsection{Ground states}

\begin{figure}[t]
\begin{center}
\includegraphics[page=1, width=0.7\linewidth]{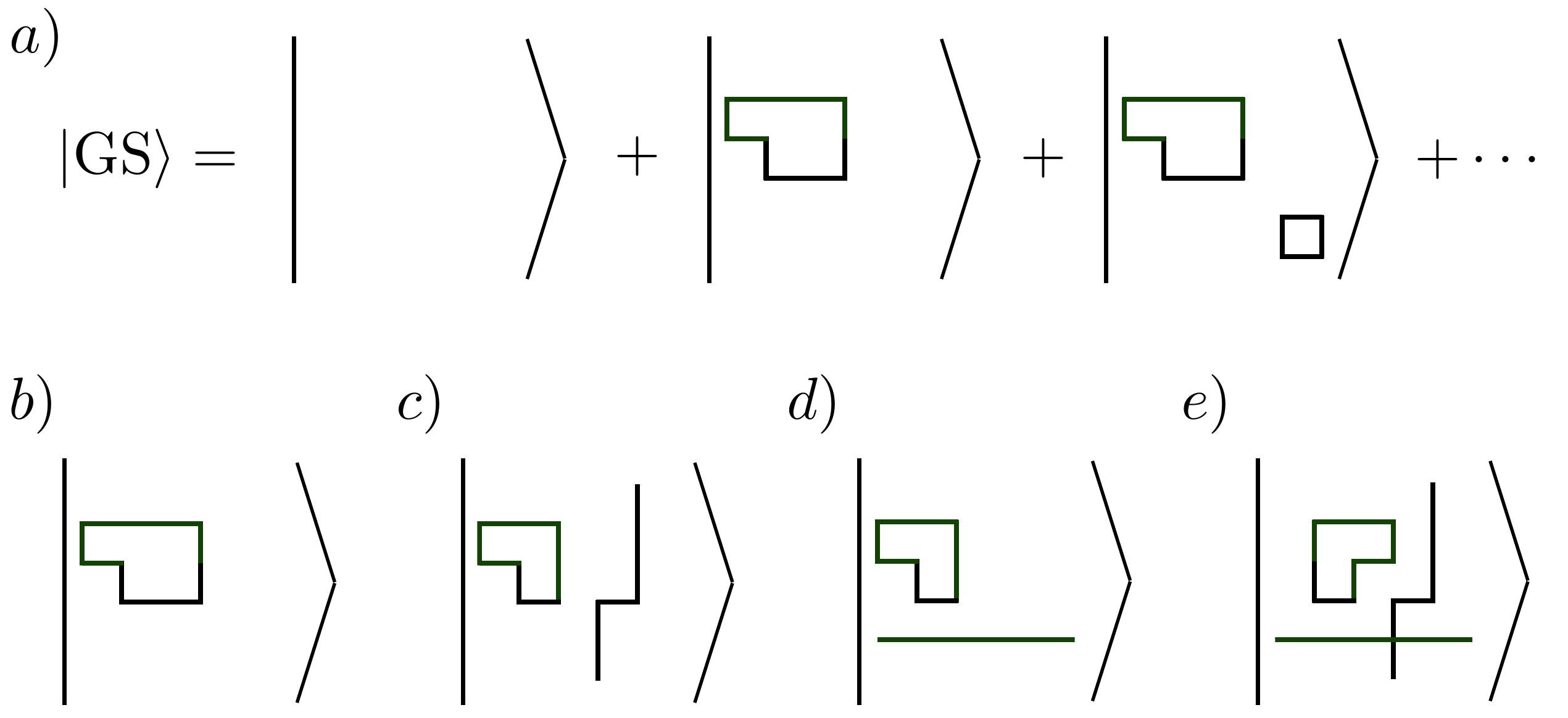}
\caption{Visualization of the toric code ground states on the torus. a) The toric code ground state is the equal amplitude superposition of all closed loop configurations. b)-e) Four base configurations $|c\rangle$ entering Eq.~\eqref{eq: toric code GS} that yield topologically distinct ground states on the torus.}
\label{fig: toric code GS}
\end{center}
\end{figure}

The Hamiltonian is defined as a sum over these commuting projectors
\begin{equation}
H=-J_{\mathrm{e}}\sum_sA_s-J_{\mathrm{m}}\sum_pB_p,
\label{eq: Toric code Hamiltonian}
\end{equation}
where the sums run over all stars $s$ and plaquettes $p$ of the lattice. Let us assume that both $J_{\mathrm{e}}$ and $J_{\mathrm{m}}$ are positive constants. 
Then, the ground state is given by a state in which all stars $s$ and plaquettes $p$ are in an eigenstate with eigenvalue $+1$ of $A_s$ and $B_p$, respectively. (The fact that all $A_s$ and $B_p$ commute allows for such a state to exist, as we can diagonalize each of them separately.)
Let us think about the ground state in the eigenbasis of the $\sigma^x$ operators and represent by bold lines those bonds with spin up and and draw no lines along bonds with spin down. Then, $A_s$ imposes on all spin configurations with nonzero amplitude in the ground state the constraint that an even number of bold lines meets at the star $s$. In other words, we can think of the bold lines as connected across the lattice and they may only form closed loops. Bold lines that end at some star (``open strings") are not allowed in the ground state configurations; they are excited states. Having found out which spin configurations are allowed in the ground state, we need to determine their amplitudes. This can be inferred from the action of the $B_p$ operators on these closed loop configurations. The $B_p$ flips all bonds around the plaquette $p$. Since $B_p^2=1$, given a spin configuration $|c\rangle$ in the $\sigma^x$-basis, we can write an eigenstate of $B_p$ with eigenvalue $1$ as
\begin{equation}
\frac{1}{\sqrt{2}}\left(|c\rangle+B_p|c\rangle\right),
\end{equation}
for some fixed $p$. This reasoning can be extended to all plaquettes so that we can write for the ground state
\begin{equation}
|\mathrm{GS}\rangle=\left(\prod_{p}\frac{1+B_p}{\sqrt{2}}\right)|c\rangle,
\label{eq: toric code GS}
\end{equation}
where $|c\rangle$ is a closed loop configuration [see Fig.~\ref{fig: toric code GS}~a)]. Is $|\mathrm{GS}\rangle$ independent of the choice of $|c\rangle$? In other words, in the ground state unique? We will see that the answer depends on the topological properties of the manifold on which the lattice is defined and thus reveals the topological order imprinted in $|\mathrm{GS}\rangle$. 

To answer these questions, let us consider the system on two topologically distinct manifolds, the torus and the sphere.
To obtain a torus, we consider a square lattice with $L_x\times L_y$ sites and impose periodic boundary conditions. This lattice hosts $2L_xL_y$ spins (2 per unit cell for they are centered along the bonds). Thus, the Hilbert space of the model has dimension $2^{2L_xL_y}$. There are $L_x L_y$ operators $A_s$ and just as many $B_p$. Hence, together they impose $2L_xL_y$ constraints on the ground state in this Hilbert space. However, not all of these constraints are independent. The relations
\begin{equation}
1=\prod_s A_s,\qquad 1=\prod_p B_p
\label{eq: redundant conditions}
\end{equation}
make two of the constraints redundant, yielding $(2L_xL_y-2)$ independent constraints. The ground state degeneracy (GSD) is obtained as the quotient of the Hilbert space dimension and the subspace modded out by the constraints
\begin{equation}
\mathrm{GSD}=\frac{2^{2L_xL_y}}{2^{2L_xL_y-2}}=4.
\end{equation}
The four ground states on the torus are distinguished by having an even or an odd number of loops wrapping the torus in the $x$ and $y$ direction, respectively. Four configurations $|c\rangle$ that can be used to build the four degenerate ground states are shown in Fig.~\ref{fig: toric code GS}~b)-e). This constitutes a set of ``topologically degenerate'' ground states and is a hallmark of the topological order in the model. 

Let us contrast this with the ground state degeneracy on the sphere. Since we use a zero correlation length model, we might as well use the smallest convenient lattice with the topology of a sphere. We consider the model~\eqref{eq: Toric code Hamiltonian} defined on the edges of a cube. The same counting as above yields that there are 12 degrees of freedom (the spins on the 12 edges), 8 constraints from the $A_s$ operators defined on the corners and 6 constraints from the $B_p$ operators defined on the faces. Subtracting the 2 redundant constraints~\eqref{eq: redundant conditions} yields $12-(8+6-2)=0$ remaining degrees of freedom. Hence, the model has a unique ground state on the sphere. 

On a  general manifold, we have
\begin{equation}
\mathrm{GSD}=2^{\text{number of noncontractible loops}}.
\end{equation}
An important property of the topologically degenerate ground states is that any local operator has vanishing off-diagonal matrix elements between them  in the thermodynamic limit. Similarly, no local operator can be used to distinguish between the ground states. We can, however, define \emph{nonlocal} operators that transform one topologically degenerate ground state into another and that distinguish the ground states by topological quantum numbers. (Notice that such operators may not appear in any physical Hamiltonian due to their nonlocality and hence the degeneracy of the ground states is protected.) On the torus, we define two pairs of so-called Wilson loop operators as
\begin{equation}
W^{\mathrm{e}}_{x/y}:=
\prod_{j\in l^{\mathrm{e}}_{x/y}} \sigma^z_j,
\qquad
W^{\mathrm{m}}_{x/y}:=
\prod_{j\in l^{\mathrm{m}}_{x/y}} \sigma^x_j.
\end{equation}
Here, $l^{\mathrm{e}}_{x/y}$ are the sets of spins on bonds parallel to a straight line wrapping the torus once along the $x$- and $y$-direction, respectively. 
The $l^{\mathrm{m}}_{x/y}$ are the sets of spins on bonds perpendicular to a straight line that connects the centers of plaquettes and wraps the torus once along the $x$ and $y$-direction, respectively.
We note that the $W^{\mathrm{e}}_{x/y}$ and $W^{\mathrm{m}}_{x/y}$ commute with all $A_s$ and $B_p$
\begin{equation}
\left[W^{\mathrm{e}/\mathrm{m}}_{x/y},A_s\right]=\left[W^{\mathrm{e}/\mathrm{m}}_{x/y},B_p\right]=0,
\end{equation}
and thus also with the Hamiltonian. Furthermore, they obey 
\begin{equation}
W^{\mathrm{e}}_{x}W^{\mathrm{m}}_{y}
=
-W^{\mathrm{m}}_{y}W^{\mathrm{e}}_{x}.
\label{eq: W algebra}
\end{equation}
This algebra must be realized in any eigenspace of the Hamiltonian. However, due to Eq.~\eqref{eq: W algebra}, it cannot be realized in a one-dimensional subspace. We conclude that all eigenspaces of the Hamiltonian, including the ground state, must be degenerate. In the $\sigma^x$ basis that we used above, $W^{\mathrm{m}}_{x/y}$ measures whether the number of loops wrapping the torus is even or odd in the $x$ and $y$ direction, respectively, giving 4 degenerate ground states. In contrast, $W^{\mathrm{e}}_{x/y}$ changes the number of loops wrapping the torus in the $x$ and $y$ direction between even and odd.

\subsubsection{Topological excitations}

\begin{figure}[t]
\begin{center}
\includegraphics[page=2, width=0.7\linewidth]{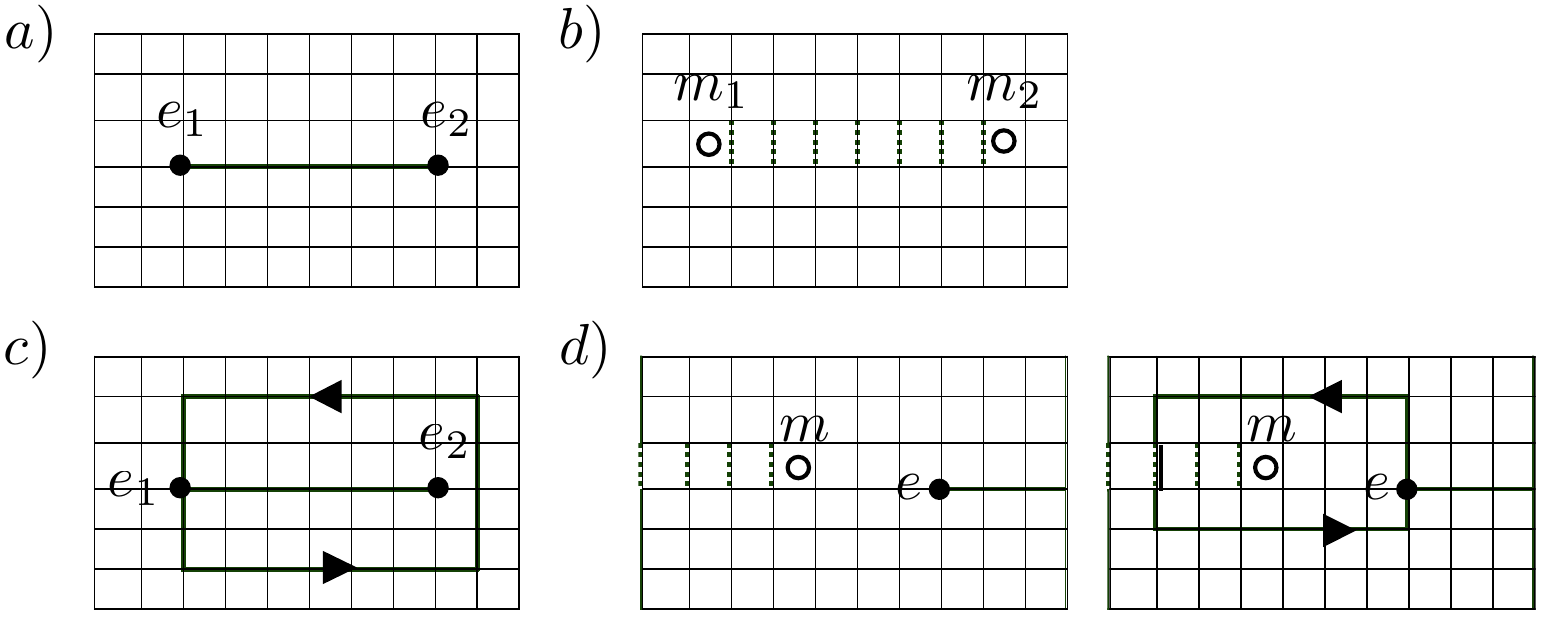}
\caption{ Visualization of operations to compute the braiding statistics of toric code anyons. 
a) Two $e$ excitations above the ground state.
b) Two $m$ excitations above the ground state.
c) Loop created by braiding $e_1$ around $e_2$. 
c) Loop created by braiding $e$ around $m$. A phase of $-1$ results for this process because there is a single bond on which both a $\sigma^x$ operator (dotted line) and a $\sigma^z$ operator (bold line) act. }
\label{fig: toric code excitations}
\end{center}
\end{figure}

To find the topological excitations of the system above the ground state, we ask which are the lowest energy excitations that we can build. Excitations are a violation of the rule that all stars $s$ are eigenstates of $A_s$ and all plaquettes $p$ are eigenstates of $B_p$.  Let us first focus on star excitations which we will call $e$. They appear as the end point of open strings, i.e., if the closed loop condition is violated. Since any string has two end points, the lowest excitation of this type is a pair of $e$. They can be created by acting on the ground state with the operator
\begin{equation}
W^{\mathrm{e}}_{l^{\mathrm{e}}}:=\prod_{j\in l^{\mathrm{e}}} \sigma^z_j,
\end{equation}
where $l^{\mathrm{e}}$ is a string of bonds connecting the two excitations $e_1$ and $e_2$ [see Fig.~\ref{fig: toric code excitations}~a)]. The state
\begin{equation}
|e_1,e_2\rangle:=W^{\mathrm{e}}_{l^{\mathrm{e}}}|\mathrm{GS}\rangle
\end{equation}
has energy $4J_{\mathrm{e}}$ above the ground state energy. Similarly, we can define an operator
\begin{equation}
W^{\mathrm{m}}_{l^{\mathrm{m}}}:=\prod_{j\in l^{\mathrm{m}}} \sigma^x_j,
\end{equation}
that creates a pair of plaquette defects $m_1$ and $m_2$ connected by the string $l^{\mathrm{m}}$ of perpendicular bonds [see Fig.~\ref{fig: toric code excitations}~b)]. (Notice that the operator $W^{\mathrm{m}}_{l^{\mathrm{m}}}$ does not flip spins when the ground state is written in the $\sigma^x$ basis. Rather, it gives weight $+1/-1$ to the different loop configurations in the ground state, depending on whether an even or an odd number of loops crosses $l^{\mathrm{m}}$.)
The state
\begin{equation}
|m_1,m_2\rangle:=W^{\mathrm{m}}_{l^{\mathrm{m}}}|\mathrm{GS}\rangle
\end{equation}
has energy $4J_{\mathrm{m}}$ above the ground state energy.
Notice that the excited states $|e_1,e_2\rangle$ and $|m_1,m_2\rangle$ only depend on the positions of the excitations and not on the particular choice of string that connects them. Furthermore, the energy of the excited state is independent of the separation between the excitations. The excitations are thus ``deconfined'', i.e., free to move independent of each other. 

\begin{subequations}
\label{eq: emf fusion rule}
It is also possible to create a combined defect when a plaquette hosts a $m$ excitation and one of its corners hosts a $e$ excitation. We call this combined defect $f$ and formalize the relation between these defects in a so-called fusion rule
\begin{equation}
e\times m= f.
\end{equation}
When two $e$-type excitations are moved to the same star, the loop $l^{\mathrm{e}}$ that connects them becomes a closed loop and the state returns to the ground state. For this, we write the fusion rule
\begin{equation}
e\times e= 1,
\end{equation}
where 1 stands for the ground state or vacuum. Similarly, moving two $m$-type excitations to the same plaquette creates a closed loop $l^{\mathrm{m}}$, which can be absorbed in the ground state, i.e.,
\begin{equation}
m\times m= 1.
\end{equation}
Superimposing the above processes yields the remaining fusion rules
\begin{equation}
m\times f= e,\qquad e\times f=m, \qquad f\times f=1.
\end{equation} 
\end{subequations}

It is now imperative to ask what type of quantum statistics these emergent excitations obey. We recall that quantum statistics are defined as the phase by which a state changes if two identical particles are exchanged. Rendering the exchange operation as an adiabatically slow evolution of the state, in three and higher dimensions only two types of statistics are allowed between point particles: that of bosons with phase $+1$ and that of fermions with phase $-1$. In 2D, richer possibilities exist and the exchange phase $\theta$ can be \emph{any} complex number on the unit circle, opening the way for \emph{any}ons.
While the exchange is only defined for quantum particles of the same type, the double exchange (braiding) is well defined between any two deconfined anyons. 
We can compute the braiding phases of the anyons $e$, $m$, and $f$ that appear in the toric code one by one. Let us start with the phase resulting from braiding $e_1$ with $e_2$.  The initial state is $W^{\mathrm{e}}_{l^{\mathrm{e}}}|\mathrm{GS}\rangle$ depicted in Fig.~\ref{fig: toric code excitations}~a). Moving $e_1$ around $e_2$ leaves a loop of flipped $\sigma^x$ bonds around $e_2$ [see Fig.~\ref{fig: toric code excitations}~c)]. This loop is created by applying $B_p$ to all plaquettes enclosed by the loop $l^{\mathrm{e}}_{e_1}$ along which $e_1$ moves. We can thus write the final state as
\begin{equation}
\begin{split}
\left(\prod_{p\in l^{\mathrm{e}}_{e_1}} B_p\right)
W^{\mathrm{e}}_{l^{\mathrm{e}}}
|\mathrm{GS}\rangle
=&\,
W^{\mathrm{e}}_{l^{\mathrm{e}}}
\left(\prod_{p\in l^{\mathrm{e}}_{e_1}} B_p\right)
|\mathrm{GS}\rangle
\\
=&\,
W^{\mathrm{e}}_{l^{\mathrm{e}}}
|\mathrm{GS}\rangle.
\end{split}
\end{equation}  
Flipping the spins in a closed loop does not alter the ground state as it is the equal amplitude of all loop configurations. We conclude that the braiding of two $e$ particles gives no phase. 
Similar considerations can be used to conclude that the braiding of two $m$ particles is trivial as well. In fact, not only the braiding, but also the exchange of two $e$ particles and two $m$ particles is trivial. (We have not shown that here.)  

More interesting is the braiding of $m$ with $e$. Let the initial state be
$W^{\mathrm{m}}_{l^{\mathrm{m}}}W^{\mathrm{e}}_{l^{\mathrm{e}}}|\mathrm{GS}\rangle$
and move the $e$ particle located on one end of the string $l^{\mathrm{e}}_{\mathrm{in}}$ around the magnetic particle $m$ on one end of the string 
$l^{\mathrm{m}}$. Again this is equivalent to applying $B_p$ to all plaquettes enclosed by the path $l^{\mathrm{e}}_{e}$ of the $e$ particle, so that the final state is given by
\begin{equation}
\begin{split}
\left(\prod_{p\in l^{\mathrm{e}}_{e}} B_p\right)
W^{\mathrm{m}}_{l^{\mathrm{m}}}W^{\mathrm{e}}_{l^{\mathrm{e}}}|\mathrm{GS}\rangle
=&\,
-
W^{\mathrm{m}}_{l^{\mathrm{m}}}
\left(\prod_{p\in l^{\mathrm{e}}_{e}} B_p\right)
W^{\mathrm{e}}_{l^{\mathrm{e}}}|\mathrm{GS}\rangle
\\
=&\,
-
W^{\mathrm{m}}_{l^{\mathrm{m}}}W^{\mathrm{e}}_{l^{\mathrm{e}}}|\mathrm{GS}\rangle.
\end{split}
\end{equation}  
The product over $B_p$ operators anticommutes with the path operator $W^{\mathrm{m}}_{l^{\mathrm{m}}}$, because there is a single bond on which a single $\sigma^x$ and a single $\sigma^z$ act at the crossing of $l^{\mathrm{m}}$ and $l^{\mathrm{e}}_{e}$ [see Fig.~\ref{fig: toric code excitations}~d)]. As a result, the initial and final state differ by a $-1$, which is the braiding phase of $e$ with $m$. Particles with this braiding phase are called (mutual) semions.

Notice that we have moved the particles on contractible loops only. If we create a pair of $e$ or $m$ particles, move one of them along a noncontractible loop on the torus, and annihilate the pair, we have effectively applied the operators $W^{\mathrm{e}}_{x/y}$ and $W^{\mathrm{m}}_{x/y}$ to the ground state (although in the process we have created finite energy states). The operation of moving anyons on noncontractible loops thus allows to operate on the manifold of topologically degenerate groundstates. This exposes the intimate connection between the presence of fractionalized excitations and topological groundstate degeneracy in topologically ordered systems. 

From the braiding relations of $e$ and $m$ we can also conclude the braiding and exchange relations of the composite particle $f$. This is most easily done in a pictorial way by representing the particle worldlines as moving upwards. For example, we represent the braiding relations of $e$ and $m$ as 
\begin{equation}
\vcenter{\hbox{\includegraphics[page=3, scale=0.45]{ToricCode.pdf}}}\qquad\qquad
\vcenter{\hbox{\includegraphics[page=4, scale=0.45]{ToricCode.pdf}}}.
\label{eq: em braiding relations}
\end{equation}
The exchange of two $f$, each of which is composed of one $e$ and one $m$ is then 
\begin{equation}
\vcenter{\hbox{\includegraphics[page=5, scale=0.45]{ToricCode.pdf}}}
\end{equation}
Notice that we have used Eq.~\eqref{eq: em braiding relations} to manipulate the crossing in the dotted rectangles.
Exchange of two $f$ thus gives a phase $-1$ and we conclude that $f$ is a fermion. 

In summary, we have used the toric code model to illustrate topological ground state degeneracy and emergent anyonic quasiparticles as hallmarks of topological order. We note that the toric code model does not support topologically protected edge states.

\subsection{The two-dimensional p-wave superconductor}

The second example of a 2D system with anyonic excitations that we want to discuss here is the chiral $p$-wave superconductor. 
Unlike the toric code, due to its chiral nature, it is a model with nonzero correlation length. The vortices of the chiral $p$-wave superconductor  
exhibit anyon excitations which have exotic non-Abelian statistics.\cite{volovik1999,read2000,ivanov2001} (The anyons in the toric code are Abelian, we will see below what that distinction refers to.)
For the system to be topologically ordered, these vortices should appear as emergent, dynamical excitations. This requires to treat the electromagnetic gauge field quantum-mechanically. (In fact, since the fermion number conservation is spontaneously broken down to the conservation of the fermion parity in the superconductor, the relevant gauge theory involves only a $\mathbb{Z}_2$ instead of a U(1) gauge field.) However, the topological properties that we want  to discuss here can also be seen if we model the gauge field and vortices as static defects, rather than within a fluctuating $\mathbb{Z}_2$ gauge theory. This allows us to study a models very similar to the ``noninteracting'' topological superconductor in 1D and still expose the non-Abelian statistics.
 
For pedagogy we will use both lattice and continuum models of the chiral superconductor. We begin with the lattice Hamiltonian defined on a square lattice
\begin{equation}
\begin{split}
H=&\sum_{m,n}
\left\{
-t\left(c^{\dagger}_{m+1,n}c^{\pdag}_{m,n}+c^{\dagger}_{m,n+1}c^{\pdag}_{m,n}+{\textrm{h.c.}}\right)
-(\mu-4t) c^{\dagger}_{m,n}c^{\pdag}_{m,n}\right.\\ 
&\qquad+\left.\left(\Delta c^{\dagger}_{m+1,n}c^{\dagger}_{m,n}+\mathrm{i}\Delta c^{\dagger}_{m,n+1}c^{\dagger}_{m,n}+{\textrm{h.c.}}
\right)\right\}.
\label{eq:chiralpwavelattice}
\end{split}
\end{equation}
\noindent 
The fermion operators $c_{m,n}$ annihilate fermions on the lattice site $(m,n)$ and we are considering spinless (or equivalently spin-polarized) fermions. We set the lattice constant  $a=1$ for simplicity. The pairing amplitude is anisotropic and has an additional phase of $\mathrm{i}$ in the $y$-direction compared to the pairing in the $x$-direction. Because the pairing is not on-site, just as in the lattice version of the $p$-wave wire, the pairing terms will have momentum dependence. We can write this Hamiltonian in the Bogoliubov-deGennes form and, assuming that  $\Delta$ is translationally invariant, can Fourier transform the lattice model to get
\begin{eqnarray}
H_{\mathrm{BdG}}=\frac{1}{2}\sum_{\bs{p}}\Psi^{\dagger}_{\bs{p}}\left(\begin{array}{cc}\epsilon(p) & 2\mathrm{i}\Delta(\sin p_x+\mathrm{i} \sin p_y)\\-2\mathrm{i}\Delta^{\ast}(\sin p_x-\mathrm{i}\sin p_y)& -\epsilon(p)\end{array}\right)\Psi^{\pdag}_{\bs{p}},
\end{eqnarray}\noindent 
where $\epsilon(p)=-2t(\cos p_x+\cos p_y)-(\mu-4t)$ and $\Psi_{\bs{p}}=\left(c^{\pdag}_{\bs{p}}\;\; c^{\dagger}_{-\bs{p}}\right)^{\mathsf{T}}.$ For convenience we have shifted the chemical potential by the constant $4t.$ As a quick aside we note that the model takes a simple familiar form in the continuum limit ($\bs{p}\to 0$):
\begin{eqnarray}
H^{(\mathrm{cont})}_{\mathrm{BdG}}=\frac{1}{2}\sum_{\bs{p}}\Psi^{\dagger}_{\bs{p}}\left(\begin{array}{cc}\frac{p^2}{2m}-\mu  &2\mathrm{i}\Delta(p_x+\mathrm{i} p_y)\\ -2\mathrm{i}\Delta^{\ast}(p_x-\mathrm{i} p_y) &  -\frac{p^2}{2m}+\mu\end{array}\right)\Psi^{\pdag}_{\bs{p}}\label{eq:chiralpwavecontinuum}
\end{eqnarray}\noindent where $m\equiv 1/2t$ and $p^2=p^{2}_{x}+p^{2}_{y}.$ We see that the continuum limit has the characteristic $p_x+\mathrm{i} p_y$ chiral form for the pairing potential. The quasiparticle spectrum of $H^{(\mathrm{cont})}_{\mathrm{BdG}}$ is $E_{\pm}=\pm\sqrt{(p^2/2m-\mu)^2+4\vert\Delta\vert^2p^2}$, which, with a nonvanishing pairing amplitude, is gapped across the entire BZ as long as $\mu\neq 0.$ This is unlike some other types of $p$-wave pairing terms [e.g., $\Delta(p)=\Delta p_{x}$] which can have gapless \emph{nodal} points or lines in the BZ for $\mu>0$. In fact, nodal superconductors, having gapless quasiparticle spectra, are not topological superconductors by definition (i.e., a bulk excitation gap does not exist).

We recognize the form of $H^{(\mathrm{cont})}_{\mathrm{BdG}}$ as a massive 2D Dirac Hamiltonian, and indeed Eq.~\eqref{eq:chiralpwavelattice} is just a lattice Dirac Hamiltonian which is what we will consider first. In the first quantized notation, the single particle Hamiltonian for a superconductor is equivalent to that of an insulator with an additional particle-hole symmetry. It is thus placed in class D of Tab.~\ref{table: symm classes} and admits a $\mathbb{Z}$ topological classification in 2D. Thus, we can classify the eigenstates of Hamiltonian~\eqref{eq:chiralpwavelattice} by a Chern number -- but due to the breaking of U(1) symmetry, the Chern number does not have the interpretation of Hall conductance. However, it is still a topological invariant. 

 We expect that $H_{\mathrm{BdG}}$ will exhibit several phases as a function of $\Delta$ and $\mu$ for a fixed $t>0.$ For simplicity let us set $t=1/2$ and make a gauge transformation $c_{\bs{p}} \to e^{\mathrm{i}\theta/2}c_{\bs{p}},\; c^{\dagger}_{\bs{p}}\to e^{-\mathrm{i}\theta/2}c^{\dagger}_{\bs{p}}$ where $\Delta=\vert\Delta\vert e^{i\theta}.$ The Bloch Hamiltonian for the lattice superconductor is then 
\begin{equation}
\mathcal{H}_{\mathrm{BdG}}(\bs{p})=\left(2-\mu-\cos p_x-\cos p_y\right)\sigma_z-2\vert\Delta\vert\sin p_x \sigma_{y}-2\vert\Delta\vert\sin p_y\sigma_{x},
\end{equation}
\noindent where the $\sigma_i,\ i=x,y,z,$ are the Pauli matrices in the particle/hole basis. Assuming $\vert\Delta\vert \neq 0$, this Hamiltonian has several fully-gapped superconducting phases separated by gapless critical points. The quasi-particle spectrum for the lattice model is 
\begin{equation}
E_{\pm}=\pm\sqrt{\left(2-\mu-\cos p_x-\cos p_y\right)^2+4\vert\Delta\vert^2 \sin^2 p_x+4\vert\Delta\vert^2 \sin^2 p_y}
\end{equation} 
and is gapped (under the assumption that $\vert\Delta\vert\neq 0$) unless the prefactors of all three Pauli matrices vanish simultaneously. As a function of $(p_x,p_y,\mu)$ we find three critical points. The first critical point occurs at $(p_x,p_y,\mu)=(0,0,0).$ The second critical point has two gap-closings in the BZ for the same value of $\mu:$ $(\pi,0,2)$ and $(0,\pi,2).$ The third critical point is again a singly degenerate point at $(\pi,\pi,4).$ We will show that the phases for $\mu<0$ and $\mu>4$ are trivial superconductors while the phases $0<\mu<2$ and $2<\mu<4$ are topological superconductors with opposite chirality. In principle one can define a Chern number topological invariant constructed from the eigenstates of the lower quasi-particle band to characterize the phases. We will show this calculation below, but first we make some physical arguments as to the nature of the phases, following the discussion in Ref.~\onlinecite{AndreiBook}.

We will first consider the phase transition at $\mu=0$. The low-energy physics for this transition occurs around $(p_x,p_y)=(0,0)$ and so we can expand the lattice Hamiltonian around this point; this is nothing but Eq.~\eqref{eq:chiralpwavecontinuum}. One way to test the character of the $\mu<0$ and $\mu>0$ phases is to make an interface between them. If we can find a continuous interpolation between these two regimes which is always gapped then they are topologically equivalent phases of matter. If we cannot find such a continuously gapped interpolation then they are topologically distinct. A simple geometry to study is  a domain wall where $\mu=\mu(x)$ such that $\mu(x)=-\mu_0$ for $x<0$ and $\mu(x)=+\mu_0$ for $x>0$ for a positive constant $\mu_0.$ This is an interface which is translationally invariant along the $y$-direction, and thus we can consider the momentum $p_y$ as a good quantum number to simplify the calculation. What we will now show is that there exist gapless, propagating fermions bound to the interface which prevent us from continuously connecting the $\mu<0$ phase to the $\mu>0$ phase. This is one indication that the two phases represent topologically distinct classes. 

The single-particle Hamiltonian in this geometry is 
\begin{eqnarray}
\mathcal{H}_{\mathrm{BdG}}(p_y)=\frac{1}{2}\left(\begin{array}{cc}-\mu(x) & 2\mathrm{i}\vert\Delta\vert\left(-\mathrm{i}\frac{d}{dx}+\mathrm{i}p_y\right) \\ -2\mathrm{i}\vert\Delta\vert\left(-\mathrm{i}\frac{d}{dx}-\mathrm{i} p_y\right) & \mu(x)\end{array}\right),
\label{eq:chiralquasi1d}
\end{eqnarray}\noindent 
where we have ignored the quadratic terms in $p$, and $p_y$ is a constant parameter, not an operator. This is a quasi-1D Hamiltonian that can be solved for each value of $p_y$ independently. We propose an ansatz for the gapless interface states:
\begin{eqnarray}
\vert\psi_{p_y}(x,y)\rangle=e^{\mathrm{i}p_y y}\exp\left(-\frac{1}{2\vert\Delta\vert}\int_{0}^{x}\mu(x')dx'\right)\vert\phi_0\rangle
\end{eqnarray}\noindent 
for a constant, normalized spinor $\vert\phi_0\rangle.$  The secular equation for a zero-energy mode at $p_y=0$ is 
\begin{equation}
\mathcal{H}_{\mathrm{BdG}}(p_y)\vert\psi_0(x,y)\rangle =0
\qquad
\implies\left(\begin{array}{cc} -\mu (x) & -\mu(x)\\  \mu(x) & \mu(x)\end{array}\right)\vert\phi_0\rangle=0.
\end{equation} 
The constant spinor which is a solution of this equation is $\vert\phi_0\rangle=1/\sqrt{2}\left(1,-1\right)^{\mathsf{T}}.$ This form of the constant spinor immediately simplifies the solution of the problem at finite $p_y.$ We see that the term proportional to $p_y$ in Eq.~\eqref{eq:chiralquasi1d} is $-2\vert\Delta\vert p_y\sigma_x.$ Since $\sigma_x\vert\phi_0\rangle=-\vert\phi_0\rangle,$ i.e., the solution $\vert\phi_0\rangle$ is an eigenstate of $\sigma_x,$ we conclude that $\vert\psi_{p_y}(x,y)\rangle$ is an eigenstate of $\mathcal{H}_{\mathrm{BdG}}(p_y)$ with energy $E(p_y)=-2\vert\Delta\vert p_y.$ Thus, we have found a normalizable bound state solution at the interface of two regions with $\mu<0$ and $\mu>0$ respectively. This set of bound states, parameterized by the conserved quantum number $p_y$ is gapless and chiral, i.e., the group velocity of the quasiparticle dispersion is always negative and never changes sign (in this simplified model). The chirality is determined by the sign of the ``spectral" Chern number mentioned above which we will calculate below. 

These gapless edge states have quite remarkable properties and are not the same chiral complex fermions that propagate on the edge of integer quantum Hall states, but chiral real (Majorana) fermions. Using Clifford algebra representation theory it can be shown that the so-called chiral Majorana (or Majorana-Weyl) fermions can only be found in spacetime dimensions $(8k+2)$, where $k=0,1,2,\cdots.$ Thus, we can only find chiral-Majorana states in $(1+1)$ dimensions or in $(9+1)$ dimensions (or higher!). 
In condensed matter, we are stuck with $(1+1)$ dimensions where we have now seen that they appear as the boundary states of chiral topological superconductors.  The simplest interpretation of such chiral Majorana fermions is as half of a conventional chiral fermion, i.e., its real or imaginary part. To show this, we will consider the edge state of a Chern number 1 quantum Hall system for a single edge
\begin{equation}
\mathcal{H}^{(\mathrm{QH})}_{\mathrm{edge}}=\hbar v\sum_{p}p\,\eta^{\dagger}_{p}\eta^{\pdag}_{p},
\end{equation}\noindent 
where $p$ is the momentum along the edge. 
The fermion operators satisfy $\left\{\eta^{\dagger}_{p},\eta^{\pdag}_{p'}\right\}=\delta_{pp'}.$ Similar to the discussion on the 1D superconducting wire we can decompose these operators into their real and imaginary Majorana parts
\begin{equation}
\eta_{p}=\frac{1}{2}(\gamma_{1,p}+\mathrm{i}\gamma_{2,p}),\qquad
\eta^{\dagger}_{p}=\frac{1}{2}(\gamma_{1,-p}-\mathrm{i}\gamma_{2, -p}),
\end{equation}
  where $\gamma_{a,p}$ ($a=1,2$)  are Majorana fermion operators satisfying $\gamma^{\dagger}_{a,p}=\gamma_{a,-p}$ and $\left\{\gamma_{a,-p},\gamma_{b,p^{'}}\right\}=2\delta_{ab}\delta_{pp^{'}}.$ The quantum Hall edge Hamiltonian now becomes 
\begin{equation}
\begin{split}
\mathcal{H}^{(\mathrm{QH})}_{\mathrm{edge}}
=&\hbar v\sum_{p\geq 0}p(\eta^{\dagger}_{p}\eta_{p}-\eta^{\dagger}_{-p}\eta_{-p})
\\
=&\frac{\hbar v}{4}\sum_{p\geq 0}p\left\{(\gamma_{1,-p}-\mathrm{i}\gamma_{2,-p})(\gamma_{1,p}+\mathrm{i}\gamma_{2,p})-(\gamma_{1,p}-\mathrm{i}\gamma_{2,p})(\gamma_{1,-p}+\mathrm{i}\gamma_{2,-p})\right\}
\\
=&\frac{\hbar v}{4}\sum_{p\geq 0}p\left(\gamma_{1,-p}\gamma_{1,p}+\gamma_{2,-p}\gamma_{2,p}-\gamma_{1,p}\gamma_{1,-p}-\gamma_{2,p}\gamma_{2,-p}\right)
\\ 
=&\frac{\hbar v}{2}\sum_{p\geq 0}p\left(\gamma_{1,-p}\gamma_{1,p}+\gamma_{2,-p}\gamma_{2,p}-2\right).
\end{split}
\end{equation}
\noindent 
Thus
\begin{equation}
\mathcal{H}^{(\mathrm{QH})}_{\mathrm{edge}}=\frac{\hbar v}{2}\sum_{p\geq 0}p\left(\gamma_{1,-p}\gamma_{1,p}+\gamma_{2,-p }\gamma_{2,p }\right)
\end{equation} 
up to a constant shift of the energy. This Hamiltonian is exactly two copies of a chiral Majorana Hamiltonian. The edge/domain-wall fermion Hamiltonian of the chiral $p$-wave superconductor will be 
\begin{equation}
\mathcal{H}^{(p-\mathrm{wave})}_{\mathrm{edge}}=\frac{\hbar v}{2}\sum_{p\geq0}p\gamma_{-p}\gamma_{p}.
\end{equation}

Finding gapless states on a domain wall of $\mu$ is an indicator that the phases with $\mu>0$ and $\mu<0$ are distinct. If they were the same phase of matter we should be able to adiabatically connect these states continuously. However, we have shown a specific case of the more general result that any interface between a region with $\mu>0$ and a region with $\mu<0$ will have gapless states that generate a discontinuity in the interpolation  between the two regions. The question remaining is: Is $\mu>0$ or $\mu<0$  non-trivial? The answer is that we have a trivial superconductor for $\mu<0$ (adiabatically continued to $\mu\to-\infty$) and a topological superconductor for $\mu>0$. Remember that for now we are only considering $\mu$ in the neighborhood of $0$ and using the continuum model expanded around $(p_x,p_y)=(0,0).$ We will now define a bulk topological invariant for 2D superconductors that can distinguish the trivial superconductor state from the chiral topological superconductor state. For the spinless Bogoliubov-deGennes Hamiltonian, which is of the form 
\begin{subequations}
\begin{eqnarray}
H_{\mathrm{BdG}}&=&\frac{1}{2}\sum_{\bs{p}}\Psi^{\dagger}_{\bs{p}}\,\left[\bs{d}(\bs{p},\mu)\cdot\bs{\sigma}\right]\Psi^{\pdag}_{\bs{p}},\\
\bs{d}(\bs{p},\mu)&=&\left(-2\vert\Delta\vert p_y,-2\vert\Delta\vert p_x,p^2/2m-\mu\right),
\end{eqnarray}
\end{subequations}
\noindent the topological invariant is the spectral Chern number defined in Eq.~\eqref{eq: Chern number 1}, which simplifies, for this Hamiltonian, to the winding number
\begin{eqnarray}
\mathsf{C}^{(1)}=\frac{1}{8\pi}\int \mathrm{d}^2 \bs{p}\; \epsilon^{ij}\;\hat{\bs{d}}\cdot\left(\partial_{p_i}\hat{\bs{d}}\times \partial_{p_j}\hat{\bs{d}}\right)
=\frac{1}{8\pi}\int \mathrm{d}^2 \bs{p}\; \frac{\epsilon^{ij}}{\vert {\bs{d}}\vert^3}\;\bs{d}\cdot\left(\partial_{p_i}\bs{d}\times \partial_{p_j}\bs{d}\right).
\end{eqnarray}
\noindent
We defined the unit vector $\hat{\bs{d}}=\bs{d}/\vert {\bs{d}}\vert$, which is possible since $|\bs{d}|\neq 0$ due to the existence of a gap. This integral has a special form and is equal to the degree of the mapping from momentum space onto the $2$-sphere $S^2$ given by $\hat{d}_{1}^{2}+\hat{d}_{2}^{2}+\hat{d}_{3}^{2}=1.$ As it stands, the degree of the mapping $\hat{\bs{d}}: \mathbb{R}^2\to S^2$ is not well-defined because the domain is not compact, i.e., $(p_x,p_y)$ is only restricted to lie in the Euclidean plane ($\mathbb{R}^2$). However, for our choice of the map $\hat{\bs{d}}$ we can define the winding number by choosing an equivalent, but compact, domain. To understand the necessary choice of domain we can simply look at the explicit form of $\hat{\bs{d}}(\bs{p})$
\begin{equation}
\hat{\bs{d}}(\bs{p})=\frac{\left(-2\vert\Delta\vert p_y,-2\vert\Delta\vert p_x,p^2/2m-\mu \right)}{\sqrt{4\vert\Delta\vert^2 p^2+\left(p^2/2m-\mu\right)^2}}.
\end{equation}
\noindent 
We see that $\lim_{\vert \bs{p}\vert\to\infty}\;\hat{\bs{d}}(\bs{p})\;=(0,0,1)$ and it does not depend on the direction in which we take the limit in the 2D plane. Because of the uniqueness of this limit we are free to perform the \emph{one-point compactification} of $\mathbb{R}^2$ which amounts to including the point at infinity in our domain. The topology of $\mathbb{R}^2\cup \{\infty\}$ is the same as $S^2$ and thus we can consider the degree of our map from the compactified momentum space ($S^2$) to the unit $\hat{\bs{d}}$-vector space ($S^2$). 
Using the explicit form of the $\hat{\bs{d}}$-vector for this model, we find
\begin{eqnarray}
\mathsf{C}^{(1)}=\frac{1}{\pi}\int \mathrm{d}^2 \bs{p} \frac{\vert\Delta\vert^2\left(\frac{p^2}{2m}+\mu\right)}{\left[4\vert\Delta\vert^2p^2+\left(\frac{p^2}{2m}-\mu\right)^2\right]^{3/2}}.
\end{eqnarray}
\noindent 
The evaluation of this integral can be easily carried out numerically. The result is $\mathsf{C}^{(1)}=0$ for $\mu<0$ and $\mathsf{C}^{(1)}=1$ for $\mu>0$, i.e., there are two different phases separated by a quantum critical point at $\mu=0.$ Thus we have identified the phase which is in the chiral superconductor state to be $\mu>0$. 

\subsubsection{Argument for the existence of Majorana bound states on vortices}
 
\begin{figure}
 \begin{center}
\includegraphics[page=3,width=0.80\textwidth]{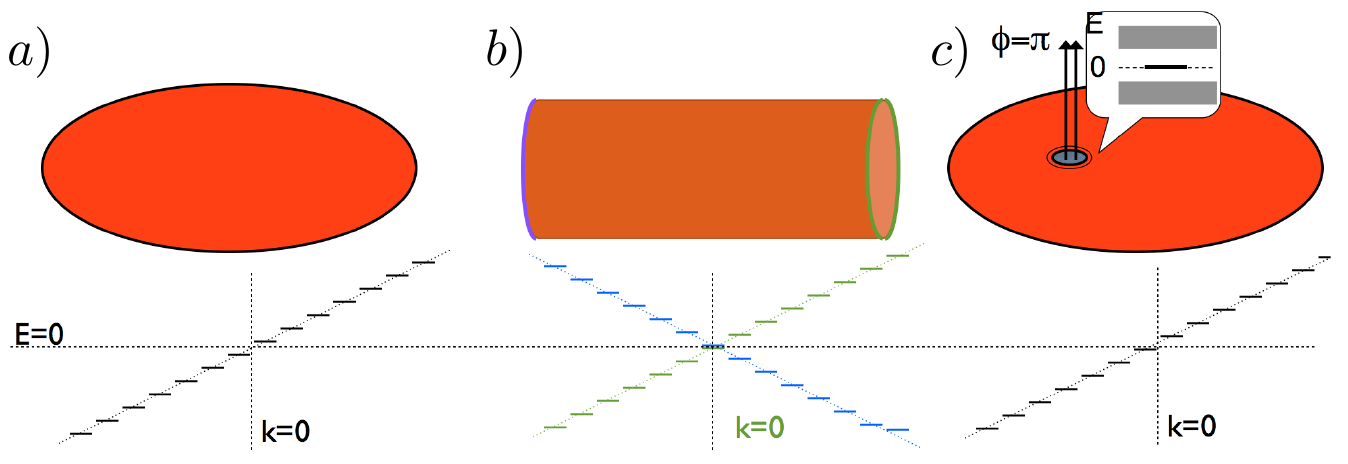}
\caption{
Spectra of a Chiral superconductor in different geometries: (a) disk, (b) cylinder, and (c) disk with flux defect. Shown are the spectra of the chiral topological boundary modes including their finite-size quantization with level spacing $v 2\pi/L$. If a $\pi$ flux is inserted in the disk geometry (c), it binds an isolated zero-energy state. At the same time, a single zero-energy state appears on the edge.}
\label{fig:MajoranaBoundVortexArgument}
\end{center}
\end{figure}

 A simple but rigorous argument can show us the presence of zero energy bound states in the core of vortices in a superconductor.  Assume we have a chiral $(p+\mathrm{i}p)$ superconductor in two geometries: a disk with an edge and a cylinder with two edges. Since it is a topological superconductor, the system will have chiral dispersing (Majorana) gapless modes along the edges. In Fig.~\ref{fig:MajoranaBoundVortexArgument}, the spectra are plotted versus the momentum along the edge, and they are qualitatively very different in the two cases. For an edge of length $L$, the smallest difference between two momenta along the edge is $2 \pi/L$. The energy difference between two levels is $v 2 \pi /L$, where $v$ is the velocity of the edge mode.

In a single particle superconducting Hamiltonian the number of total single-particle eigenvalues is always even. This is clear from the fact that whatever the spinor of the nonsuperconducting Hamiltonian is, when superconductivity is added, we have a doubled spectrum, so that every energy state at $E>0$ comes with a counterpart at energy $-E$. When labeled by momentum quantum number, for a system with just one edge, like the disk, there cannot be a single state at momentum $p=0$ at energy $E=0$. If such a state was there, the spectrum would contain an odd number of states. Hence the spectrum of the linearized edge mode cannot have a state at $E=0,$ $p=0$ on the disk. The one way to introduce such a state is to have antiperiodic boundary conditions, with the spectrum of the edge being at momenta $\pi(2n +1)/L$, $n\in\mathbb{Z}$. On the cylinder, as two edges are present, periodic boundary conditions are allowed (as are antiperiodic, which can be obtained by threading a flux through the cylinder). 

We now add a single vortex inside the disk, far away from the edge of the disk. What is the influence of the vortex on the edge? The vortex induces a phase $2 \pi$ in the units of the superconducting quantum $hc/2e$, which means that the phase of $\Delta$ changes by $2\pi$, and that of the electronic operators by $\pi$ upon a full rotation around the edge. This implies that the antiperiodic boundary conditions on the edge without vortex changes to periodic boundary conditions in the presence of the vortex. The spectrum on the edge then is translated by $\pi/L$ compared to the case without the vortex, making it have an energy level at $p=0, E=0$. This would mean that the spectrum has an odd number of levels. However, this cannot be true, as we explained above, since the number of levels is always even. We are hence missing one unpaired level. Where is it? Since the only difference from the case with no vortex is the vortex itself, we draw the conclusion that the missing level is associated with the vortex, and is a bound state on the vortex. We also draw the conclusion that, since it is unpaired and really bound to the vortex, it has to rest exactly at $E=0$, thereby showing that chiral superconductors have Majorana zero modes in their vortex core.

\subsubsection{Bound states on vortices in two-dimensional chiral $p$-wave superconductors}

Let us explicitly show that a vortex in a chiral superconductor will contain a zero mode.\cite{kopnin1991,volovik1999,read2000}  For this calculation, which is a variant of our calculation for the existence of a Majorana mode at the interface between a topological and a trivial superconductor, we follow the discussion in Ref.~\onlinecite{AndreiBook}.
For this construction 
consider a disk of radius $R$ which has $\mu>0$ surrounded by a region with $\mu<0$ for $r>R$. We know from our previous discussion that there will be a single branch of chiral Majorana states localized near $r=R$, but no exact zero mode. If we take the limit $R\to 0$ this represents a vortex and all the low-energy modes on the interface will be pushed to higher energies. If we put a $\pi$-flux inside the trivial region it will change the boundary conditions such that even in the $R\to 0$ limit there will be a zero-mode in the spectrum localized on the vortex. 

Now let us take the Bogoliubov-deGennes Hamiltonian in the Dirac limit ($m\to\infty$) and solve the Bogoliubov-deGennes equations in the presence of a vortex located at $r=0$ in the disk geometry in polar coordinates. Let $\Delta(r,\theta)=\vert\Delta(r)\vert e^{\mathrm{i}\alpha(\bs{r})}.$ The profile $\vert\Delta(r)\vert$ for a vortex will depend on the details of the model, but must vanish inside the vortex core region, e.g., for an infinitely thin core we just need $\vert\Delta(0)\vert=0.$ We take the phase $\alpha(\bs{r})$ to be equal to the polar angle at $\bs{r}$.

The first step in the solution of the bound state for this vortex profile is to gauge transform the phase of $\Delta(r,\theta)$ into the fermion operators via $\Psi(\bs{r})\rightarrow e^{\mathrm{i}\alpha({\bs{r}})/2}\Psi(\bs{r}).$ This has two effects: (i) it simplifies the solution of the Bogoliubov-deGennes differential equations and (ii) converts the boundary conditions of $\Psi(\bs{r})$ from periodic to anti-periodic around the vortex position $\bs{r}=0.$ In polar coordinates the remaining single-particle Bogoliubov-deGennes Hamiltonian is simply
\begin{equation}
\mathcal{H}_{\mathrm{BdG}}=\frac{1}{2}\left(\begin{array}{cc}-\mu & 2\vert\Delta(r)\vert e^{\mathrm{i}\theta}\left(\frac{\partial}{\partial r}+ \frac{\mathrm{i}}{r}\frac{\partial}{\partial\theta}\right)\\ -2\vert\Delta(r)\vert e^{-\mathrm{i}\theta}\left(\frac{\partial}{\partial r}- \frac{\mathrm{i}}{r}\frac{\partial}{\partial\theta}\right) &\mu\end{array}\right).
\end{equation}
\noindent We want to solve $\mathcal{H}_{\mathrm{BdG}}\Psi=E\Psi=0$ which we can do with the ansatz
\begin{eqnarray}
\Psi_0(r,\theta)=\frac{\mathrm{i}}{\sqrt{r}\cal{N}}\exp\left[-\frac{1}{2}\int_{0}^{r}\frac{\mu(r')}{\vert\Delta(r')\vert}\mathrm{d}r'\right]\left(\begin{array}{c} -e^{\mathrm{i}\theta/2}\\ e^{-\mathrm{i}\theta/2}\end{array}\right)\equiv \mathrm{i}g(r)\left(\begin{array}{c}-e^{\mathrm{i}\theta/2}\\e^{-\mathrm{i}\theta/2}\end{array}\right),
\end{eqnarray}
\noindent where ${\cal{N}}$ is a normalization constant.  The function $g(r)$ is localized at the location of the vortex. We see that $\Psi_{0}(r,\theta+2\pi)=-\Psi_{0}(r,\theta)$ as required. From an explicit check one can see that $\mathcal{H}_{\mathrm{BdG}}\Psi_{0}(r,\theta)=0.$ The field operator which annihilates fermion quanta in this localized state is
\begin{eqnarray}
\gamma=\int r\mathrm{d}r\mathrm{d}\theta \; \mathrm{i}g(r)\left[-e^{\mathrm{i}\theta/2}c(r,\theta)+e^{-\mathrm{i}\theta/2}c^{\dagger}(r,\theta)\right],
\end{eqnarray}\noindent 
from which we can immediately see that $\gamma=\gamma^{\dagger}.$ Thus the vortex traps a single Majorana bound state at zero-energy. 

\subsubsection{Non-Abelian statistics of vortices in chiral $p$-wave superconductors}
 \begin{figure}\begin{center}
\includegraphics[page=2,width=0.40\textwidth]{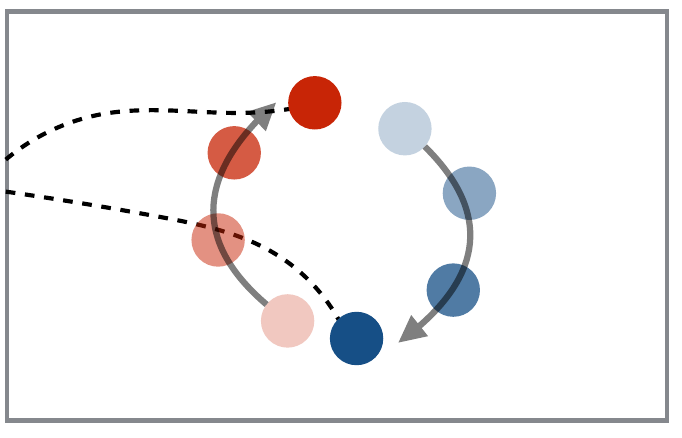}
\caption{
Illustration of the exchange of two vortices in a chiral $p$-wave superconductor. The dotted lines represent branch cuts across which the phase of the superconducting order parameter jumps by $2\pi.$}\label{fig:vortexexchange}
\end{center}
\end{figure}
We have shown in the last Section that on each vortex in a spinless chiral superconductor there exists a single Majorana bound state. If we have a collection of $2N$ vortices which are well-separated from each other, a low-energy subspace is generated which in the thermodynamic limit leads to a ground state degeneracy of $2^{N}$.~\cite{moore1991,nayak1996} For example, two vortices give a degeneracy of $2$, which can be understood by combining the two localized Majorana bound states into a single complex fermion state which can be occupied or un-occupied, akin to the end states of the superconducting wire. From $2N$ vortices one can form $N$ complex fermion states giving a degeneracy of $2^{N}$, which can be broken up into the subspace of $2^{N-1}$ states with even fermion parity and the $2^{N-1}$ states with odd fermion parity.  As an aside, since we have operators that mutually anti-commute and square to $+1$ we can define a  Clifford algebra operator structure using the set of $2N$ $\gamma_i.$

To illustrate the statistical properties of the vortices under exchanges we closely follow the work of Ivanov~\cite{ivanov2001} and the discussion in Ref.~\onlinecite{AndreiBook}. Let us begin with a single pair of vortices which have localized Majorana operators $\gamma_1$ and $\gamma_2$ respectively and are assumed to be well separated. We imagine that we adiabatically move the vortices in order to exchange the two Majorana fermions. If we move them slow enough then the only outcome of exchanging the vortices is a unitary operator acting on the two degenerate states which make up the ground state subspace. If we exchange the two vortices then we have $\gamma_1\to\gamma_2$ and $\gamma_2\to\gamma_1.$ However if we look at Fig.~\ref{fig:vortexexchange} we immediately see there is a complication. In this figure we have illustrated the exchange of two vortices and the dotted lines represent branch cuts across which the phase of the superconductor order parameter jumps by $2\pi.$ Since our solution of the Majorana bound states used the gauge transformed fermion operators we see that the bound state on the red vortex, which passes through the branch cut of the blue vortex, picks up an additional minus sign upon exchange. Thus the exchange of two vortices is effected by
\begin{equation}
\gamma_{1}\to \gamma_{2},\qquad
\gamma_{2}\to -\gamma_{1}.
\label{eq:majexchange}
\end{equation} 
In general, if we have $2N$ vortices, we can think of the different exchange operators $T_{ij}(\gamma_a)$ which for our choice of conventions send $\gamma_{i}\to\gamma_{j},$  $\gamma_{j}\to -\gamma_{i},$ and $\gamma_{k}\to\gamma_{k}$ for all $k\neq i,j.$  We can construct a representation of this exchange process on the Hilbert space by finding a $\tau(T_{ij})$ such that $\tau(T_{ij})\gamma_a\tau^{-1}(T_{ij})=T_{ij}(\gamma_a).$ Such a representation is given by
\begin{eqnarray}
\tau(T_{ij})=\exp\left(\frac{\pi}{4}\gamma_{j}\gamma_{i}\right)=\frac{1}{\sqrt{2}}\left(1+\gamma_{j}\gamma_{i}\right).
\end{eqnarray}
 Let us prove this by showing an explicit example for $T_{12}$, which will have the transformation given in Eq.~\eqref{eq:majexchange},
\begin{subequations}
\begin{eqnarray}
\tau(T_{12})\gamma_{1}\tau^{-1}(T_{12})&=&\frac{1}{2}\left(\gamma_{1}-\gamma_{1}\gamma_{2}\gamma_{1}+\gamma_{2}-\gamma_{1}\right)=\gamma_{2},
\\
\tau(T_{12})\gamma_{2}\tau^{-1}(T_{12})&=&\frac{1}{2}\left(\gamma_{2}-\gamma_{1}+\gamma_{2}\gamma_{1}\gamma_{2}-\gamma_{2}\right)=-\gamma_{1},
\\
\tau(T_{12})\gamma_{3}\tau^{-1}(T_{12})&=&\gamma_{3}\tau(T_{12})\tau^{-1}(T_{12})=\gamma_3.
\end{eqnarray} 
\end{subequations}

Now that we have this representation we can illustrate the non-Abelian statistics. We start with four vortices with Majorana operators $\gamma_{1},\gamma_{2},\gamma_{3},\gamma_{4}.$ To illustrate the action of the exchange operators on the four-fold degenerate ground state space we need to pair these Majorana operators into complex fermions
\begin{equation}
\begin{split}
a&=\frac{1}{2}(\gamma_{1}+\mathrm{i}\gamma_{2}),\;\;\;\;\; a^{\dagger}=\frac{1}{2}(\gamma_{1}-\mathrm{i}\gamma_{2}),\\
b&=\frac{1}{2}(\gamma_{3}+\mathrm{i}\gamma_{4}),\;\;\;\;\;b^{\dagger}=\frac{1}{2}(\gamma_{3}-\mathrm{i}\gamma_{4}).
\end{split}
\end{equation}
\noindent The basis vectors of the ground state subspace can now be written as 
\begin{equation}
\{\vert 0\rangle_{a}\otimes\vert 0\rangle_{b},\vert 1\rangle_{a}\otimes\vert 1\rangle_{b},\vert 1\rangle_{a}\otimes\vert 0\rangle_{b},\vert 0\rangle_{a}\otimes\vert 1\rangle_{b}\},
\label{eq: basis vortex}
\end{equation}
where we have ordered the basis so that states of the same fermion parity are together. The notation $\vert n\rangle_{a,b}$ means, $a^{\dagger}a^{\pdag}\vert n\rangle_{a}=n\vert n\rangle_{a}$ and $b^{\dagger}b^{\pdag}\vert n\rangle_{b}=n\vert n\rangle_{b} .$ The set of statistical exchanges is generated by $T_{12},T_{23},T_{34}$ and we want to understand how these exchanges act on the ground state subspace. We can rewrite these three operators as
\begin{subequations}
\begin{eqnarray}
\tau(T_{12})&=&\frac{1}{\sqrt{2}}(1+\gamma_{2}\gamma_{1})=\frac{1}{\sqrt{2}}\left[1-\mathrm{i}(aa^{\dagger}-a^{\dagger}a)\right]\\
\tau(T_{23})&=&\frac{1}{\sqrt{2}}\left[1-\mathrm{i}(ba-ba^{\dagger}+b^{\dagger}a-b^{\dagger}a^{\dagger})\right]\\
\tau(T_{34})&=&\frac{1}{\sqrt{2}}\left[1-\mathrm{i}(bb^{\dagger}-b^{\dagger}b)\right]
\end{eqnarray}
\end{subequations}
\noindent Taking matrix elements in our chosen ground state basis Eq.~\eqref{eq: basis vortex} we find
\begin{subequations}
\begin{eqnarray}
\tau(T_{12})&=&\frac{1}{\sqrt{2}}\left(\begin{array}{cccc} 
(1-\mathrm{i})&0&0&0\\
0&(1+\mathrm{i})&0&0\\
0&0&(1+\mathrm{i})&0\\
0&0&0&(1-\mathrm{i})\end{array}\right)\\
\tau(T_{23})&=&\frac{1}{\sqrt{2}}\left(\begin{array}{cccc} 
1&-\mathrm{i}&0&0\\
\mathrm{i}&1&0&0\\
0&0&1&-\mathrm{i}\\
0&0&\mathrm{i}&1\end{array}\right)\\
\tau(T_{34})&=&\frac{1}{\sqrt{2}}\left(\begin{array}{cccc} 
(1+\mathrm{i})&0&0&0\\
0&(1-\mathrm{i})&0&0\\
0&0&(1-\mathrm{i})&0\\
0&0&0&(1+\mathrm{i})\end{array}\right)
\end{eqnarray}
\end{subequations}
\noindent 
We see that with our basis choice $T_{12}$ and $T_{34}$ are Abelian phases acting on each state, while $T_{23}$ exhibits non-trivial mixing terms between the states with the same fermion parity. Thus, the form of $T_{23}$ represents non-Abelian statistics. Given an initial state $\vert\psi_{\mathrm{in}}\rangle=\vert 0\rangle_{a}\otimes \vert 0\rangle_{b},$ if we take vortex $2$ around vortex $3$ the final state is $\vert\psi_{\mathrm{f}}\rangle=\frac{1}{\sqrt{2}}\left(\vert 0\rangle_{a}\otimes \vert 0\rangle_{b}+\mathrm{i}\vert 1\rangle_{a}\otimes \vert 1\rangle_{b}\right).$  In principle one must also keep track of the Berry phase contribution to the statistical phase. Here we have only considered the wave function monodromy, however it can be proven that the Berry phase does not contribute in this case. The field of topological quantum computation is built on the idea that such exchange or braiding operations will lead to non-trivial quantum evolutions of the ground state which can be used for quantum computations. 
\index{Superconductors!p-wave!2d chiral superconductor|)}\index{Chiral superconductor!lattice model|)}\index{Majorana bound states!2d chiral superconductor vortices!non-Abelian statistics|)}\index{Chiral superconductor!Majorana vortex bound states!non-Abelian statistics|)}\index{Non-Abelian statistics|)}

\subsubsection{The 16-fold way}
\label{sec: Kitaev 16-fold}

We have now noticed that there are two characterizations of a topological superconductor, but they are seemingly different. First, the spectral Chern number is an integer $\mathsf{C}^{(1)} \in \mathbb{Z}$. Directly related to it is the number of chiral Majorana modes on the edge, which in turn is related to an experimental observable, the thermal conductivity on the edge. Hence the system has a $\mathbb{Z}$ index, which becomes obvious when an edge exists. We then saw that a $(p+\mathrm{i}p)$ superconductor (i.e., a topological superconductor with Chern number equal to one) with a vortex threaded through it exhibits a Majorana zero energy mode at the core of the vortex. A $(d+\mathrm{i}d)$ superconductor, with Chern number equal to 2, would exhibit two Majorana modes in the core of the vortex. However, those two Majorana modes would be unstable towards single particle hybridization terms, which would push them away from zero energy, and leave the core of the vortex with no states in it. The generalization tells us that an even Chern number topological superconductor has no Majorana zero modes in the vortex while an odd Chern number topological superconductor has one Majorana zero mode in its core. This shows that the defects (vortices) in a topological superconductor are classified by a $\mathbb{Z}_2$ number ($\mathsf{C}^{(1)} \mod 2$).

We now show that there is a third classification related to the idea of topological order.~\cite{Kitaev05} In the absence of an edge and in the absence of vortex defects, there is a $\mathbb{Z}_{16}$ classification of topological superconductors indexed by $\mathsf{C}^{(1)}\mod 16$, which can be put on solid grounds by the formalism of topological quantum field theory (TQFT) that we will introduce in the next Section. This shows that the edge-bulk correspondence needs revisiting -- the bulk does not know if we add $16$ edge modes or not, and hence that the edge contains more information than the bulk.~\cite{Cano} We first give a simple argument for the existence of a $\mathbb{Z}_{16}$ classification. 

We ask how we can classify the system in the absence of an edge. One way would be to compute the phases that wavefunctions can ackquire upon taking particles or quasiparticles around each other. However, the system is made out of electrons (its a superconductor), so usually nothing special can happen to phases of electrons. The only ``special" excitation of the superconductor is a vortex, so we will look at the phase that two vortices acquire upon exchange. We can calculate this with an argument. Take two copies of the $(p+\mathrm{i}p)$ superconductor governed by the Hamiltonian
\beq
H= \frac{\mathrm{i}}{4} \sum_{j,k} A_{jk} (\gamma_{1,j} \gamma_{1,k} + \gamma_{2,j} \gamma_{2,k} ),
\eneq 
written in terms of Majorana operators $\gamma_{1,j}$ for one copy and $\gamma_{2,j}$ for the other copy.
These operators can be combined into an complex fermion $c_j = (\gamma_{1,j}+ \mathrm{i} \gamma_{2,j})/2$ in terms of which the Hamiltonian becomes
\beq
H= \mathrm{i}\sum_{j,k} A_{jk} c_j^\dagger c_k.
\eneq 
This Hamiltonian has a ``fake'' U(1) symmetry given by our choice of $A_{jk}$ for both Hamiltonians. (Since the system is gapped, we expect our universal conclusions to hold even when this symmetry is stripped away). Thus, the system is a quantum Hall state of Hall conductance $\mathsf{C}^{(1)}$ (in units of $e^2/h$) if each of the superconductors had Chern number $\mathsf{C}^{(1)}$. We now ask what happens when we thread a superconducting vortex $h/2e$, which is equal to $\pi$. Threading a flux $2\pi$ in a quantum Hall state of Chern number $\mathsf{C}^{(1)}$ pulls $\mathsf{C}^{(1)}$ electron charges to the vortex core through the Hall effect, hence a $\pi$ flux pulls $\mathsf{C}^{(1)}/2$ electron charges towards the core. We then try to compute the phase acquired when a vortex is exchanged with another vortex. This is an exchange process, which is half a braid. A braid of two vortices is equivalent to $\mathsf{C}^{(1)}/2$ electrons braided with a $\pi$ vortex, giving rise to a phase $\pi \mathsf{C}^{(1)}/2$ upon a braid, and $\pi \mathsf{C}^{(1)}/4$ under exchange. Since this is the phase for exchange of vortices in two exactly identical superimposed superconductors, the  phase for exchange in one of them is half that, $\pi \mathsf{C}^{(1)}/8 = 2\pi \mathsf{C}^{(1)}/16$. This shows that the phase for vortex exchange is defined only mod $16$.  

We will show this more rigorously within the framework of TQFT that we will introduce axiomatically in the next Section. Before doing so, let us summarize what we have learned about the vortices in chiral superconductors with odd Chern number in a language that anticipates the formalism that we will introduce.   
We have seen that well-separated vortices hold a Majorana zero mode at their core. When these vortices come together, the two Majorana modes hybridize and split, giving rise to two states which differ by their fermion parity. Let us call the Bogoliubov-deGennes vacuum $1$ and the Bogoliubov quasiparticle $\psi$, and the Majorana fermion of the vortex $\sigma$. We can then formalize the fusion of two vortices by writing down a fusion rule
\beq
\sigma \times \sigma = 1 + \psi,
\label{eq: fusion of sigma sigma}
\eneq which basically tells us that combining two Majoranas can either go to a state with no fermion or at one with a fermion -- the fermion parity (and density) would be different for the two states. Which one it is depends on the microscopics of the model. Hence a quantum state of two Majoranas has to be described by another quantum number, which describes the ``fusion channel" of those two Majoranas -- either the vacuum or the Bogoliubov quasiparticle. 
The fusion rule~\eqref{eq: fusion of sigma sigma} allows for multiple fusion channels unlike the fusion rules~\eqref{eq: emf fusion rule} that we deduced for the toric code. This difference is a manifestation of the fact that the Majoranas are non-Abelian anyons, while the toric code anyons are Abelian.
When two Bogoliubov quasiparticles fuse, they condense (form a Cooper pair) and go to the vacuum
\beq
\psi \times \psi = 1,
\eneq
while the fusion of a Bogoliubov and a Majorana quasiparticle basically creates another Majorana
\beq
\psi \times \sigma = \sigma.
\eneq  
This can be rationalized by thinking of the complex Bogoliubov quasiparticle as made out of two Majoranas which then couple to the third Majorana. The Hamiltonian is a $3 \times 3$ antisymmetric matrix that necessarily has a zero eigenvalue which is another Majorana fermion coming as a result of the fusion.

\section{Category theory}
\label{sec: Category theory}

So far, we tried to gain some intuition about topologically ordered phases of matter 
by ways of examples. In this Section, we are going to define a framework that describes 
topological order in 2D space in a unified and axiomatic way. 
At the same time, our description strips all nonuniversal details off the problem. 
A field theory with these properties is known as a topological field theory. 
It does not contain any information about energy scales of the problem. 

The topological field theory that we study is based on the mathematical concepts of category theory.~\cite{Kitaev05,Nayak-etal,Bonderson07}
We will, however, try to keep the description as light as possible.
For our purpose, we can view category theory as a generalization of group theory which is based on the fusion rules between anyon species that we have already encountered in examples. Consistent implementation of fusion defines a fusion category. Subsequently, we can impose more structure on the fusion category which elevates it to a braiding category, or a braided tensor fusion category. Our presentation will follow Refs.~\onlinecite{Kitaev05,Bonderson07}, while giving more examples of the use of the theory.

\subsection{Fusion Category}

A fusion category is based on a finite number of topological sectors (also called anyons, topological charges, or simply particles) which we will label
\begin{equation}
a, b, c, \cdots.
\end{equation}
For every charge $a$ there exists a unique conjugate charge or anti-particle, that we denote by $\bar{a}$. It is possible that an anyon is its own antiparticle $a=\bar{a}$ (for example the Majorana $\sigma$). There exists a unique vacuum sector denoted 1 (or sometimes 0).
The fusion category is defined by its fusion rules
\begin{equation}
a\times b=\sum_cN^c_{ab}\, c,
\end{equation}
where $N^c_{ab}\in \mathbb{Z}_+$ are nonnegative integers. 
We have already encountered two examples of fusion rules in the previous Section, namely the toric code with charges $1$, $e$, $m$, and $f$,
\begin{equation}
\begin{split}
&1\times e=e,\qquad
1\times m=m,\qquad
1\times f=f,\\
&e\times m=f, \qquad
e\times f= m, \qquad
m\times f=e,\\
&e\times e=1, \qquad
m\times m= 1, \qquad
f\times f=1,
\end{split}
\label{eq: Toric code}
\end{equation}
and the so-called Ising anyon theory that we found realized by Majorana fermions in a chiral $p$-wave superconductor with charges $1$, $\sigma$, and $\Psi$,
\begin{equation}
\begin{split}
&1\times \sigma=\sigma,\qquad
1\times \psi=\psi
\\
&\sigma\times \sigma=1+\psi, \qquad
\sigma\times \psi= \sigma, \qquad
\psi\times \psi=1.
\end{split}
\label{eq: Ising}
\end{equation}
A principal difference between Eq.~\eqref{eq: Toric code} and Eq.~\eqref{eq: Ising} is that the former has always only one fusion product on the righthand side, while the fusion of two $\sigma$ in the latter produces two outcomes. Hence, the $\times$-product in the toric code can still be thought of as a group operation, while this is not possible in the Ising theory. We will see that this distinction coincides with the notion of an Abelian theory (toric code) and a non-Abelian theory (Ising). 

Does any choice of fusion rules, i.e., $N^c_{ab}\in \mathbb{Z}_+$, define a permissible fusion category? The answer to this question is negative, as we have to impose the following conditions on a fusion category.
\begin{itemize}
\item The  fusion rules be \emph{commutative} 
\begin{equation}
a\times b= b\times a\qquad \Rightarrow\qquad
N^c_{ab}=N^c_{ba}.
\end{equation}
\item The  fusion rules be \emph{associative} 
\begin{equation}
(a\times b)\times c= a\times (b\times c)
\qquad \Rightarrow\qquad
\sum_m N^m_{ab}N^n_{mc}=\sum_m N^n_{am}N^m_{bc}.
\end{equation}
If we define the matrix $N_a$ with matrix elements $(N_a)_{bc}=N^c_{ab}$, this relation 
becomes a vanishing commutator
\begin{equation}
[N_a,N_c]=0,
\label{eq: N commute}
\end{equation}
which implies that all fusion matrices $N_a$ are diagonalized by the same eigenvectors. 
We will exploit this fact later.
\item Fusion with the identity leaves any anyon unchanged 
\begin{equation}
a\times 1=a.
\end{equation}
\item The fusion product of $a$ with its antiparticle $\bar{a}$ contains the vacuum with prefactor $1$, i.e.,
\begin{equation}
a\times\bar{a}=1+\sum_{c\neq 1}N^c_{a\bar{a}} c.
\end{equation}
\item There exists a solution the a consistency condition called \emph{pentagon equation}, which we will discuss below.
\end{itemize}

\subsubsection{Diagrammatics}

Before we turn to the pentagon equation, we want to introduce a diagrammatic language that will facilitate computations within the fusion and braiding categories. 
In this formalism, we denote anyon $a$ traveling forward in time as an upward oriented line. It is the same as the associated anti-particle traveling backward in time
\begin{equation}
\vcenter{\hbox{\includegraphics[page=3, scale=0.45]{BraidingFigures-cropped.pdf}}}\ .
\end{equation}

A diagram with open anyon worldlines at the top and bottom represents a state in a Hilbert space that depends on the number and types of open anyon worldlines. A diagram without open worldlines represents an amplitude or complex number. The simplest nontrivial Hilbert space is the \emph{fusion space} $V_{ab}^c$. Its dimension is given by the number of ways that anyons $a$ and $b$ can fuse into $c$, i.e., $\mathrm{dim}\,V_{ab}^c=N_{ab}^c$. A basis in $V_{ab}^c$ is denoted by
\begin{equation}
\langle a,b;c,\mu|=:\left(\frac{d_c}{d_a d_b}\right)^{1/4}
\vcenter{\hbox{\includegraphics[page=1, scale=0.45]{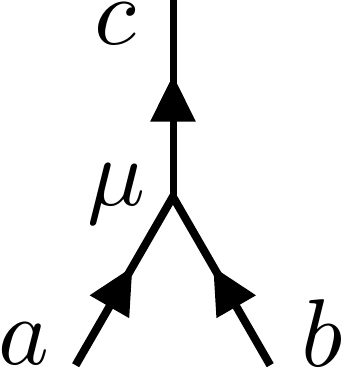}}}
\in V_{ab}^c.
\end{equation}
Here, $\mu=1,\cdots, N_{ab}^c$ labels the fusion multiplicity and the real positive prefactor $(d_c/d_a\,d_b)^{1/4}$ should be understood as a normalization constant at this stage. We adopt the normalization of Ref.~\onlinecite{Bonderson07}. 
Likewise, we define the \emph{splitting space} $V^{ab}_c$ of the same dimension $\mathrm{dim}\,V^{ab}_c=N_{ab}^c$ and write a basis as
\begin{equation}
| a,b;c,\mu\rangle=:\left(\frac{d_c}{d_a d_b}\right)^{1/4}
\vcenter{\hbox{\includegraphics[page=2, scale=0.45]{BraidingFigures-cropped.pdf}}}
\in V^{ab}_c.
\end{equation}

Two propagating particles $a$ and $b$ live in a vector space $V^{ab}_{ab}=\bigoplus_c V^{ab}_c\otimes V_{ab}^c$. The identity element $I_{ab}$ in $V^{ab}_{ab}$ is then represented by the \emph{completeness} relation
\begin{equation}
I_{ab}=\sum_{c,\mu} | a,b;c,\mu\rangle \langle a,b;c,\mu|
\end{equation}
which we represent pictorially as
\begin{equation}
\vcenter{\hbox{\includegraphics[page=4, scale=0.45]{BraidingFigures-cropped.pdf}}}
=
\sum_{c,\mu}
\sqrt{\frac{d_c}{d_a d_b}}
\vcenter{\hbox{\includegraphics[page=5, scale=0.45]{BraidingFigures-cropped.pdf}}}.
\label{eq: completeness rel}
\end{equation}

The basis vectors in $V^{ab}_c$ and  $V_{ab}^c$ furthermore satisfy the \emph{orthogonality} relation
\begin{equation}
\langle a,b;c,\mu|a,b;c',\mu'\rangle =\delta_{c,c'}\delta_{\mu\mu'}
\end{equation}
which we represent pictorially as
\begin{equation}
\vcenter{\hbox{\includegraphics[page=7, scale=0.45]{BraidingFigures-cropped.pdf}}}
=
\delta_{c,c'}\delta_{\mu\mu'}
\sqrt{\frac{d_a d_b}{d_c}}\ 
\vcenter{\hbox{\includegraphics[page=6, scale=0.45]{BraidingFigures-cropped.pdf}}}.
\label{eq: ortho rel}
\end{equation}

In particular, we can choose $c=1$ (a dashed worldline) to obtain
\begin{equation}
\vcenter{\hbox{\includegraphics[page=9, scale=0.45]{BraidingFigures-cropped.pdf}}}
=
\vcenter{\hbox{\includegraphics[page=10, scale=0.45]{BraidingFigures-cropped.pdf}}}
=\sqrt{\frac{d_a d_{\bar{a}}}{d_1}}.
\end{equation}
We can use this relation to determine the normalization constants $d_a$, which are called \emph{quantum dimensions}. We have the freedom to choose $d_1=1$ and note that for all examples discussed here $d_a=d_{\bar{a}}$. It follows that 
\begin{equation}
d_a=\vcenter{\hbox{\includegraphics[page=11, scale=0.45]{BraidingFigures-cropped.pdf}}}.
\end{equation}

\subsubsection{F-moves and the pentagon equation}
As noted above, we need to impose a further consistency condition to complete the definition of a fusion category. 
For this, we generalize the notion of associativity which we imposed on the fusion coefficients by imposing associativity on the basis. The consistency equations basically say that observables only depend on the state of the particles at the beginning (the fusion channel) and at the end. Nothing in between can matter, up to phases and rotations in possibly degenerate spaces. We consider the Hilbert space of particle $d$ splitting into three (not two) particles $a$, $b$, $c$
\begin{equation}
V^{abc}_d 
=\sum_eV^{ab}_e \otimes V^{ec}_d 
=\sum_fV^{af}_d \otimes V^{bc}_f.
\end{equation}
There is hence a unitary transformation $F^{abc}_d$ (``$F$-move'') between the two vector spaces
\begin{equation}
|a,b;e,\alpha\rangle\otimes |e,c;d,\beta\rangle 
=\sum_{f,\mu,\nu}
\left[F^{abc}_d\right]_{(e,\alpha,\beta),(f,\mu,\nu)} 
|b,c;f,\mu\rangle\otimes |a,f;d,\nu\rangle 
\label{eq: def F symbol}
\end{equation}
which reads diagrammatically
\begin{equation}
\vcenter{\hbox{\includegraphics[page=12, scale=0.45]{BraidingFigures-cropped.pdf}}}
=\sum_{f,\mu,\nu}
\left[F^{abc}_d\right]_{(e,\alpha,\beta),(f,\mu,\nu)} 
\vcenter{\hbox{\includegraphics[page=13, scale=0.45]{BraidingFigures-cropped.pdf}}}.
\end{equation}
For a fusion category to be \emph{unitary}, we require that $(F^{abc}_d)^\dagger = (F^{abc}_d)^{-1}$ (as a
matrix). Notice that $F^{abc}_d$ is trivial if any of $a, b, c = 1$. 

\begin{figure}[t]
\begin{center}
\includegraphics[page=14, scale=0.45]{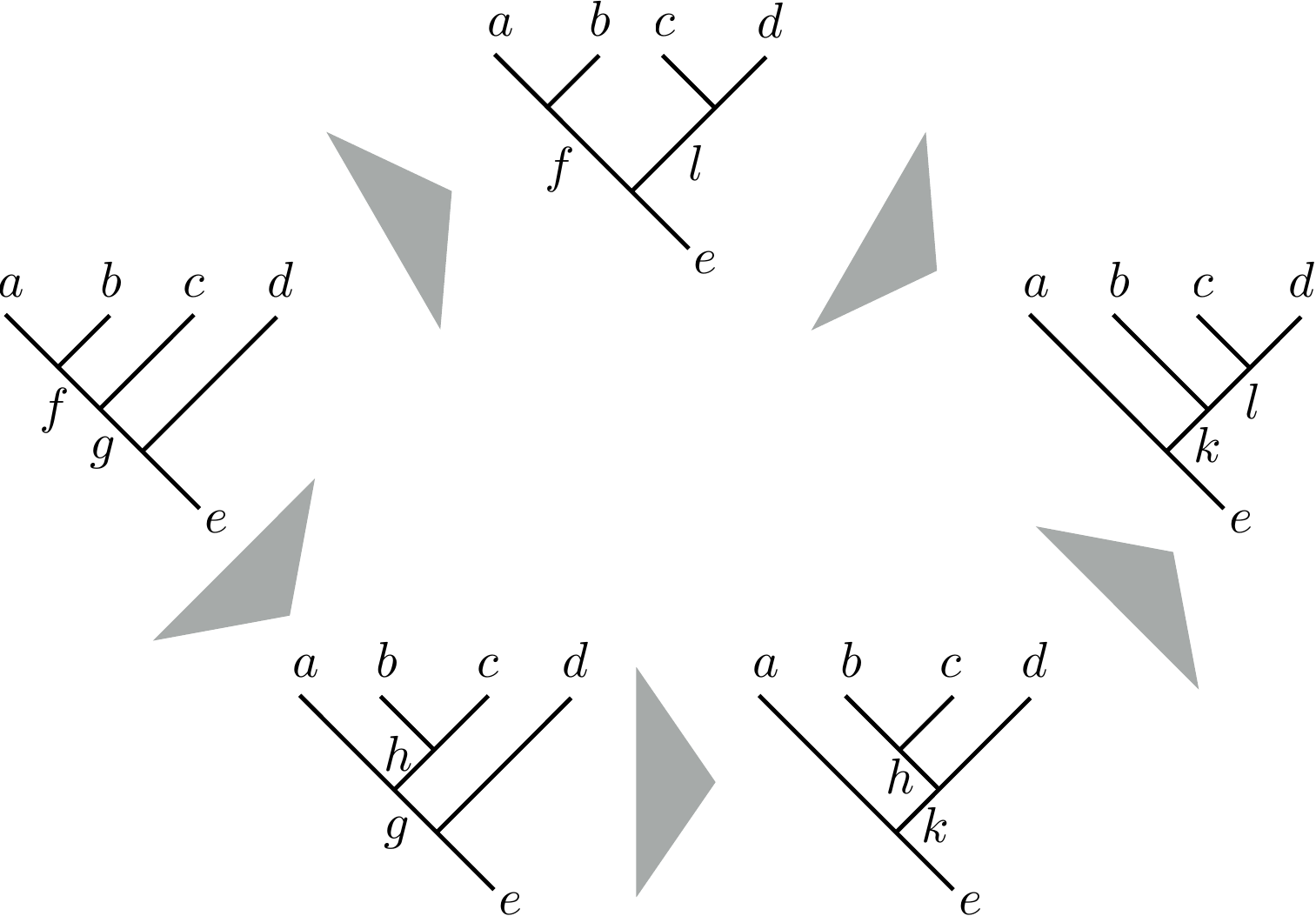}
\caption{The pentagon equation defines a consistency relation that has to be imposed on a fusion category.}
\label{fig: pentagon}
\end{center}
\end{figure}

The pentagon equation is the consistency condition diagrammatically represented in Fig.~\ref{fig: pentagon}. It shows that there are two ways to build the same mapping between two vector spaces out of $F$-moves. Each distinct solution, up to gauge freedom, of the pentagon equation
\begin{equation}
\left[F^{fcd}_e\right]_{gl}\left[F^{abl}_e\right]_{fk}
=
\sum_h
\left[F^{abc}_g\right]_{fh}\left[F^{ahd}_e\right]_{gk}\left[F^{bcd}_k\right]_{hl}
\end{equation}
is a distinct fusion category (with the same fusion rules). 
Here and below we have suppressed the Greek indices that correspond to the fusion multiplicities. We will focus on theories of multiplicity 1 only, i.e., all $N^c_{ab}$ fusion coefficients will be either 0 or 1. As the $F$-moves relate different basis states, not all of them are gauge invariant (see below). However, there are some $F$ symbols which are gauge invariant. Those are related to invariants of the theory called Frobenius Schur indicators.

\subsubsection{Gauge freedom and its fixing}

Consider a gauge transformation on the basis states $|a,b;c\rangle$ 
\be
	\ket{a,b;c}' = u^{ab}_c \ket{a,b;c}.
\end{equation}
We only consider the case without multiplicities, i.e., $N^{ab}_c =0,1$, for which $u^{ab}_c \in\mathbb{C}$ are scalars with $|u^{ab}_c|=1$. Likewise, if $N^{bc}_f =0,1$  the $F$ symbols are scalars with $\left|\left[F^{abc}_{d}\right]_{ef}\right|=1$.
In view of the definition Eq.~\eqref{eq: def F symbol}, the $F$ symbols are not invariant under the gauge transformation. 
They transform as
\be
\left[F^{abc}_{d}\right]^\prime_{ef} = \left[F^{abc}_{d}\right]_{ef} \frac{u^{ab}_e u^{ec}_d}{u^{af}_d u^{bc}_f}.
\ee
Furthermore, we need to set 
\be 
u^{1b}_c = \delta_{bc}, 
\ee
since the fusion of the identity particle can be added at any point in time to the worldline of any particle $b$ without changing the state.

From this, we can conclude that the following $F$ symbols are gauge invariant
\beq
\left[F^{1bc}_{d}\right]_{ef} = \left[F^{1bc}_{d}\right]_{bd},\qquad
\left[F^{a1c}_{d}\right]_{ef} = \left[F^{a1c}_{d}\right]_{ac},\qquad 
\left[F^{ab1}_d\right]_{ef} = \left[F^{ab1}_{d}\right]_{d b}.
\eneq 
In a theory with no multiplicity, all these $F$ symbols are equal to unity because they represent identity maps from spaces $\ket{b,c;d}$ into the same space
\be
\left[F^{1bc}_{d}\right]_{ef} = \left[F^{1bc}_{d}\right]_{bd}=
\left[F^{a1c}_{d}\right]_{ef} = \left[F^{a1c}_{d}\right]_{ac}= 
\left[F^{ab1}_d\right]_{ef} = \left[F^{ab1}_{d}\right]_{d b}=1.
\label{eq: gauge fixed F symbols}
\ee

\subsubsection{Quantum dimensions and Frobenius Schur indicators}

Having completed the definition of a fusion category, we now explore its structure. 
First, we shall properly define the quantum dimension $d_a$ of an anyon $a$ that has already entered several
relations as a normalization factor. 
Physically the definition of $d_a$ can be obtained from imposing isotopy invariance, which means the ability to remove bends in particle worldlines. This should be possible as long as lines are not crossed and end points are not moved. Bending a line slightly (so that the line always flows upward) is a trivial allowed move, but a complication arises when a line is bent so much that is acquires a turning point. The $F$-move associated with this type of bending is 
\begin{equation}
\vcenter{\hbox{\includegraphics[page=19, scale=0.45]{BraidingFigures-cropped.pdf}}}
=
\left[F^{a\bar{a}a}_a\right]_{1,1} 
\vcenter{\hbox{\includegraphics[page=20, scale=0.45]{BraidingFigures-cropped.pdf}}}
=d_a
\left[F^{a\bar{a}a}_a\right]_{1,1} 
\vcenter{\hbox{\includegraphics[page=21, scale=0.45]{BraidingFigures-cropped.pdf}}}.
\end{equation}
Notice that the symbol $\left[F_a^{a\bar{a}a}\right]_{11}$ is gauge invariant. Hence its value is a topological invariant. Since we know that the line in the left diagram is isotopically equivalent to a line going up, it should be, up to a phase, equal to a line going up. We conclude that $\left[F_a^{a\bar{a}a}\right]_{11}= \chi_a/d_a$, where $\chi_a$ is a phase called the Frobenius Schur indicator. If $a$ is its own antiparticle, $\chi_a$ has to equal either $+ 1$ or $-1$. Since it can take different values, it is a topological invariant characterizing the fusion category. Interestingly, there are other Frobenius Schur indicators that characterize the theory as topological invariants. For example, one which we will not further elaborate on is connected to the trivalent vertex.~\cite{Bonderson07}

We now know how to compute the quantum dimension through the $F$-symbols. However, there is another, easier way of computing the quantum dimensions. This is obvious once the space of states has been endowed with a completeness and an orthonormality relation, that we gave in Eq.~\eqref{eq: completeness rel} and Eq.~\eqref{eq: ortho rel}, respectively. 
One can use them to show the identity
\beq
d_a d_b =\sum_c N_{ab}^c d_c,
\label{eq: eigenvalue eq N}
\eneq
for
\begin{equation}
\begin{split}
d_a d_b
=
&\,\vcenter{\hbox{\includegraphics[page=15, scale=0.45]{BraidingFigures-cropped.pdf}}}
=
\sum_{c}
\sqrt{\frac{d_c}{d_a d_b}}\
\vcenter{\hbox{\includegraphics[page=16, scale=0.45]{BraidingFigures-cropped.pdf}}}
=
\sum_{c}
\sqrt{\frac{d_c}{d_a d_b}}\ 
\vcenter{\hbox{\includegraphics[page=17, scale=0.45]{BraidingFigures-cropped.pdf}}}
\\
=
&\,
\sum_{c}
N^c_{ab}
\vcenter{\hbox{\includegraphics[page=18, scale=0.45]{BraidingFigures-cropped.pdf}}}
=
\sum_{c}
N^c_{ab}
d_c.
\end{split}
\end{equation}

Equation~\eqref{eq: eigenvalue eq N} is key to understand how the quantum dimensions follow from the fusion rules. It is again useful to render the fusion coefficients in a matrix form $(N_a)_{bc}=N^c_{ab}$ ($b$ and $c$ are the indices of the matrix $N_a$). Then, Eq.~\eqref{eq: eigenvalue eq N} is nothing but an eigenvalue equation for $N_a$. 
We see that $d_a$ is an eigenvalue of $N_a$ and its eigenvector is the vector that contains all quantum dimensions $d_c$. The existence of the real positive eigenvalue $d_a$ is a highly nontrivial fact. The Perron-Frobenius theorem, proved by Oskar Perron (1907) and Georg Frobenius (1912), asserts, in its weak version, that a real square matrix with nonnegative entries has a largest positive eigenvalue and that the corresponding (possibly degenerate) eigenvector has nonnegative components. 
We would like to use it to show that the eigenvalue $d_a$ is the largest eigenvalue of the matrix $N_a$. By assumption, the vector with entries $d_c$ has only strictly positive components. Suppose we have another eigenvector $v$ of $N_a$ with nonnegative components $v_c\geq0$ and eigenvalue $\mu_a$. Then the strict equality 
\begin{equation}
\sum_c v_c d_c>0
\end{equation}
holds since all $d_c$ are strictly positive and at least one $v_c$ is strictly positive as well. From
\begin{equation}
d_a \left(\sum_c d_c v_c\right)
=\sum_{b,c}d_b N_{ab}^c v_b
=\mu_a \left(\sum_c d_c v_c\right),
\end{equation} 
it thus follows that the eigenvalues $d_a$ and $\mu_a$ are equal. In other words, \emph{any} nonnegative eigenvector of $N_{a}$ has the same eigenvalue $d_a$. This includes the eigenvector of the largest eigenvalue of $N_{a}$, which is nonnegative due to the Perron-Frobenius theorem. Hence, $d_a$ is the largest eigenvalue of  $N_{a}$.


Using the fact that $d_a$ is the largest eigenvalue, we can now give a more physical interpretation of the quantum dimension. For that, consider the fusion of some anyon $a$ with itself $n$ times
\beq
\underbrace{a\times a \times  \cdots \times a}_{n}  = 
\sum_{c_1, c_2, \cdots, c_{n-1}} (N_a)_{ a c_1} \,(N_a)_{c_1 c_2}\,   \times\cdots \times (N_a)_{c_{n-2} c_{n-1} }\, c_{n-1}.
\eneq 
The righthand side contains the $(n-1)$-th power of the matrix $N_a$. 
Hence, approximating $N_a$ by its highest eigenvalue, we conclude the the dimension of the fusion space of $n$ anyons of type $a$ is dominated by the quantum dimension $\mathrm{dim}(\bigoplus_{c} V_{a\ldots a}^{c} )\sim d_a^n$ for large $n$. In other words, the quantum dimension tells us how fast the Hilbert space for a particle grows! Any non-Abelian particle has a quantum dimension strictly larger than unity (Abelian particles have quantum dimension unity). 

\subsubsection{Examples}

Before moving on to impose more structure on the fusion category in order to obtain a braiding category, we shall briefly follow up on the two examples of semion TQFT and Ising TQFT. 

\emph{Semion TQFT} --- The simplest nontrivial TQFT has one particle $s$ besides the identity and the semion fusion rules
\begin{equation}
s\times s=1, \qquad 1\times s=s.
\label{eq: semion fusion rules}
\end{equation}
The theory is Abelian, i.e., $d_s=1$. 
Let us solve the pentagon equation for this theory. There is only one $F$-symbol, which is not entirely determined by gauge fixing alone, namely $\left[F^{sss}_s\right]_{11}$. We can deduce its allowed values from the pentagon equation
\begin{equation}
\left[F^{1ss}_1\right]_{s1}\left[F^{ss1}_1\right]_{1s}
=
\left[F^{sss}_s\right]_{11}\left[F^{s1s}_1\right]_{ss}\left[F^{sss}_s\right]_{11}
\end{equation}
which, using Eq.~\eqref{eq: gauge fixed F symbols}, yields the two possibilities
\be
\left[F^{sss}_s\right]_{11}=\pm1.
\ee
In fact, $\left[F^{sss}_s\right]_{11}$ is equal to the Frobenius Schur indicator mentioned above and the two values $\pm1$ distinguish two different fusion categories. The choice $+1$ is trivial, while the $-1$ is what is commonly called the semion TQFT. For example, it is realized in the $\nu=1/2$ Laughlin state of bosons in the fractional quantum Hall effect. 

\emph{Ising TQFT} --- We gave the fusion rules of the non-Abelian Ising TQFT in Eq.~\eqref{eq: Ising}. We now want to use Eq.~\eqref{eq: eigenvalue eq N} to compute the quantum dimensions of the anyons. For that we note that the fusion matrix [in the basis ($1$, $\sigma$, $\psi$)] of the $\sigma$ anyon is given by
\be
N_\sigma=
\begin{pmatrix}
0&1&0\\1&0&1\\0&1&0
\end{pmatrix}.
\ee 
Its eigenvalues are given by $\pm\sqrt{2}$ and $0$, the largest of which $d_\sigma=\sqrt{2}$ is the quantum dimension of the $\sigma$ particle. Its corresponding eigenvector $(d_1,d_\sigma, d_\psi)=(1,\sqrt{2},1)$ indeed contains the quantum dimensions of all anyons. 
The quantum dimension $\sqrt{2}$ is compatible with our explicit calculation of the degeneracy resulting from Majorana bound states in the vortices of a $p$-wave superconductor. For example, 2 vortices with one Majorana state each gave rise to a degeneracy $\sqrt{2}^2=2$ of the state.

\subsection{Braiding Category}

A bare fusion category has no means to relate the two fusion spaces $V^{ab}_c$ and $V^{ba}_c$. The physical operation that corresponds to such a relation is an exchange between the particles $a$ and $b$.
In a braiding category, exchange is a map $R_{ab} : V^{ab}_c \rightarrow V^{ba}_c$. 
A double exchange is an automorphism in a given fusion space $R_{ba} R_{ab} : V^{ab}_c \rightarrow V^{ab}_c$. For fusion processes without multiplicities, for which the space is one-dimensional, $R_{ba} R_{ab}$ is thus a phase (it is represented by a matrix if the multiplicity $N_{ab}^c$ is larger than 1). Upon braiding the state changes by flipping the particles, by convention,
\beq
R_{ab} \ket{a,b;c,\mu} = \sum_\nu [R_c^{ab}]_{\mu \nu} \ket{b,a;c,\nu}.
\eneq
Diagrammatically, this operation (``$R$ move") is represented as
\begin{equation}
\vcenter{\hbox{\includegraphics[page=22, scale=0.45]{BraidingFigures-cropped.pdf}}}
=
R_c^{ab}
\vcenter{\hbox{\includegraphics[page=23, scale=0.45]{BraidingFigures-cropped.pdf}}},
\end{equation}
in the case without multiplicities. 

\begin{figure}[t]
\begin{center}
\includegraphics[page=24, scale=0.45]{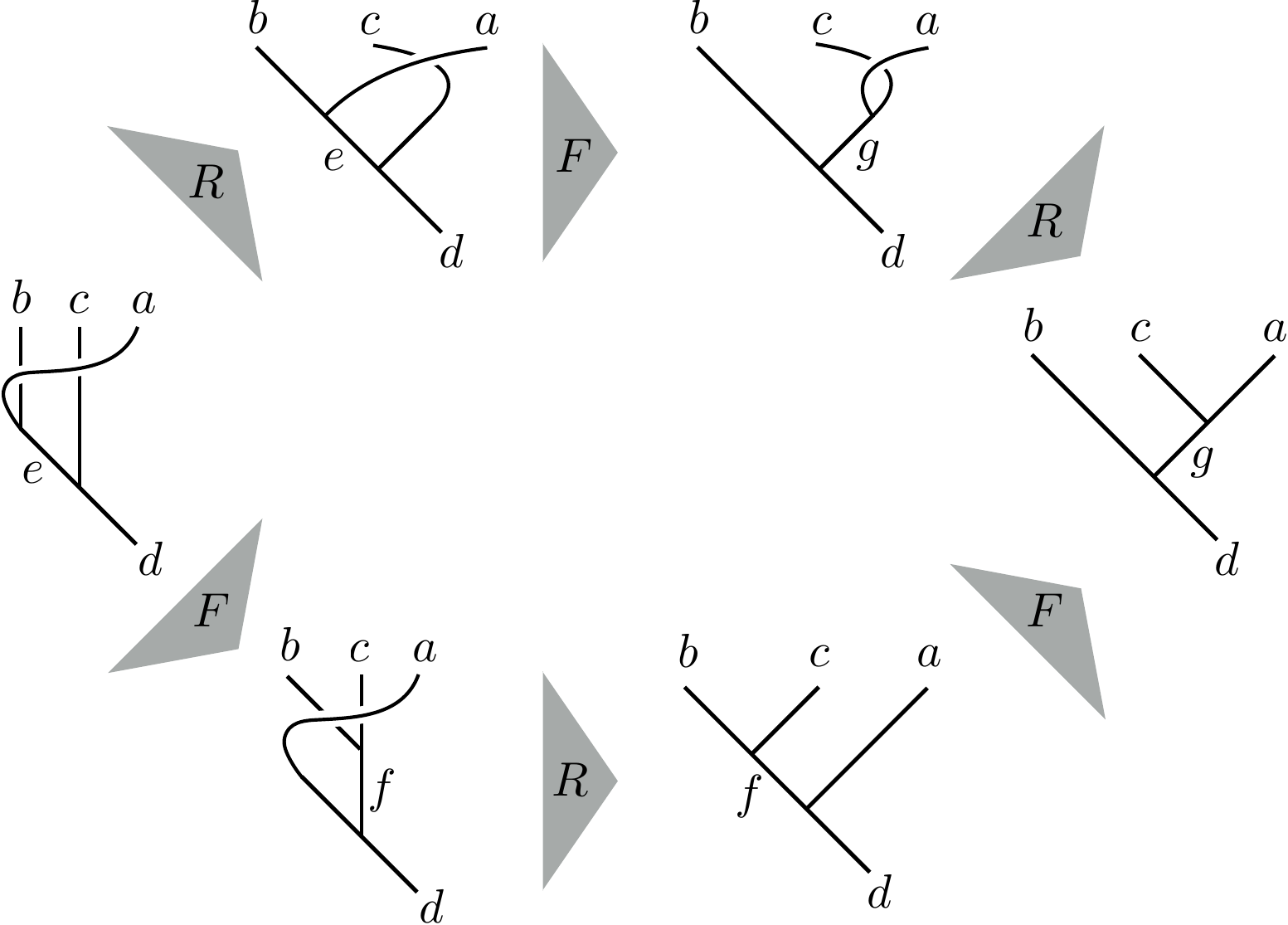}
\caption{The hexagon equation defines a consistency relation that has to be imposed on a braiding category.}
\label{fig: hexagon}
\end{center}
\end{figure}

For this braiding relation to define a consistent braiding category, the $R$ and $F$ symbols have to satisfy 
two consistency  relations called hexagon equations. (The two hexagon equations differ only in the directions of the braids involved; one uses $R$, the other one uses $R^{-1}$. For simplicity, we only present the equation for $R$ here.) As with the pentagon equation for the $F$ symbols alone, the hexagon equation represents a way to mix $F$ and $R$ moves to get between the same two diagrams in two different ways. Analytically, it reads
 \beq
 R_e^{ab} \left[F_d^{bac}\right]_{eg} R_g^{ac} = \left[F_d^{abc}\right]_{ef}R_d^{af} \left[F_d^{bca}\right]_{fg}
 \eneq
 and they are shown diagrammatically in Fig~\ref{fig: hexagon}. The matrix $R_c^{ab}$ is unitary, and if any of $a$ or $b$ is the identity particle then the $R$ matrix is unity (braiding with the vacuum is trivial)
 \be
R^{1b}_c=R^{a1}_c=1. 
 \ee

\subsubsection{Topological spin}

Every quasiparticle type $a$ in a braiding category carries another topological quantum number besides its quantum dimension, namely the complex-valued topological spin $\theta_a$ with $|\theta_a|=1$.
Physically, the topological spin can be interpreted as the phase resulting from a rotation of the particle around its own axis by $2 \pi$. It is $-1$ for fermions, but can be fractional for other particles. Diagrammatically, we can define $\theta_a$ as
\begin{equation}
\vcenter{\hbox{\includegraphics[page=25, scale=0.45]{BraidingFigures-cropped.pdf}}}
=
\theta_a
\vcenter{\hbox{\includegraphics[page=26, scale=0.45]{BraidingFigures-cropped.pdf}}}
=
\vcenter{\hbox{\includegraphics[page=27, scale=0.45]{BraidingFigures-cropped.pdf}}}\ ,
\qquad
\vcenter{\hbox{\includegraphics[page=28, scale=0.45]{BraidingFigures-cropped.pdf}}}
=
\theta_a^*
\vcenter{\hbox{\includegraphics[page=26, scale=0.45]{BraidingFigures-cropped.pdf}}}
=
\vcenter{\hbox{\includegraphics[page=29, scale=0.45]{BraidingFigures-cropped.pdf}}}\ .
\label{eq: first def of theta a}
\end{equation}
We bend the world line of a particle $a$, bring it down, then bring it back to itself, this time braided, then take it to infinity. This corresponds, up to a phase, to the configuration without bending.
 
To see how the topological spin is related to the $R$ symbols, we want to evaluate a diagram that represents an amplitude, i.e., it is closed. For that, we simply connect up the open worldlines in Eq.~\eqref{eq: first def of theta a} with an $\bar{a}$ worldline and obtain the definition
\be
\theta_a:=\frac{1}{d_a}\ 
\vcenter{\hbox{\includegraphics[page=30, scale=0.45]{BraidingFigures-cropped.pdf}}}\ .
\label{eq: def pictorially topo spin}
\ee
We evaluate this diagram by inserting an identity and applying an $R$ move (by the definition of the $R$ symbols, exchange is only defined between worldlines that are in a definite fusion channel)
\be
\begin{split}
\vcenter{\hbox{\includegraphics[page=30, scale=0.45]{BraidingFigures-cropped.pdf}}}
=&\,
\sum_c\sqrt{\frac{d_c}{d_a^2}}\
\vcenter{\hbox{\includegraphics[page=31, scale=0.45]{BraidingFigures-cropped.pdf}}}
=
\sum_c\sqrt{\frac{d_c}{d_a^2}}\
R^{aa}_c
\vcenter{\hbox{\includegraphics[page=32, scale=0.45]{BraidingFigures-cropped.pdf}}}
\\
=&\,
\sum_c\sqrt{\frac{d_c}{d_a^2}}\
R^{aa}_c
\vcenter{\hbox{\includegraphics[page=33, scale=0.45]{BraidingFigures-cropped.pdf}}}
=
\sum_c\sqrt{\frac{d_c}{d_a^2}}\
R^{aa}_c\ 
\vcenter{\hbox{\includegraphics[page=34, scale=0.45]{BraidingFigures-cropped.pdf}}}
\\
=&\,
\sum_c
R^{aa}_c\ 
\vcenter{\hbox{\includegraphics[page=18, scale=0.45]{BraidingFigures-cropped.pdf}}}
=\sum_c R^{aa}_c d_c
\end{split}
\label{eq: computation topo spin}
\ee

Combining Eqs.~\eqref{eq: def pictorially topo spin} and~\eqref{eq: computation topo spin}, we obtain the definition of $\theta_a$ in terms of the $R$ symbol
\be
\theta_a:=\frac{1}{d_a}
 \sum_c d_c \mathrm{Tr}[R_c^{aa}].
\ee
Here, the matrix trace Tr is relevant for theories with multiplicities, in which case $R_c^{aa}$ is a square matrix with dimension equal to the multiplicity of the fusion channel $c$ of $a\times a$.

Another identity, which requires bending to be proven, is
 \beq
R_1^{\bar{a}a} =\theta_a^* \chi_a
\eneq 
where $\chi_a$ is the Frobenius Schur indicator. 

\subsubsection{Ribbon equation}

We are now in good shape to prove an important identity relating topological spin to the $R$ matrices that is called the ribbon equation
\beq
R_c^{ab} R_c^{ba} = \frac{\theta_c}{\theta_a \theta_b} I,
\label{eq: ribbon equation}
\eneq 
where $I$ is the identity element in the space $V_c^{ab}$. 
Physically, it equates the operation of twisting each worldline in a splitting diagram by $2\pi$ with the operation of braiding the splitting products. 
We can prove the ribbon equation via diagrammatic manipulations. For that, we evaluate the same diagram in two different ways. For one, we can use the topological spin of the split particle $c$
\be
\begin{split}
\vcenter{\hbox{\includegraphics[page=35, scale=0.45]{BraidingFigures-cropped.pdf}}}
=&\,
\vcenter{\hbox{\includegraphics[page=36, scale=0.45]{BraidingFigures-cropped.pdf}}}
=
\theta_c\ 
\vcenter{\hbox{\includegraphics[page=37, scale=0.45]{BraidingFigures-cropped.pdf}}}
\end{split}
\label{eq: Ribbon proof 1}
\ee
On the other hand, we can use a combination of topological spins and $R$ moves
\be
\begin{split}
\vcenter{\hbox{\includegraphics[page=35, scale=0.45]{BraidingFigures-cropped.pdf}}}
=&\,
\theta_b\
\vcenter{\hbox{\includegraphics[page=38, scale=0.45]{BraidingFigures-cropped.pdf}}}
=
\theta_b\ 
\vcenter{\hbox{\includegraphics[page=39, scale=0.45]{BraidingFigures-cropped.pdf}}}
=
\theta_a\theta_b\ 
\vcenter{\hbox{\includegraphics[page=40, scale=0.45]{BraidingFigures-cropped.pdf}}}
=
\theta_a\theta_b\, R^{ab}_c\ 
\vcenter{\hbox{\includegraphics[page=41, scale=0.45]{BraidingFigures-cropped.pdf}}}
\\
=&\,
\theta_a\theta_b\, R^{ab}_c\,R^{ba}_c\ 
\vcenter{\hbox{\includegraphics[page=37, scale=0.45]{BraidingFigures-cropped.pdf}}}.
\end{split}
\label{eq: Ribbon proof 2}
\ee
Together, Eq.~\eqref{eq: Ribbon proof 1} and Eq.~\eqref{eq: Ribbon proof 2} yield the ribbon equation~\eqref{eq: ribbon equation}.

\subsubsection{Vafa's Theorem}

An important theorem relating the topological spin and the structure constants $N_{ab}^c$ is Vafa's theorem.~\cite{Vafa} It shows that the topological spin is a rational number. We will not derive it here but point the reader to Ref.~\onlinecite{Kitaev05} for an easy derivation. Vafa's theorem proceeds by writing down in matrix form the two hexagon equations, one for $R$ and one for $R^{-1}$,  dividing them and using $R_c^{ab} R_c^{ba} = \theta_c/(\theta_a \theta_b) I$ to obtain
\beq
\prod_c \left(\frac{\theta_c}{\theta_a \theta_b}\right)^{N_{ab}^c N_{cd}^e} 
\prod_f \left(\frac{\theta_f}{\theta_a \theta_d}\right)^{N_{bf}^e N_{a d}^f} 
= \prod_r \left(\frac{\theta_e}{\theta_a \theta_r}\right)^{N^r_{ b d} N^e_{a r} }.
\label{eq: vafa}
\eneq
As an example, we will later use Vafa's theorem to deduce the spin of the $\sigma$ particle in the Ising TQFT.

\subsection{Modular matrices}

In the last Sections, we have constructed the fusion and braiding category from the $F$ and $R$ moves and deduced the universal data (quantum dimensions $d_a$ and topological spins $\theta_a$) that characterizes the anyons in this category. 

For TQFT in physical systems, i.e., quantum liquid ground states of matter, 
it is important to address how the universal information about the TQFT can in general be accessed. We know that topological properties cannot be accessed by local measurements. 
In contrast, it turns out that there is a set of global measurements that can be used, namely the action of automorphisms on the manifold over which the system is defined. Automorphisms are transformations that map the manifold back to itself and form the so-called mapping class group of the manifold. Here, we will explore the most standard case, namely $(2+1)$-dimensional systems with periodic boundary conditions, in which case the manifold is a torus. Automorphisms on the torus form the modular group, which has two generators. The first generator $S$ exchanges the two coordinate axes. The second generator $T$ changes the angle between the coordinate axes.
 
A TQFT on the torus exhibits a topological ground state degeneracy, where the number of ground states is equal to the number of anyons in the theory (including the identity). It is the representation of the $S$ and $T$ operations in this ground-state manifold that reveals information about the nature of the TQFT. We call these representations the $S$ and $T$ matrices. Instead of deriving the $S$ and $T$ matrices from the action of the respective transformations, we will simply give their definitions within the TQFT here and subsequently explore their properties (see Ref.~\onlinecite{Mulligan} for a more complete discussion of the connection).

\subsubsection{ The $S$ matrix}

In the TQFT, the $S$ matrix is defined diagrammatically as
\be
S_{ab}
:=
\frac{1}{D}
\vcenter{\hbox{\includegraphics[page=42, scale=0.45]{BraidingFigures-cropped.pdf}}},
\label{eq: def S matrix}
\ee
where $D$ is the total quantum dimension of the TQFT
\be
D=\sqrt{\sum_a d_a^2}.
\ee
(We choose the same convention for the diagrammatic definition of $S_{ab}$ here as in Ref.~\cite{Bonderson07} which is different from the one chosen in Ref.~\cite{Kitaev05})
We can relate it to the topological spins and fusion coefficients via the following diagrammatic manipulations
\be
\begin{split}
\vcenter{\hbox{\includegraphics[page=42, scale=0.45]{BraidingFigures-cropped.pdf}}}
=&\,
\sum_{c,\mu}
\sqrt{\frac{d_c}{d_a d_b}}
\vcenter{\hbox{\includegraphics[page=43, scale=0.45]{BraidingFigures-cropped.pdf}}}
=
\sum_{c,\mu,\nu}
\sqrt{\frac{d_c}{d_a d_b}}
\left[R^{ab}_c\right]_{\mu\nu}\left[R^{ba}_{c}\right]_{\nu\mu}\ 
\vcenter{\hbox{\includegraphics[page=45, scale=0.45]{BraidingFigures-cropped.pdf}}}
\\
=&\,
\sum_{c,\mu,\nu}
\left[R^{ab}_c\right]_{\mu\nu}\left[R^{ba}_{c}\right]_{\nu\mu}\,
\sqrt{\frac{d_c}{d_a d_b}}\ 
\vcenter{\hbox{\includegraphics[page=46, scale=0.45]{BraidingFigures-cropped.pdf}}}
\\
=&\,
\sum_{c}
N_{ab}^c\mathrm{Tr}\left(R^{ab}_cR^{ba}_{c}\right) d_c.
\end{split}
\ee
From here, we use the ribbon Eq.~\eqref{eq: ribbon equation} and Eq.~\eqref{eq: def S matrix}
to arrive at the final expression for the $S$ matrix
\be
S_{ab}=\frac{1}{D}\sum_c N^c_{ab} \frac{\theta_c}{\theta_a \theta_b} \, d_c.
\label{eq: def S matrix equation}
\ee

\subsubsection{Verlinde Formula}

We now derive a fundamental formula in both TQFT and in conformal field theory. This formula relates the $S$-matrix to the fusion coefficients. It allows us to find the set of braiding phases among the anyons, but not always the topological spin of every particle. 

We start by showing that 
\be
S_{1x} = \frac{d_x}{D}.
\label{eq: S1x}
\ee 
To prove this, we first observe that 
\be
\vcenter{\hbox{\includegraphics[page=47, scale=0.45]{BraidingFigures-cropped.pdf}}}\ 
=
\frac{S_{ax}}{S_{1x}}\ 
\vcenter{\hbox{\includegraphics[page=48, scale=0.45]{BraidingFigures-cropped.pdf}}},
\label{eq: relation for Verlinde formula}
\ee
holds, because we can close the $x$ worldlines in this diagram to obtain amplitudes and an identity loop can be added without changing the value of the diagram. 
Relation~\eqref{eq: relation for Verlinde formula} can be used to obtain the line of equalities:
\be
D\,S_{ax}=
\vcenter{\hbox{\includegraphics[page=49, scale=0.45]{BraidingFigures-cropped.pdf}}}
=
\frac{S_{ax}}{S_{1x}}
\vcenter{\hbox{\includegraphics[page=50, scale=0.45]{BraidingFigures-cropped.pdf}}}
=
\frac{S_{ax}}{S_{1x}}
d_x\, ,
\ee
which yields Eq.~\eqref{eq: S1x}.
Equation~\eqref{eq: S1x} says that the first column of the $S$ matrix contains only positive numbers, the quantum dimensions of the theory.

The next step is to derive the all-important Verlinde formula. 
Staring from two copies of Eq.~\eqref{eq: relation for Verlinde formula}, we can perform the following set of diagrammatic manipulations
\be
\begin{split}
\frac{S_{ax}}{S_{1x}}
\frac{S_{bx}}{S_{1x}}\ 
\vcenter{\hbox{\includegraphics[page=48, scale=0.45]{BraidingFigures-cropped.pdf}}}
=&\,\ 
\vcenter{\hbox{\includegraphics[page=51, scale=0.45]{BraidingFigures-cropped.pdf}}}\ 
=
\sum_c
\sqrt{\frac{d_c}{d_a d_b}}\ 
\vcenter{\hbox{\includegraphics[page=52, scale=0.45]{BraidingFigures-cropped.pdf}}}\ 
=
\sum_c
\sqrt{\frac{d_c}{d_a d_b}}\ 
\vcenter{\hbox{\includegraphics[page=53, scale=0.45]{BraidingFigures-cropped.pdf}}}
\\
=&\,
\sum_c
N_{ab}^c
\vcenter{\hbox{\includegraphics[page=54, scale=0.45]{BraidingFigures-cropped.pdf}}}
=\sum_c
N_{ab}^c
\frac{S_{cx}}{S_{1x}}\,
\vcenter{\hbox{\includegraphics[page=48, scale=0.45]{BraidingFigures-cropped.pdf}}}.
\end{split}
\ee
yielding the Verlinde formula
\beq
\sum_c N_{ab}^c S_{cx} = S_{bx} \frac{S_{ax}}{S_{1x}}.
\eneq
This is a remarkable equality, which takes again the form of an eigenvalue equation for the fusion matrices $N_a$. 
We had already encountered an eigenvalue equation for $N_a$ when solving for the quantum dimensions, i.e., the first column of the $S$ matrix. 
The Verlinde formula says that the $S$ matrix contains both the eigenvectors and the eigenvalues of the $(N_a)_{bc}$ matrices. 
If the $S$ matrix is unitary (and thus has an inverse), we call the theory a unitary modular category.
In this case we have a simple relation between the fusion coefficients and the $S$ matrix
\beq
N_a = S D_a S^{-1},
\eneq 
where we have defined the diagonal matrices
\beq
(D_a)_{mn} = S_{am}/S_{1m} \delta_{mn}.
\eneq
This also implies that all the $N_a$ matrices are commuting [which we already knew from Eq.~\eqref{eq: N commute}], as they are diagonalized by the same eigenvectors.

\subsubsection{Obstruction for theories with multiplicities}

In this Subsection, we want to illustrate how we can use the structure of the braiding category to discard certain fusion 
categories as unphysical (or at least not unitary). As an example, we ask which theories are possible with only two particles, the identity 1 and a particle $s$. For $s$ to have an inverse, all possible fusion rules 
\be
s\times s=1+ m\, s
\ee
are labelled by a nonnegative integer $m$. The case $m=0$ is the semion TQFT of Eq.~\eqref{eq: semion fusion rules}. For $m=1$ we have the non-Abelian Fibonacci fusion rules. 

We would like to answer the question for which $m>1$ we can define a consistent modular braiding category. 
Observe that the fusion matrix 
\be
N_s=
\begin{pmatrix}
0&1\\
1&m
\end{pmatrix}
\ee
yields the quantum dimension
\be
d_s=\frac{m+\sqrt{m^2+4}}{2}.
\ee

From the definition of the $S$ matrix, we have that
\beq
S_{s1} = \frac{1}{D} \sum_{c} N_{s1}^c \frac{\theta_c}{\theta_s} d_c  = \frac{1}{D} d_s = S_{1s}; \;\;\;\;\;
S_{11} = \frac{1}{D};\;\;\;\;\;
S_{ss} = \frac{1}{D} \frac{1}{\theta_s^2} (1 + m \theta_s d_s).
\eneq 
It is already possible to see that something will go wrong for large enough $m$ if we demand a unitary theory, i.e., a theory with a unitary $S$ matrix.  For a unitary matrix, all the matrix elements have to be less or equal to $1$. When $m$ is large, $d_s$ is proportional to $m$ (and so is $D$) but the $S_{ss}$ matrix element has a $m d_s/ D$ which is proportional to $m$ in the large $m$ limit. Hence we forsake unitarity.  

Lets us calculate the exact $m$ where unitarity breaks down. Imposing unitarity of the $S$ matrix yields
\be
\begin{split}
 S^\dagger S
=&\, \frac{1}{D^2} \left(
\begin{array}{cc}
 D^2 & d_s \left(1+  \frac{1 + m \theta_s d_s}{\theta_s^2}   \right)\\
 d_s\left(1+ \frac{1 + m \theta_s^* d_s}{\left(\theta_s^*\right)^2}\right)  & d_s^2+ ( 1 + m \theta_s^* d_s)( 1 + m \theta_s d_s)
\end{array}
\right),
\end{split}
\ee
and we find
\beq
 ( 1 + m \theta_s^* d_s)( 1 + m \theta_s d_s)=1,
\eneq 
which reduces to 
\beq
\theta_s+ \theta_s^* + m d_s =0,
\eneq 
where we have used that the value of $d_s$ is positive. Now, $m d_s$ is a positive number growing with $m$, while $\theta_s$ is a phase so the sum with its conjugate cannot be smaller than $-2$. Hence 
\beq
m d_s \le 2 \;\;  \rightarrow \;\;   m^2+ m \sqrt{m^2+ 4} \le 4 
\eneq 
with the solutions $m=0$ and $m=1$. We can see that already $m=2$ gives a left side too large. 
We conclude that the semion and the Fibonacci TQFT are the only allowed modular unitary theories with one nontrivial anyon $s$.

\subsubsection{ The $T$ matrix}

The $T$-matrix is diagonal and simply given by
\be
T_{ab}=\theta_a\delta_{ab}.
\ee

\subsection{Examples: The 16-fold way revisited}

To motivate our study of TQFTs and as an example of topologically ordered phases, we have 
studied in Sec.~\ref{sec: Kitaev 16-fold} Kitaev's 16-fold way of classifying topological superconductors as gauge theories from their bulk properties. Equipped with category theory understanding of TQFTs, we now want to revisit this classification, as it provides us with several examples of TQFTs. In particular, we want to characterize the theories from their $S$ and $T$ matrices. 

\subsubsection{Case: $\mathsf{C}^{(1)}$ odd}

If we couple an odd number of layers of spinless chiral $p$-wave superconductors, the core of each vortex still carries
an unpaired Majorana state. For that reason, the vortices will have Ising fusion rules~\eqref{eq: Ising} in this case. Using the Verlinde formula, we can compute the $S$ matrix in the basis (1, $\sigma$, $\psi$)
\be
S=
\begin{pmatrix}
1&\sqrt{2}&1\\
\sqrt{2}&0&-\sqrt{2}\\
1&-\sqrt{2}&1
\end{pmatrix}.
\ee
If, in contrast, we compute the $S$ matrix via its definition~\eqref{eq: def S matrix equation}, we find in particular for the ($\sigma$,$\sigma$) matrix element
\be
S_{\sigma\sigma}=\frac{1+\theta_\psi}{2\theta_\sigma^2}.
\ee
For this to vanish, we conclude that $\theta_\psi=-1$, that is, $\psi$ is a fermion. 
We cannot obtain $\theta_\sigma$ from similar relations using only the $S$ matrix. Rather, we can use Vafa's theorem~\eqref{eq: vafa} to deduce the spin of the $\sigma$ particle. If we take $a=b=d=e =\sigma$ we find
\beq
\prod_c (\frac{\theta_c}{\theta_\sigma \theta_\sigma})^{N_{\sigma \sigma}^c N_{c\sigma}^\sigma} \prod_f (\frac{\theta_f}{\theta_\sigma \theta_\sigma})^{N_{\sigma f}^\sigma N_{\sigma \sigma}^f} = \prod_r (\frac{\theta_\sigma}{\theta_\sigma \theta_r})^{N^r_{ \sigma \sigma} N^\sigma_{\sigma r} } 
\eneq or
\beq
\frac{1}{\theta_\sigma^2} \frac{\theta_\psi}{\theta_\sigma^2} \frac{1}{\theta_\sigma^2} \frac{\theta_\psi}{\theta_\sigma^2} = \frac{1}{\theta_\psi},
\eneq 
which gives 
\beq 
\theta_\sigma^8 = \theta_\psi^3.
\eneq 
Since $\theta_\psi=-1$, we have
\beq 
\theta_\sigma^8 = \theta_\psi=-1,
\label{eq: minus for spin sigma}
\eneq 
i.e., the phase of the topological spin of the Majorana is an odd-integer multiple of $1/16$.
We can thus discriminate $8$ different TQFTs with Ising fusion rules by the values
\be
\theta_\sigma=e^{2\pi\mathrm{i}\,\frac{\mathsf{C}^{(1)}}{16}},
\ee
where $\mathsf{C}^{(1)}$ is the [odd, as needed by Eq.~\eqref{eq: minus for spin sigma}] Chern number or the number of stacked spinless chiral $p$-wave superconductors. 

\subsubsection{Case $\mathsf{C}^{(1)}=2$ mod 4}

If we stack $\mathsf{C}^{(1)}=2$ mod $4$ layers of chiral $p$-wave superconductors, we had argued before, heuristically,  that the system can be described 
as one species of spinless Dirac fermions with Chern number $\tilde{\mathsf{C}}^{(1)}=\mathsf{C}^{(1)}/2$. 
In this system the $2\pi$ flux $\psi$ binds odd integer charge $\tilde{\mathsf{C}}^{(1)}$, that is, the $2\pi$ flux is a fermion $\psi$.
However, unlike in the quantum Hall effect, the superconducting $\pi$ or $(-\pi)$ fluxes are allowed topological excitations that bind half-integer charge. Let us denote the $\pi$ flux with charge $\tilde{\mathsf{C}}^{(1)}/2$ by $a$. Fusing such a $\pi$ flux with the fermion (or $2\pi$ flux) gives another excitation, a $(-\pi)$ flux, with $3\tilde{\mathsf{C}}^{(1)}/2$ charge that we call $\tilde{a}$. (We note that $4\pi$ flux is identified with zero flux, as this corresponds to a charge $2\tilde{\mathsf{C}}^{(1)}$ object, i.e.,  $\tilde{\mathsf{C}}^{(1)}$ Cooper pairs that can be absorbed by the condensate.) Two fluxes of either type $a$ or $\tilde{a}$ thus fuse into $\psi$. This motivates the following fusion rules
\be
\begin{split}
a\times  \tilde{a}=1,
\qquad
a\times\psi=\tilde{a},
\qquad
\tilde{a}\times\psi=a,\\
a\times a=  \tilde{a}\times  \tilde{a}=\psi,
\qquad
\psi\times\psi=1.
\end{split}
\ee
These fusion rules are Abelian, so that the quantum dimensions are $d_1=d_a=d_{\tilde{a}}=d_\psi=1$.
Given the fusion matrices, we can compute the $S$ matrix as the matrix of their simultaneous eigenvectors. In the basis $(1,a,\tilde{a},\psi)$ it can take one of two forms

\be
S^{(1)}=\frac12
\begin{pmatrix}
1&1&1&1\\
1&-\mathrm{i}&\mathrm{i}&-1\\
1&\mathrm{i}&-\mathrm{i}&-1\\
1&-1&-1&1
\end{pmatrix},
\qquad
S^{(2)}=\frac12
\begin{pmatrix}
1&1&1&1\\
1&\mathrm{i}&-\mathrm{i}&-1\\
1&-\mathrm{i}&\mathrm{i}&-1\\
1&-1&-1&1
\end{pmatrix}.
\ee
From $S_{aa}=S_{\tilde{a}\tilde{a}}=\pm\mathrm{i}/2$, we conclude
\be
\theta_a^2=\theta_{\tilde{a}}^2=\mp\mathrm{i},
\ee
while $S_{a\psi}=S_{\tilde{a}\psi}=-1/2$ gives
\be
\theta_a=\theta_{\tilde{a}},
\ee
 with the help of Eq.~\eqref{eq: def S matrix equation}.
We conclude that there are four possible theories with topological spins
\be
\theta_a=\theta_{\tilde{a}}=e^{2\pi\mathrm{i}\,\frac{\mathsf{C}^{(1)}}{16}}.
\ee

\subsubsection{Case $\mathsf{C}^{(1)}=0$ mod 4}
 If we couple 4 layers of chiral $p$-wave superconductors, the $\pi$ superconducting vortices (let us denote them $e$) bind a full electron charge. However, the electrons exist also as free fermionic quasiparticles $f$ in the theory. Hence, the fusion of 
$e$ with $f$ should yield a new excitation $m$ that is a $\pi$ vortex stripped of its charge. Together, $e$, $m$, and $f$ obey the toric code fusion rules Eq.~\eqref{eq: Toric code}. The $S$ matrix of the theory in the basis $(1,e,m,f)$ can take one of two forms
\be
S^{(1)}=\frac12
\begin{pmatrix}
1&1&1&1\\
1&-1&1&-1\\
1&1&-1&-1\\
1&-1&-1&1\\
\end{pmatrix},
\qquad
S^{(2)}=\frac12
\begin{pmatrix}
1&1&1&1\\
1&1&-1&-1\\
1&-1&1&-1\\
1&-1&-1&1\\
\end{pmatrix}
.
\ee
If we assume that $\theta_f=-1$, we find from $S^{(1)}$ that $\theta_e^2=\theta_m^2=\theta_e\theta_m=-1$, while
the theory with $S^{(2)}$ has $\theta_e^2=\theta_m^2=\theta_e\theta_m=1$. In total, we have four possibilities
\be
\theta_e=\theta_{m}=e^{2\pi\mathrm{i}\,\frac{\mathsf{C}^{(1)}}{16}}
\ee
with $\mathsf{C}^{(1)}=0\,\mathrm{mod}\,4$ as allowed theories. This concludes Kitaev's 16-fold way.~\cite{Kitaev05}

\section*{Acknowledgements}

The authors were supported by NSF CAREER DMR-095242, ONR-N00014-11-1-0635, MURI-130-6082, MERSEC Grant, the Packard Foundation, and Keck grant.

\end{document}